\documentclass[10pt,aps,prb,twocolumn,amsmath,amssymb,superscriptaddress,nobibnotes]{revtex4-1}
\usepackage{amsmath}
\usepackage{graphicx}
\usepackage{color}
\usepackage{multirow}
\usepackage[caption = false]{subfig}
\usepackage{array}
\usepackage{threeparttable}
\usepackage[version=4]{mhchem}
\usepackage{mathtools}
\usepackage{siunitx}
\usepackage{centernot}
\usepackage{tikz}
\usepackage{hyperref}
\usepackage{xurl}
\hypersetup{breaklinks=true}
\usepackage[capitalise]{cleveref}
\tikzset{font={\fontsize{11pt}{12}\selectfont}}
\usetikzlibrary{shapes,arrows}
\usepackage{lipsum}
\usepackage[normalem]{ulem}
\makeatletter

\renewcommand*{\p@subsection}{}

\renewcommand*{\p@subsubsection}{}
\makeatother

\usepackage{hyperref}
\hypersetup{
	linkcolor=blue,          
	citecolor=blue,        
	urlcolor=blue,           
}

\usepackage{acro}
\DeclareAcronym{ahc}{
short = AHC, 
long =  Allen-Heine-Cardona,
short-indefinite = an,
long-indefinite = an,
single-style = long,
}
\DeclareAcronym{aimd}{
short = AIMD, 
long =  \textit{ab initio} molecular dynamics,
short-indefinite = an,
long-indefinite = an,
single-style = long,
}
\DeclareAcronym{bte}{
short = BTE, 
long =  Boltzmann transport equation,
short-indefinite = a,
long-indefinite = a,
single-style = long
}
\DeclareAcronym{bse}{
short = BSE, 
long =  Bethe–Salpeter equation,
short-indefinite = a,
long-indefinite = a,
single-style = long
}
\DeclareAcronym{dft}{
short = DFT, 
long = density functional theory,
short-indefinite = a,
long-indefinite = a,
single-style = long
}
\DeclareAcronym{dfpt}{
short = DFPT, 
long = density functional perturbation theory,
short-indefinite = a,
long-indefinite = a,
single-style = long
}
\DeclareAcronym{dmft}{
short = DMFT, 
long = dynamical mean field theory,
short-indefinite = a,
long-indefinite = a,
single-style = long
}
\DeclareAcronym{dmpt}{
short = DMPT, 
long = density matrix perturbation theory,
short-indefinite = a,
long-indefinite = a,
single-style = long
}
\DeclareAcronym{eda}{
short = EDA, 
long = energy decomposition analysis,
short-indefinite = an,
long-indefinite = an,
single-style = long
}
\DeclareAcronym{epi}{
short = EPI, 
long = electron-phonon interaction,
short-indefinite = an,
long-indefinite = an,
single-style = long
}
\DeclareAcronym{gga}{
short = GGA, 
long = generalized gradient approximation,
short-indefinite = a,
long-indefinite = a,
single-style = long
}
\DeclareAcronym{gto}{
short = GTO, 
long = Gaussian-type orbital,
short-indefinite = a,
long-indefinite = a,
single-style = long
}
\DeclareAcronym{lda}{
short = LDA, 
long = local density approximation,
short-indefinite = an,
long-indefinite = a,
single-style = long
}
\DeclareAcronym{md}{
short = MD, 
long = molecular dynamics,
short-indefinite = a,
long-indefinite = a,
single-style = long
}
\DeclareAcronym{pes}{
short = PES, 
long =  potential energy surface,
short-indefinite = a,
long-indefinite = a,
single-style = long
}
\DeclareAcronym{pw}{
short = PW, 
long =  plane wave,
short-indefinite = a,
long-indefinite = a,
single-style = long
}
\DeclareAcronym{sc}{
short = SC, 
long =  strong coupling,
short-indefinite = a,
long-indefinite = a,
single-style = long
}
\DeclareAcronym{vdmft}{
short = VDMFT, 
long =  vibrational \acs{dmft},
short-indefinite = a,
long-indefinite = a,
single-style = long
}
\DeclareAcronym{wc}{
short = WC, 
long =  weak coupling,
short-indefinite = a,
long-indefinite = a,
single-style = long
}
\DeclareAcronym{wf}{
short = WF, 
long = Wannier function,
short-indefinite = a,
long-indefinite = a,
single-style = long
}

\newcommand{\Harvard}{\affiliation{Department of Chemistry and Chemical Biology, Harvard University, Cambridge, MA, USA}}

\begin{document}
\title{Condensed-Phase Quantum Chemistry}
\author{Paul J. Robinson}
\thanks{These authors contributed equally to this work}
\author{Adam Rettig}
\thanks{These authors contributed equally to this work}
\author{Hieu Q. Dinh}
\thanks{These authors contributed equally to this work}
\author{Meng-Fu Chen}
\thanks{These authors contributed equally to this work}
\author{Joonho Lee}
\email{joonholee@g.harvard.edu}
\Harvard

\begin{abstract}
Molecular quantum chemistry has seen enormous progress 
in the last few decades thanks to the 
more advanced and sophisticated numerical techniques and computing power. 
Following the recent interest in extending these capabilities to condensed-phase problems, 
we summarize basic knowledge of condensed-phase quantum chemistry 
for ones with experience in molecular quantum chemistry. 
We highlight recent efforts 
in this direction, including solving the electron repulsion integrals bottleneck and 
implementing hybrid density functional theory and wavefunction methods, and lattice dynamics for periodic systems within atom-centered basis sets. Many computational techniques presented
here are inspired
by the extensive method developments 
rooted in quantum chemistry.
In this Focus Article, we selectively focus on the computational techniques rooted in molecular quantum chemistry, emphasize some challenges, and point out open questions.
We hope our perspectives will encourage researchers 
to pursue 
this exciting and promising research avenue. 

\end{abstract}
\maketitle

\section{Introduction}
Quantum chemical simulations of more complex and larger systems 
have been an 
ever-growing endeavor
in quantum chemistry.~\cite{Head-Gordon2008Apr}
In the very early years of quantum chemistry, 
the focus was on leveraging the principle of quantum mechanics 
to elucidate the spectroscopic features of small molecules.~\cite{Pople1999Jul} 
The emergence of computer programs in the 1950s 
has enabled quantum chemistry calculations to have more predictive power 
over previously intractable systems. Since then, the field has grown rapidly due to 
fast-paced advances in computing architectures 
and the introduction of many algorithmic developments.
For example, liquid-state systems \cite{Gross2022Jan}, solid-state systems,~\cite{Dovesi2005Jan,Hirata2009} macromolecules, \cite{Hirata2009} 
and biomolecules \cite{albuquerque2021quantum} have all been simulated using quantum chemistry methods. 
Other methods based on quantum mechanical modeling, 
such as quantum mechanics/molecular mechanics (QM/MM), allow for ever larger multi-scale modeling.~\cite{Keith2021Jul} 
New computing paradigms, such as quantum computing \cite{cao2019quantum} and machine learning, \cite{westermayr2021perspective} also promise to accelerate the traditional approach to the electronic structure method.  As the cost of quantum chemical simulation continues to drop, quantum chemistry methods have become indispensable in understanding chemical processes and guiding experimental research.\cite{norskov2009towards} 

Recently, there has been a surge of interest in the quantum chemistry community to leverage decades of method development in molecular systems to extended/solid-state systems. Recent works 
include
hybrid density functional theory,~\cite{Qin2020Jun,Lee2022Dec,Sharma2022Nov,rettig_even_2023,Bussy2024Feb}
random phase approximation,~\cite{Ren2012Nov,Grundei2017Feb,Yeh2023Sep}
M{\o}ller-Plesset perturbation theory,~\cite{Pisani2008Oct, del2012second}
quantum Monte Carlo,~\cite{Foulkes2001Jan,Booth2013Jan,Motta2019Jul,Malone2020Jul}
and coupled-cluster theory.~\cite{McClain2017Mar,Gruber2018May,Zhang2019Jun}
The use of molecular quantum chemistry tools, primarily developed based on atom-centered basis sets such as Gaussian orbitals, offers new opportunities in tackling condensed-phase problems with the hope of achieving better accuracy at a cheaper cost. However, extending the molecular code to simulate condensed-phase problems demands additional consideration. 

\begin{figure}
    \centering
    \includegraphics{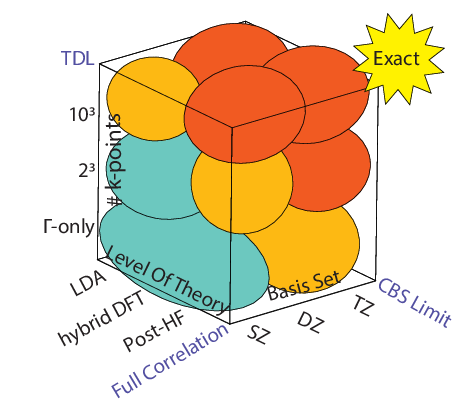}
    \caption{Three dimensional Pople diagram illustrating the general trend of computational expense for solid-state calculations. The three axes consider the approach towards the thermodynamic limit (TDL), the limit of full correlation, and the complete basis set (CBS) limit. The green shaded areas indicate where calculations are currently routine, orange regions indicate calculations that may be possible on small to intermediate-sized systems, and red indicates the most expensive calculations.}
    \label{fig:Pople_diagram}
\end{figure}

Aside from the language barrier of solid-state physics,~\cite{Hoffmann1987Sep} practical implementation requires efficient handling of the formally infinite size of solid-state systems. In practice, this is modeled through imposing translational symmetry, often handled via $\mathbf{k}$ point sampling \cite{monkhorst1976special} in the reciprocal space. The incomplete sampling introduces finite-size error \cite{Drummond2008Jun,xing2024unified} independent of the basis set incompleteness error (BSIE) one encounters in molecular cases. Therefore, converging the energy calculation for solid-state systems requires sampling many $\textbf{k}$ points (or increasing the size of supercell) (\cref{fig:Pople_diagram}), which implies significant increases in computation time. We shall cover the scaling of the methods in greater detail later. Furthermore, complex-valued orbitals can also complicate the extension because many production-level implementations have assumed real-valued orbitals and integrals and optimized for the simplified equation under that assumption. Hence, optimization for implementing more general complex-valued equations is essential for production-level solid-state simulations.  

In this Focus Article, we aim to lay out the fundamental concepts required to translate knowledge from molecular quantum chemistry to condensed-phase quantum chemistry. We aim to summarize some of the more recent progress commensurate with the surge of recent interest. Moreover, we will highlight some of the recent developments that bridge these two areas, as well as open questions and future opportunities.
Nonetheless, 
this field has seen enormous progress
over the past two decades before this recurring interest in our field.
We refer interested readers to Refs. \citenum{Foulkes2001Jan,Janesko2009,Ren2012Nov,Zhang2019Jun} for
topical reviews that are more specific and detailed than this Focus Article. We also refer readers to the book by Evarestov~\cite{evarestov2007quantum} as an additional resource.

This paper is organized as follows:
(1) we will first discuss how Gaussian-type orbitals are used in the solid-state context, (2) will then highlight some of the more widely used algorithms for evaluating electron repulsion integrals (ERIs) for solids, (3) will discuss hybrid density functionals, (4) correlated wavefunction methods, (5) lattice dynamics, and (6) will conclude.

\section{Gaussian-type orbital basis sets}
Atom-centered Gaussian-type orbital (GTO) basis functions are 
the {\it de facto} choice of discretization for molecular
quantum chemistry calculations. \cite{huzinaga2012gaussian, dunning1977gaussian}
In their Cartesian form, (unnormalized) primitive GTO basis functions are given by
\begin{align}\nonumber
\xi(\alpha, \mathbf r) =&
(x-X)^{l_x}
(y-Y)^{l_y}
(z-Z)^{l_z}
\\
&\times e^{-\alpha((x-X)^2+(y-Y)^2+(z-Z)^2)}
\label{eq:primitive}
\end{align}
where $\alpha$ is the exponent of the Gaussian function, $l_x, l_y, l_z$ are angular momentum of each cartesian component, and $X, Y, Z$ define the location of the atom center. 
In practice, we work with contracted GTOs that take a linear combination of primitive GTOs in \cref{eq:primitive} with the same center,
\begin{equation}
\phi(\mathbf r) = 
\sum_i c_i \xi(\alpha_i, \mathbf r)
\end{equation}
where $c_i$ is the contraction coefficient, which includes the overall normalization constant.

Due to the lack of derivative discontinuity at nuclear centers,
GTOs do not obey
the Kato's cusp condition,~\cite{kato1957eigenfunctions} and
often struggle to
reach the complete basis set (CBS) limit.
Nonetheless, several series of Gaussian basis sets have been developed for molecular application and are routinely used in molecular calculations.~\cite{pritchard2019new} Some examples are split-valence basis sets \cite{schafer1992fully, schafer1994fully} by Ahlrich, correlation consistent basis sets \cite{dunning1989gaussian, woon1993gaussian} by Dunning, polarization consistent basis sets \cite{jensen2001polarization} by Jensen, and atomic natural orbitals (ANO) \cite{almlof1991atomic} by \text { Almlöf } and Taylor. Typically, these basis sets are augmented with polarization functions (i.e., high-angular momentum functions) or diffuse functions. These basis sets have been benchmarked extensively for molecules at different levels of theory. \cite{neese2011revisiting, kirschner2020performance}

\begin{figure}
    \centering
\includegraphics[width=0.99\columnwidth]{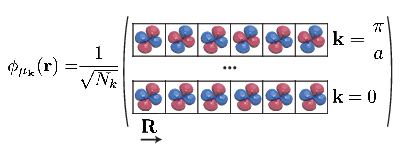}
    \caption{Illustration of \cref{eq: Bloch equation} demonstrating the nature of orbitals at the different k-points of a 1D lattice. Here $a$ is the lattice constant.}
    \label{fig:figure_eq}
\end{figure}

For solid-state applications, one works with Bloch crystalline orbitals,
\begin{equation} \label{eq: Bloch equation}
\phi_{\mu_{\mathbf k}}(\mathbf r)
=
\frac{1}{\sqrt{N_k}}
\sum_{\mathbf R}
\exp(i\mathbf k \cdot \mathbf R) \phi_\mu(\mathbf r - \mathbf R)
\end{equation}
where $N_k$ is the number of sampled k-points, $\mathbf k$ is the crystalline momentum, $\mathbf R$ is a displacement vector pointing to each unit cell, and the summation over $\mathbf R$ is usually referred to as the lattice summation.
This central equation is pictorially illustrated in Fig. \ref{fig:figure_eq}.
One can easily see that \cref{eq: Bloch equation} satisfies Bloch's theorem,
$
\phi_{\mu_{\mathbf k}}(\mathbf r+\mathbf R)
=
\phi_{\mu_{\mathbf k}}(\mathbf r) \exp(i\mathbf k\cdot\mathbf R).
$
These atomic Bloch crystalline orbitals are used to obtain band wavefunctions,
$\{\psi_{p_{\mathbf k}}\}$,
\begin{equation}
    \psi_{p_{\mathbf k}}(\mathbf r)
    =
    \sum_\mu C_{\mu p}^{\mathbf k}
    \phi_{\mu_{\mathbf k}}(\mathbf r),
\end{equation}
where $C_{\mu p}^{\mathbf k}$ is the band coefficient or the molecular orbital coefficient, which normally is determined through a self-consistent field (SCF) procedure. \cite{evarestov2007quantum} We note that in the basis of the crystalline orbitals, many fundamental quantities, including one-body terms (Fock matrix) and the ERIs, become block diagonal. These block diagonal quantities are reminiscent of the point-group symmetry in molecular symmetry and will reduce the computational cost significantly. 

Using the GTO basis sets optimized for molecules in the condensed phase poses significant challenges due to the linear dependency of the basis functions.~\cite{klahn1977completeness, peintinger2013consistent} Especially when the interatomic distance is shorter (commonly found in condensed-phase problems), GTOs become even more overcomplete, creating serious numerical issues when approaching the thermodynamic limit.~\cite{lee2021approaching} Because molecular-optimized GTOs often contain diffuse functions that can extend many unit cells away, this numerical issue could be serious if one uses molecular GTOs for solid-state applications. Therefore, in practice, it is quite common to optimize basis functions with limited diffusivity \cite{} and even penalize the condition number of the overlap \cite{li2021optimized, vandevondele2007gaussian, daga2020gaussian}
\begin{equation}
    \mathcal{L} = E + \gamma \ln \left[\mathrm{cond} (\mathbf{S})\right],
\end{equation}
where $\mathbf{S}$ is the overlap matrix, and $E$ is the correlated (mean-field) energy. The optimization is performed on the Gaussian exponents and the contraction coefficients with small $\gamma$, from $10^{-3}$ to $10^{-4}$. 
Recently, several basis sets, such as the def2 series \cite{daga2020gaussian} and the correlation consistent series, \cite{laun2018consistent, peintinger2013consistent, ye2022correlation} have been reoptimized for use in correlated solid-state calculations. Despite the effort to numerically stabilize the basis functions, one may still run into near-linear dependencies.
This can be handled via canonical orthogonalization \cite{lee2021approaching, szabo1996modern} although the linear dependency threshold will ultimately limit the numerical precision.
Furthermore, the transferability of these basis sets to a wide range of correlated methods is unclear, and more extensive benchmarks are necessary to understand the characteristics of these basic sets.  In addition, how accurate these basis sets are for molecular systems or molecular-like problems still needs to be assessed.

Historically, computational solid-state physics has mainly employed plane wave basis as the fundamental discretization.~\cite{kresse1996efficient}
From a theoretical point of view, this is natural because plane waves automatically satisfy Bloch's theorem. 
From a practical point of view, a plane wave basis provides a simple and numerically well-behaved way to reach the basis set limit without suffering from basis set superposition error (BSSE) when calculating non-covalent interaction energies.
Furthermore, their use in pure density functional theory (DFT) calculations has been made exceptionally fast. \cite{hafner2008ab}

\begin{figure*}
    \centering
    \includegraphics[width=0.6\textwidth]{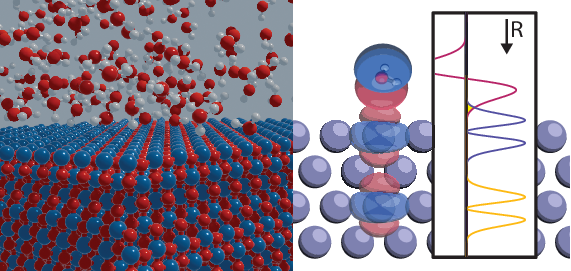}
    \caption{Left: A metal oxide surface in bulk water solution. Right: illustration of the sparsity built into surface-adsorbate interactions when applying a localized Gaussian basis.}
    \label{fig:surface_gaussian}
\end{figure*}

The central motivation for using GTOs over plane waves is the hope that they provide a more compact representation and there is also favorable error cancellation that yields accurate relative energies. Moreover, it offers natural connections to chemical origin without having to perform any localization procedures. See \cref{fig:surface_gaussian} for illustration. The compactness of GTOs may not help to speed up pure functionals.
However, for hybrid functionals and wavefunction methods,
the locality of GTOs could offer significant computational benefits over plane-wave bases.~\cite{Booth2016Aug} 
As we shall see, further computational cost savings can be achieved by leveraging algorithmic advances proposed in the molecular quantum chemistry community. 

\section{Computational Strategies for ERIs}
In molecular quantum chemistry, the computation of electron-repulsion integrals (ERIs) has been highly optimized as it is often the rate-limiting step in Hartree-Fock (HF), hybrid DFT, and some of the simpler correlated wavefunction methods due to the relatively large $\mathcal{O}(N^4)$ number of ERIs, where $N$ is the number of basis functions. If GTOs are employed as the underlying basis set, this asymptotic scaling is reduced to  $\mathcal{O}(N^2)$ because of the locality of the Gaussian functions. \cite{helgaker2013molecular}

For periodic systems, the ERIs are more complicated, 
\begin{align}\nonumber
    &(\mu_{\mathbf{k_1}} \nu_{\mathbf{k_2}} | \lambda_{\mathbf{k_3}} \sigma_{\mathbf{k_4}}) = \\ \label{eq:ERIs}
    &\iint \frac{\phi_{\mu_{\mathbf{k_1}}}(\mathbf{r_1})^* \phi_{\nu_{\mathbf{k_2}}}(\mathbf{r_1}) \phi_{\lambda_{\mathbf{k_3}}}(\mathbf{r_2})^* \phi_{\sigma_{\mathbf{k_4}}}(\mathbf{r_2})}{|\mathbf{r_1} - \mathbf{r_2}|} \mathrm d\mathbf{r_1}\mathrm d\mathbf{r_2}.
\end{align}
For crystalline orbital bases, each basis function has an additional $\mathbf{k}$-point dependence, increasing the size of the already large ERI tensor to $\mathcal{O}(N_k^3 N^4)$, where conservation of momentum can be used to eliminate the dependence on one $\mathbf{k}$ index. Storage of the ERI tensor can be prohibitive for large-basis correlated wavefunction theory calculations.

The second complication arises from the potential divergence in evaluating Coulomb integrals in periodic systems. Upon the lattice summation, the electron repulsion, electron-nuclear, and nuclear-nuclear integrals lead to diverging quantities that exactly cancel each other in the total energy calculation. \cite{ihm1979momentum} The evaluation of these integrals is often done separately in modern electronic structure codes. Therefore, one needs to pay special attention to the handling of the singularity. In this section, we cover several prominent approaches that have been used to evaluate the ERIs efficiently while avoiding the divergence.

\subsection{Gaussian Plane Wave}
The Gaussian plane wave (GPW) approach\cite{lippert_hybrid_1997} is commonly used in many GTO periodic codes due to its speed and ease of implementation.\cite{vandevondele_quickstep_2005,kuhne_cp2k_2020,sun_recent_2020} 
This approach has been popularized by CP2K. \cite{kuhne_cp2k_2020}
This method can be considered a form of density fitting, with plane waves as the auxiliary basis set. One can systematically improve the quality of the fitting procedure by increasing the number of plane waves through the energy cutoff. 

In this approach, the pair density $\rho_{\mu_{\mathbf{k_1}}\nu_{\mathbf{k_2}}}(\mathbf{r}) = (\phi_{\mu_{\mathbf{k_1}}}(\mathbf{r}))^* \phi_{\nu_{\mathbf{k_2}}}(\mathbf{r})$ is expanded in a plane wave basis, where the wavevector of plane waves, $\mathbf G$, is proportional to the reciprocal space lattice vector, $\{\mathbf b_i\}$,
\begin{equation} 
\rho_{\mu_{\mathbf{k_1}}\nu_{\mathbf{k_2}}}(\mathbf r)
=
\sum_{\mathbf G}
\tilde{\rho}_{\mu_{\mathbf{k_1}}\nu_{\mathbf{k_2}}}(\mathbf{G})
e^{i(\mathbf G+\mathbf k_2-\mathbf k_1)\cdot\mathbf r}.
\end{equation}
The Fourier transform efficiently determines the fitting coefficients,
\begin{equation}
\Tilde{\rho}_{\mu_{\mathbf{k_1}}\nu_{\mathbf{k_2}}}(\mathbf{G}) = \frac{1}{\sqrt{\Omega_0}} \int_{\Omega_0} 
\rho_{\mu_{\mathbf{k_1}}\nu_{\mathbf{k_2}}}(\mathbf{r}) 
e^{-i\left(\mathbf{G} + \mathbf k_2 - \mathbf k_1 \right) \cdot \mathbf{r}} d\mathbf{r},
\end{equation}
where $\Omega_0$ denotes the volume of the unit cell in real space. 
The transformation can be carried out numerically using the fast Fourier transform (FFT) with numerical packages such as FFTW3.\cite{frigo2005design} 

With the pair density and Coulomb operator in reciprocal space, the ERIs in \cref{eq:ERIs} can be evaluated in reciprocal space as follows
\begin{equation}
    (\mu_{\mathbf{k_1}} \nu_{\mathbf{k_2}} | \lambda_{\mathbf{k_3}} \sigma_{\mathbf{k_4}}) = \sum_{\mathbf{G}} \Tilde{\rho}_{\mu_{\mathbf{k_1}}\nu_{\mathbf{k_2}}}(\mathbf{G}) 
    \frac{ 4\pi}{ |\mathbf{G} + \mathbf{k_{12}}|^2} \Tilde{\rho}_{\lambda_{\mathbf{k_3}}\sigma_{\mathbf{k_4}}}(\mathbf{-G}),
\end{equation}
where the singularity at $\mathbf{G} + \mathbf{k_{12}} = \textbf{0}$ is excluded from the summation. Alternatively, one may form the Hartree-like potential in reciprocal space, then transform it to a real space grid, and contract with the real-space pair density.

There are several drawbacks to the GPW strategy. In the GPW framework, the use of pseudopotentials is necessary since highly localized orbitals in the core region require a prohibitively large number of grid points and plane waves for converged numerical results. Even for second-row atoms, the use of a pseudopotential is necessary. \cite{goedecker1996separable, vanderbilt1990soft}
Additionally, the storage advantage of GTO basis sets is somewhat negated as a plane wave representation must be stored, leading to storage requirements similar to plane-wave codes.
\noindent
GPW is available in many periodic codes and has been used for periodic hybrid DFT,\cite{Lee2022Dec} RPA, \cite{BibEntry2015Feb}  GW, \cite{Wilhelm2017Jun} MP2,\cite{DelBen2012Sep,DelBen2013May} and coupled cluster.\cite{McClain2017Mar}

\subsection{Range Separation/Ewald}\label{subsec:ewald}
One approach that combines the strength of real space and reciprocal space evaluation is through Ewald summation.~\cite{saunders_electrostatic_1992,sun_exact_2023} In this approach, the Coulomb kernel is split into the short-range (SR) and long-range (LR) contributions, 
\begin{align}
    \frac{1}{|\textbf{r}_{12}|} = \frac{\text{erfc}(\omega \textbf{r}_{12})}{|\textbf{r}_{12}|} + \frac{\text{erf}(\omega \textbf{r}_{12})}{|\textbf{r}_{12}|},
\end{align}
where erf is the error function, erfc is the complementary error function, and $\omega$ is the range separation parameter. Physically, this separation is equivalent to adding a compensating Gaussian charge distribution to screen out long-ranged interactions, such that the long-ranged component now comes solely from the Gaussian charge distribution.~\cite{stenberg2020exact} The parameter $\omega$ controls the decay rate of the SR kernel in real space and the LR kernel in reciprocal space. In the limit $\omega =0$, one recovers the original unscreened Coulomb kernel with the SR part. On the other hand, as $\omega $ approaches $\infty$, the LR kernel recovers the unscreened Coulomb kernel. It is most efficient to evaluate the SR contribution in real space and LR contribution in reciprocal space because of their exponential decay in each space.

The real space evaluation of the short-range ERIs can be adapted from the existing molecular codes \cite{gill1989efficient, head1988method} with additional phase factors to construct the Bloch crystalline orbitals in \cref{eq: Bloch equation}. Efficient screening for lattice summation is necessary to achieve the target numerical accuracy without spending computational time on insignificant integrals. \cite{guidon2008ab, sun_exact_2023} We note that a tight distance screening scheme using the multipole expansion has been developed for short-ranged exchange in hybrid density functional theory \cite{heyd2003hybrid} and has been extended to periodic systems. \cite{izmaylov2006efficient, ye2021tight}  Furthermore, the density fitting technique can also be employed here for better performance.\cite{Lazarski2015Jun,ye_fast_2021} 
The long-ranged component can be efficiently evaluated in reciprocal space using the GPW approach.\cite{saunders_electrostatic_1992,sun2023efficient} 

While this approach may seem most natural for applications to periodic systems, 
this was also explored in the molecular context by Fusti-Molnar and Pulay.\cite{fusti2002fourier}
In the Fourier transform Coulomb (FTC) approach, one uses the Ewald approach and directly evaluates the unscreened Coulomb potential in reciprocal space. Depending on the degree of compactness of the primitive GTOs, one utilizes one of the two algorithms.

\subsection{Density Fitting or Resolution-of-the-Identity}

In molecular quantum chemistry, density fitting or resolution of the identity (RI) approximations have been widely adopted to accelerate HF, \cite{weigend2002fully} DFT, \cite{hesselmann2005density} and correlated wavefunction calculations.\cite{weigend1998ri, hattig2000cc2} In this technique, the pair density $\rho_{\mu_{\mathbf{k_1}}\nu_{\mathbf{k_2}}}(\mathbf{r})$ is expanded in a set of atom-centered auxiliary basis sets
\begin{equation}
    \rho_{\mu_{\mathbf{k_1}}\nu_{\mathbf{k_2}}}(\mathbf{r}) \approx 
    \Tilde{\rho}_{\mu_{\mathbf{k_1}}\nu_{\mathbf{k_2}}}(\mathbf{r}) 
    = \sum_P^{N_\text{aux}} C_{\mu_{\mathbf{k_1}}\nu_{\mathbf{k_2}}}^{P_{\mathbf{k_2}-\mathbf{k_1}} }\phi_P^{\mathbf{k_2}-\mathbf{k_1}}(\mathbf{r}),
\end{equation}
where $N_\text{aux}$ is the number of auxiliary basis functions, $\Tilde{\rho}_{\mu_{\mathbf{k_1}}\nu_{\mathbf{k_2}}}(\mathbf{r})$ denotes the fitted density, $\{\phi_P^{\mathbf{k_2}-\mathbf{k_1}}\}$ is the auxiliary basis functions, and $C_{\mu_{\mathbf{k_1}}\nu_{\mathbf{k_2}}}^{P_{\mathbf{k_2}-\mathbf{k_1}} }$'s are fitting coefficients. The auxiliary basis is typically 3-5 times larger than the primary basis sets \cite{eichkorn1995auxiliary} and is optimized for specific calculations, such as MP2 \cite{weigend1998ri} and exact exchange. \cite{weigend2008hartree} The auxiliary basis can also be generated automatically for different primary basis sets and types of calculations. \cite{stoychev2017automatic}

The most common approach in molecular applications to determine the fit coefficients follows a minimization of the self-interaction between residual density
\begin{equation} \label{eq: objective for fit coefficient}
    \iint \left( \rho(\mathbf{r}_1) - \Tilde{\rho}(\mathbf{r}_1) \right) \frac1{|\mathbf r_1-\mathbf r_2|} \left( \rho(\mathbf{r}_2) - \Tilde{\rho}(\mathbf{r}_2) \right) \mathrm d\mathbf{r}_1 \mathrm d\mathbf{r}_2.
\end{equation}
To avoid the Coulomb divergence, one may want to conserve charge\cite{sodt2006linear} explicitly and higher order multipole moments,\cite{van1988ab} which has been achieved in various ways.\cite{maschio2008fitting,Lazarski2015Jun,sun2017gaussian}
In the simple case where there are no additional constraints, the coefficient matrix reads
\begin{equation}
    C_{\mu_{\mathbf{k_1}}\nu_{\mathbf{k_2}}}^{P_{\mathbf{k_2}-\mathbf{k_1}} }  =  \sum_{\mathbf{k_{1} \mathbf{k_{2}}}} \sum_{Q} (P_{\mathbf{k}_{2}-\mathbf{k}_{1}} |Q_{\mathbf{k}_{2}-\mathbf{k}_{1}})^{-1} ( Q_{\mathbf{k}_{2}-\mathbf{k}_{1}} |\mu_{\mathbf{k_1}} \nu_{\mathbf{k_2}}),
\end{equation}
where these integrals need to be evaluated carefully to avoid any Coulomb divergence.
Similarly to the molecular case, one of the benefits of density fitting is the reduction of data storage from quartic to cubic scaling.
The cost of storing the three-index tensors is $\mathcal O(N_k^2 N_\text{aux} N^2)$ where $N$ is the number of primary basis functions, and $N_\text{aux}$ is the number of auxiliary basis functions.
Nonetheless, this cost makes medium-sized unit cells with hundreds of k-points impractical, so practical implementations may have to resort to integral-direct approaches.~\cite{bintrim2022integral}

Global density fitting with the unscreened Coulomb metric is highly inefficient for large molecular calculations due to the spurious contribution from distant auxiliary functions to the fitted density. \cite{jung2005auxiliary} In essence, in global density fitting, we are fitting a density with functions that may be very far away from that density. To overcome this, one may use a metric that decays faster than the Coulomb metric. Available options are the overlap metric,\cite{baerends1973self} the Yukawa Coulomb metric,\cite{gill2005decay} or the attenuated Coulomb metric. \cite{jung2005auxiliary} The overlap metric was used in early work by Baerends et al., \cite{baerends1973self} but the error was worse than the Coulomb metric. Jung et al. then used the attenuated Coulomb metric to control the trade-off between the locality and the accuracy of the density fitting. \cite{jung2005auxiliary} In the follow-up work, Reine et al. proposed a scheme for robust and variational fitting with a local and attenuated metric. \cite{reine2008variational}

An alternative approach is to employ local fitting domains where only a subset of the auxiliary basis is used for fitting in each local region. Gallant et al. divided the charge density into subsystems, independent of the size of the molecule, and performed density fitting separately.\cite{yang1995density, gallant1996linear} Sodt et al. introduced linear-scaling local atomic density fitting (or atomic resolution of identity), which employs a ``bump" function to exclude distant fitting functions without requiring predefined subsystems.\cite{sodt2006linear, sodt2008hartree} In the tight limit of the bump function, the pair density is fitted by auxiliary basis functions located on two atoms. \cite{fonseca1998towards} This fitting scheme is called concentric atomic density fitting \cite{hollman2017fast} or pair-atomic resolution of identity.\cite{manzer2015efficient, rebolini2016comparison}

The last approach we mention is the Poisson fitting approach originally proposed by Manby and co-workers~\cite{Mintmire1982Jan,manby2001poisson} for electronic structure problems and by McCurdy and co-workers~\cite{McCurdy2001Jan} for molecular scattering theory.
In this approach, one fits the electrostatic potential in Poisson's equation directly instead of density.
As a result,
the sophisticated ERI tensors can be approximated by three-center overlap and kinetic energy integrals.
Since Poisson fitting functions carry no multipoles, the resulting integrals all readily converge, and there is no need for special treatment of the Coulomb divergence.
Despite the lack of multipoles, with a large enough basis set (with multiple diffuse basis functions), 
it was shown that Poisson fitting converges to
exact results.\cite{Lambrecht2011Mar}

While these density-fitting approaches were initially developed for molecular quantum chemistry methods, many density-fitting techniques and ideas in quantum chemistry have been extended to periodic systems. For example, Maschio and Usvyat applied the Poisson density fitting to periodic systems.\cite{maschio2008fitting} Lazarski et al.~\cite{Lazarski2015Jun} combined density fitting with the continuous fast multipole method (CFMM; see \cref{sec:cfmm})for periodic systems. Another example is that Wang et al. \cite{wang2020efficient} extended the concentric atomic density fitting scheme to periodic systems. Sun et al. \cite{sun_gaussian_2017} used a mixed Gaussian and plane wave basis to fit diffuse and compact densities efficiently. Ye and Berkelbach \cite{ye_fast_2021} used the range-separation method (i.e., Ewald combined with FTC) to perform periodic density fitting. Other ideas in local density fitting seem interesting to pursue in the context of solid-state quantum chemistry method development.

\subsection{Cholesky Decomposition}
Cholesky decomposition (CD) is a well-known method in linear algebra that can be traced back to the early twentieth century.\cite{benoit1924note} It is a particular form of the LU decomposition that can be used on a symmetric semi-positive definite matrix $\mathbf{M}$ as $\mathbf{M} = \mathbf{L}\mathbf{L}^T$, where $\mathbf{L}$ is a lower triangular matrix. The use of CD on ERI matrices was first introduced by Beebe and Linderberg back in 1977.\cite{Beebe1977} In molecular quantum chemistry, one decomposes the matrix of two-electron four-center integrals $M_{\mu\nu, \lambda\sigma} = (\mu\nu|\lambda\sigma)$ using CD,
\begin{equation}
    M_{\mu\nu, \lambda\sigma}
    =
    \sum_P^{N_\text{chol}} L_{\mu\nu}^P L_{\lambda\sigma}^P,
\end{equation}
where $\mathbf L$ is the Cholesky matrix and
$N_\text{chol}$ is the number of Cholesky vectors.
The ERI tensor is
approximately low-rank, so it is possible to compress
the tensor by rank-revealing CD.
One uses the modified CD algorithm~\cite{Aquilante2011} to 
achieve this.
With a threshold of $10^{-3}$ for the maximum diagonal residual, one may obtain $N_\text{chol}$ similar to $N_\text{aux}$ of the density fitting basis set with comparable accuracy.

Similarly, for periodic ERIs, 
$
M_{\mu_{\mathbf{k}_1}\nu_{\mathbf{k}_2}, \lambda_{\mathbf{k}_3} \sigma_{\mathbf{k}_4}} = (\mu_{\mathbf{k}_1}\nu_{\mathbf{k}_2} |
\lambda_{\mathbf{k}_3} \sigma_{\mathbf{k}_4}),
$
we factorize, 
\begin{align}
    M_{\mu_{\mathbf{k}_1}\nu_{\mathbf{k}_2}, \lambda_{\mathbf{k}_3} \sigma_{\mathbf{k}_4}} = \sum_P^{N_\text{chol}}
    L^{P}_{\mu_{\mathbf{k}_1}\nu_{\mathbf{k}_2}}
    (L^{P}_{\sigma_{\mathbf{k}_4}\lambda_{\mathbf{k}_3}})^*.
    \label{eq:chol}
\end{align}
We note that the second Cholesky vector in \cref{eq:chol}
has the index ordering $\sigma$ and $\lambda$ permuted, which is because k-points ERIs are complex-valued.
 The main advantage one gains from using CD is the reduced storage cost of the periodic ERIs from $\mathcal{O}(N_k^3N^4)$ to $\mathcal{O}(N_k^2N^2N_\text{chol})$, assuming momentum conservation, just like density-fitting. 

\subsection{Tensor Hypercontraction}
Tensor hypercontraction (THC) offers a different low-rank factorization of the 4-index ERI tensor using a product of five 2-index tensors. 
Because we only deal with 2-index tensors, this could potentially lead to significant storage and computational cost reductions.
Indeed, this technique was originally developed by Martinez and co-workers for molecular systems, and it has been used to accelerate many electronic structure methods.\cite{Hohenstein2012,Parrish2012,Hohenstein2012a,Parrish2013a,Hohenstein2013a,Hohenstein2013,Parrish2014}
This factorization has its roots in canonical polyadic decomposition or CANDECOMP/PARAFAC~\cite{Hitchcock1927Apr,Carroll1970Sep} and reads
\begin{equation}
(\mu\nu|\lambda\sigma)
\simeq
\sum_{\hat{P}\hat{Q}}
X_{\mu}^{\hat{P}}
X_{\nu}^{\hat{P}}
M_{\hat P \hat Q}
X_{\lambda}^{\hat{Q}}
X_{\sigma}^{\hat{Q}},
\label{eq:thc}
\end{equation}
where $\{\hat P\}$ denotes the THC ``grid'' points, $\mathbf X$ is the THC leaf tensors,
and
$\mathbf M$ is the THC core tensor.
The factorization, by construction, limits the dimension of the THC grid to $\mathcal O(N)$, and its accuracy has been benchmarked 
through various methods and applications.\cite{Malone2018Dec,Matthews2020Jan,lee_systematically_2020,Matthews2021Apr,Zhao2023Jun}

One of the challenging aspects of using THC for broad applications was the need for generating THC grid points $\{\hat{P}\}$ for each atomic species and perhaps for each method.
This challenge was overcome by Lu and Ying in Ref. \citenum{Lu2015}.
It was recognized that the THC factorization is based on
an interpolative decomposition of the electron density,
\begin{equation}
\phi_\mu(\mathbf r)
\phi_\nu(\mathbf r)
=
\sum_{\hat{P}}
\phi_\mu(\mathbf r_{\hat{P}})
\phi_\nu(\mathbf r_{\hat{P}})
\xi_{\hat{P}}(\mathbf r),
\label{eq:isdf}
\end{equation}
where THC grid points are the interpolating points and $\xi_{\hat{P}}$ are the interpolating vectors.
These points can be obtained from rank-revealing QR decomposition~\cite{Lu2015} or centroid Voronoi Tessellation.~\cite{Dong2018Jan}
Then, the interpolating vectors are obtained from solving the least squares problem to minimize the residual in \cref{eq:isdf}.\cite{Parrish2012,Lu2015}
Finally, the THC core tensor can be calculated by
\begin{equation}
M_{\hat P \hat Q}
=
\int\mathrm d\mathbf r_1
\int\mathrm d\mathbf r_2
\frac{\xi_{\hat{P}}(\mathbf r_1)\xi_{\hat{Q}}(\mathbf r_2)}{|\mathbf r_1 - \mathbf r_2|}.
\end{equation}
This approach is called interpolative separable density fitting (ISDF) and was successfully used for molecular applications.~\cite{lee_systematically_2020}

ISDF has also been used for solid-state applications with k-point sampling.\cite{hu_interpolative_2017,Wu2022Jan,qin_interpolative_2023}
In essence, interpolating points and vectors are chosen to be k-independent,
\begin{align}
    (\phi_{\mu_{\mathbf{k_1}}}(\mathbf{r}))^*
    \phi_{\nu_{\mathbf{k_2}}}(\mathbf{r}) = \sum_{\hat P} (\phi_{\mu_{\mathbf{k_1}}}(\mathbf{r}_{\hat{P}}) )^*
    \phi_{\nu_{\mathbf{k_2}}}(\mathbf{r}_{\hat P}) \xi_{\hat P}(\mathbf{r}).
\end{align}
These interpolating points and vectors are then used to factorize the ERIs into the following form:
\begin{equation}
     (\mu_{\mathbf{k_1}} \nu_{\mathbf{k_2}} | \lambda_{\mathbf{k_3}} \sigma_{\mathbf{k_4}}) 
     \simeq \sum_{\hat P\hat Q} 
     (X_{\mu_{\mathbf{k_1}}}^{\hat P})^*
     X_{\nu_{\mathbf{k_2}}}^{\hat P} 
     M_{PQ}^{\mathbf k_1 - \mathbf k_2}
     (X_{\lambda_{\mathbf{k_3}}}^{\hat Q})^*
     X_{\sigma_{\mathbf{k_4}}}^{\hat Q},
\end{equation}
where
\begin{equation}
M_{\hat P \hat Q}^{\mathbf k}
=
\int\mathrm d\mathbf r_1
\int\mathrm d\mathbf r_2
\frac{(\xi_{\hat{P}}(\mathbf r_1))^*\xi_{\hat{Q}}(\mathbf r_2)}{|\mathbf r_1 - \mathbf r_2|}
e^{i \mathbf k (\mathbf r_2 - \mathbf r_1)}.
\end{equation}

This special case of density fitting allows the fitting coefficients to be factorized, requiring the storage of quantities only quadratic in basis set size $N$ and linear in $N_k$, which are often feasible to store in memory unlike the full ERI tensor. This factorization of the ERIs additionally enables a reduction in $N_k$ scaling in contractions over ERIs as is done for exact exchange or correlated wavefunction theories. ISDF has thus far been applied to periodic exact exchange\cite{hu_interpolative_2017, rettig_even_2023} and RPA\cite{Yeh2023Sep} using $\mathbf{k}$-point sampling. For a more in-depth review of the current state of ISDF, we refer interested readers to the excellent review by Qin et al.~\cite{qin_interpolative_2023}

\subsection{Continuous Fast Multipole Methods}\label{sec:cfmm}

An approach that is usually discussed as a comparison point for the Ewald approach (\cref{subsec:ewald}) is the fast multipole method.
The fast multipole method was originally designed for the efficient computation (i.e., $\mathcal O(N)$) of the Coulomb energy of point changes, as opposed to their more na{\"i}ve scaling of $\mathcal O(N^2)$.\cite{greengard1987fast,White1994Oct}
Later, the algorithm was generalized to continuous charge distributions as relevant for ERI evaluations.\cite{white_continuous_1994,Strain1996Jan}
Furthermore, the extension to periodic systems was made available later\cite{challacombe_periodic_1997,kudin1998fast,tymczak2005linear} and also combined with density fitting.\cite{burow_resolution_2009,Lazarski2015Jun}
Unlike the other approaches covered before, the CFMM approach only pertains to the J-build needed for HF and DFT,
\begin{equation}
J_{\mu\nu}^{\mathbf k}
=
\sum_{\mathbf q}\sum_{\lambda\sigma}
(\mu_{\mathbf k}\nu_{\mathbf k}|\lambda_{\mathbf q}\sigma_{\mathbf q})
P_{\sigma\lambda}^{\mathbf q}
\label{eq:Jmat}
\end{equation}
where $P_{\nu\mu}^{\mathbf q}$ is the density matrix defined as 
$
P_{\nu\mu}^{\mathbf q}
=
\sum_i^{N_\text{elec}}
C_{\nu i}^{\mathbf q}
(C_{\mu i}^{\mathbf q})^*
$
.

CFMM separates the ERIs into near-field (NF) and far-field (FF) contributions, similar in spirit to Ewald summation. Like the short-range component in Ewald, the NF contribution may be performed using direct lattice summation of analytical molecular ERIs. The FF contribution is then computed using multipole expansions within real space.

For the FF contribution, basis functions within neighboring cells are grouped into sets of locally interacting distributions, and a hierarchical tree of multipole expansions is formed. The tree of multipole expansions and recurrence relations are used to efficiently compute the interaction between all multipole distributions, yielding the FF contribution. The CFMM algorithm uses a parameter to specify the minimum distance at which the interaction may be computed via multipole expansion - controlling the accuracy of the method. In practice, the majority of the ERIs arise from the FF contribution, but the NF contribution accounts for the majority of the computation time. Additionally, CFMM-based approaches apply to all-electron calculations as the evaluation is done entirely in real space.

One may combine CFMM with the FTC\cite{fusti2002fourier} and J-engine techniques, \cite{White1996Feb}
and this was shown to be one of the most efficient algorithms for building
the J-matrix for DFT calculations.\cite{Fusti-Molnar2005Feb}
Exploration of this idea along with other related ones for periodic systems 
is an interesting direction.

\section{Density functional theory with hybrid functionals}\label{sec:hybrid}

The inclusion of exact exchange (i.e., Hartree-Fock exchange) in DFT has been shown to improve significantly upon local density functionals across nearly all molecular applications,~\cite{Mardirossian2017Oct} leading to hybrid functionals such as B3LYP\cite{Becke1993Jan} and range-separated hybrid functionals such as $\omega$B97X-V.~\cite{Mardirossian2014May} 
These functionals have become the {\it de facto} standard in molecular applications. 
Furthermore, for molecular calculations, the nominal scaling of the JK-build with sparsity is $\mathcal O(N^2)$, making the use of exact exchange only marginally more costly than local density functionals.

For periodic calculations, one needs to either sample many $\textbf{k}$-points or perform large supercell calculations to reach the thermodynamic limit \cite{nusspickel2022systematic}. This often increases the computational time of all computational methods by multiple orders of magnitude compared to molecular calculations of sizes similar to the unit cell. The K-build needed for hybrid functionals could be particularly costly in this setup. The K-matrix reads
\begin{equation}
    K_{\mu\nu}^{\mathbf k} = \sum_{{\mathbf k}}\sum_{\lambda \sigma} (\phi_{\mu_{\mathbf k}}\phi_{\lambda_{\mathbf k'}} | \phi_{\sigma_{\mathbf k'}} \phi_{\nu_{\mathbf k}}) P_{\lambda \sigma}^{\mathbf k'},
    \label{eq:pbcexchange}
\end{equation}
where unlike \cref{eq:Jmat} there is 
no straightforward way to avoid $\mathcal O(N_k^2)$, although THC has been applied to lower this cost to $\mathcal O(N_k)$.~\cite{Wu2022Jan,rettig_even_2023}
Furthermore, the scaling with respect to the unit cell is $\mathcal O(N^3)$. 
This should be contrasted with the cost of the J-build, $\mathcal O(N_k N)$, which is obtained assuming the sparsity of the GTO basis set and the GPW algorithm.

Plane-wave-based hybrid DFT calculations especially can become prohibitively expensive because of the extremely large basis sets - although recent developments in adaptively compressed exchange~\cite{Lin2016Apr} have made hybrid DFT plane-wave studies possible for small to medium systems. GTO-based codes benefit from much more compact bases, but complications can arise when reaching the basis set limit. Nonetheless, numerous methods for computing the exchange operator are in use today, building upon the ERI techniques from the previous section of this paper. Despite active development, hybrid DFT remains multiple orders of magnitude more expensive than local DFT for most realistic systems, leading local DFT to account for most periodic DFT applications.
Below, we discuss two main applications for which we think the design of better hybrid density functionals could be especially valuable.

\subsection{Band Gaps}
\begin{figure}[h]
    \centering
    \includegraphics[width=0.9\linewidth]{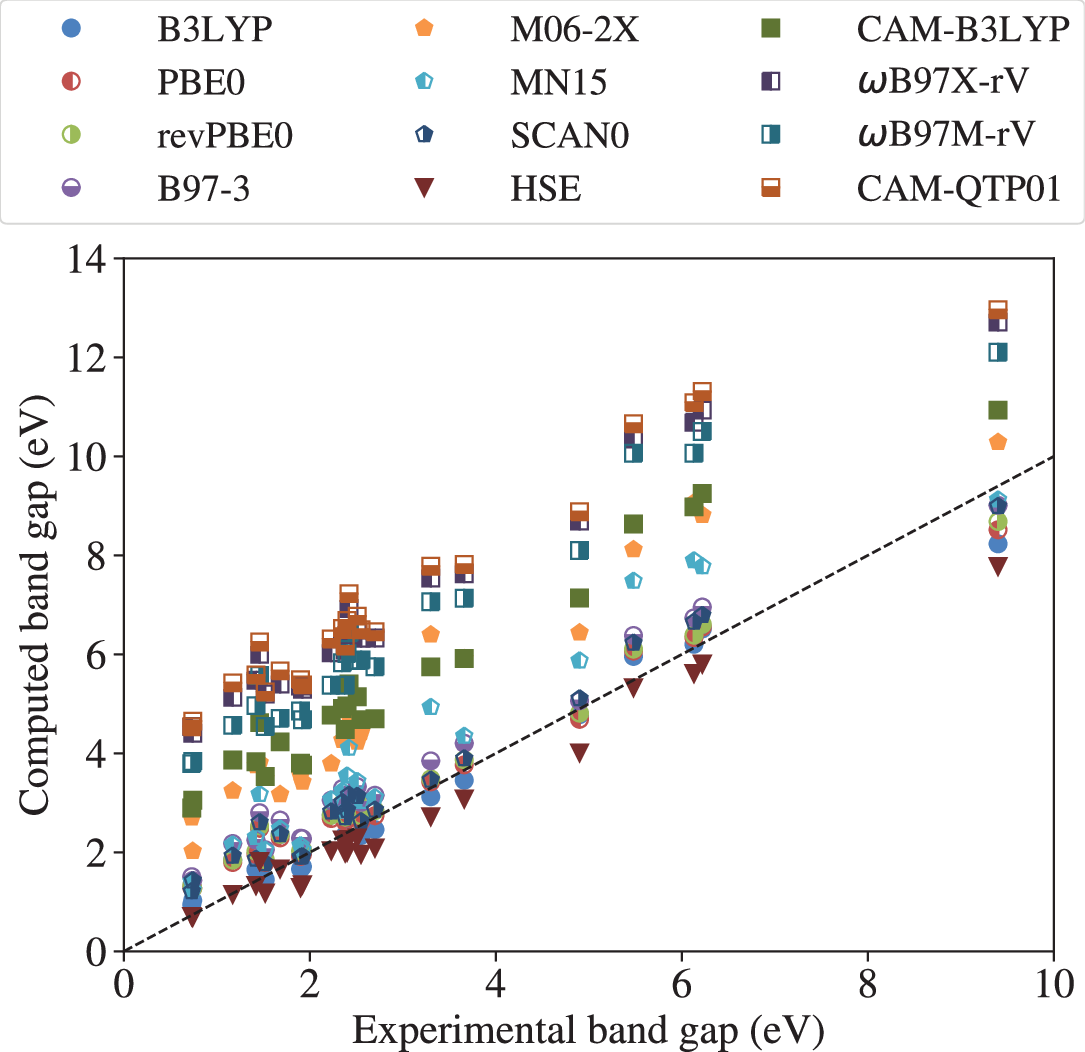}
    \caption{Scatter plot of computed band gaps (eV) versus experimental band gaps (eV). Round markers are used for GGA global hybrids, pentagons are used for mGGA global hybrids, triangles are used for a short-range hybrid functional (HSE), and squares are used for long-range corrected range separated functionals. The black dotted line is a guide for the eye. Reprinted with permission from Ref. \citenum{Lee2022Dec}. Copyright 2022 American Chemical Society.}
    \label{fig:band_gaps}
\end{figure}

Band gaps are one of the most commonly computed properties with periodic DFT due to their relevance in many active areas of materials research, such as photoabsorption, transport, etc. In molecular applications, the underestimation of the HOMO-LUMO gap is a well-documented failure of local DFT, especially for charge-transfer excitations.\cite{AndreasDreuw*2004Mar} This behavior also extends to periodic systems and can even lead to erroneous predictions of metallic behavior for prototypical semiconductors.\cite{Perdew1985Mar,lee2021approaching} Commonly, this underestimation is mitigated by using hybrid functionals - which cancels out some of the underestimation of local functionals with the overestimation of band gaps with HF. Some of the best-performing functionals use only short-range exchange (or screened exchange),\cite{Heyd2004Jul} which yields more accurate results and is computationally cheaper to compute. While hybrid DFT can often lead to adequate results, band gaps depend on the fraction of exact exchange included in the functional used, as shown in \cref{fig:band_gaps}.\cite{Lee2022Dec} Unfortunately, the ideal fraction of exact exchange is system-dependent - preventing the recommendation of a single functional for general use.

\subsection{Surface Chemistry}
The interface of molecules with solids represents another critical use case of periodic density functional theory with applications in materials research and catalysis. An accurate treatment of the system requires a functional that is accurate for both molecules and solids - a nontrivial problem as most functionals are fitted to one use case or the other. Another complication is the lack of high-quality computational references for many adsorption systems. Accurate representation of the system requires multiple surface layers along with a large amount of vacuum - leading even seemingly trivial systems to be computationally demanding. DFT studies must, therefore, rely on experimental data for comparison - making the development of systematic benchmarks much more difficult.
This is in sharp contrast to molecular functional developments.~\cite{Mardirossian2017Oct}

Recent studies have found that dispersion-corrected local functionals are often adequate for surface chemistry studies. Additionally, functionals corrected for the systematic overbinding seen by local DFT, such as RPBE, often excel in surface chemistry applications. RPBE was seen to accurately reproduce experimental adsorption energies for electrocatalytically relevant materials, \cite{lininger_challenges_2021} giving less than 0.2 eV error. A recent study of adsorption energies on transition metal surfaces found that RPBE is accurate within 0.2 eV error on average. \cite{schmidt_benchmark_2018} Thus far, the application of hybrid density functional theory to surface chemistry problems has been severely limited by the additional cost of exact exchange.\cite{Stroppa2008Jun}
Given the significant advances made in molecular density functionals for describing non-covalent interactions, many opportunities await in
testing and developing
density functionals that could potentially work for
both molecular and solid-state systems.

\section{Wavefunction methods}\label{sec:wfn}
Wavefunction methods offer systematically improvable approximations with accuracy beyond DFT.
They are more computationally demanding than DFT. Therefore, they have seen relatively limited use in modeling condensed-phase problems.
Other than the 
steep computational scaling,
wavefunction methods
pose additional challenges
that 
may be harder to overcome for
some choices of discretization.~\cite{Booth2016Aug}

Wavefunction methods
require 
unoccupied (i.e., conduction) bands to compute
correlation energy.
Formally, there are infinitely many
unoccupied bands to consider
towards the CBS limit.
Since reaching the CBS limit by brute force and increasing the basis set size is hardly possible,
extrapolation schemes are used in practice 
with a presumed asymptotic behavior.~\cite{Helgaker1997Jun,Bakowies2007Aug}
Practical basis sets should, therefore, provide consistent descriptions across different chemical elements and offer a favorable convergence to the asymptote for an accurate extrapolation to the CBS limit.
This has been one of the most important design principles in 
GTOs for both molecular~\cite{Dunning1989Jan} and solid-state applications.~\cite{laun2018consistent,ye2022correlation}
Unlike GTOs, 
plane wave or other grid-based bases
may struggle to 
provide as compact
unoccupied bands as GTOs.~\cite{Booth2016Aug}
The favorable compactness of GTOs presents opportunities for practical wavefunction calculations that are challenging to other discretization schemes.
In this Section, we present a summary of
translational symmetry adapted wavefunction methods in terms of their computational cost and any future research directions. 

As is customary in molecular quantum chemistry, we will use indices $i,j,k,\cdots$ to denote occupied orbitals, indices $a,b,c,\cdots$ to denote unoccupied (virtual) orbitals, and indices $p,q,r,\cdots$ to denote any orbitals.

\subsection{M\o ller-Plesset perturbation theory}
M\o ller-Plesset (MP) perturbation theory
adds electron correlation effects order-by-order on top of a given zeroth order single determinant wavefunction.
The most widely used MP method is the second-order MP (MP2) method, which computes
\begin{equation}
E_{c}
=
\sum_{ia}
\frac{|F_{ia}|^2}{\Delta_i^a}
+
\frac14
\sum_{iajb}
\frac{|\langle ij||ab\rangle|^2}
{\Delta_{ij}^{ab}},
\end{equation}
where $\mathbf F$ is the Fock matrix, $\langle ij||ab\rangle = (ia|jb) - (ib|ja)$, $\Delta_i^a = \epsilon_i - \epsilon_a$ and
$\Delta_{ij}^{ab} = \epsilon_i+\epsilon_j - \epsilon_a-\epsilon_b$ with $\epsilon_p = F_{pp}$.
Here, we used the spin-orbital notation and dropped the momentum dependence of orbitals for simplicity. 
The momentum dependence can be recovered by assigning each spin-orbital to a momentum and enforcing momentum conservation.
More concretely,
momentum conservation can be inferred for arbitrary-order tensors by
\begin{equation}
O_{p_{\mathbf k_p}q_{\mathbf k_q}r_{\mathbf k_r}\cdots}^{s_{\mathbf k_s}t_{\mathbf k_t}u_{\mathbf k_u}\cdots}
\propto
\delta_{\mathbf k_p + \mathbf k_q + \mathbf k_r+\cdots, \mathbf k_s + \mathbf k_t + \mathbf k_u + \cdots},
\end{equation}
where the Kronecker-delta compares the incoming and outgoing momenta modulo the reciprocal lattice vectors $\mathbf b$. 
This is quite similar to the procedure of recovering spin-explicit expressions from spin-orbital expressions, utilizing that fact that certain spin-blocks are zero by construction. 

As an example, let us consider
the MP2 correlation energy expression per unit cell with momentum indices,
\begin{align}\nonumber
\frac{E_{c}}{N_k}&=
\frac{1}{N_k}
\sum_{ia}
\sum_{
\mathbf k}
\frac{|F_{i_{\mathbf k}a_{\mathbf k}}|^2}{\Delta_{i_{\mathbf k}}^{a_{\mathbf k}}}\\
&+\frac{1}{N_k}
\frac14
\sum_{iajb}
\sum_{
\mathbf k_1,
\mathbf k_2,
\mathbf k_3
}
\frac{|\langle i_{\mathbf k_1}j_{\mathbf k_2}||a_{\mathbf k_3}b_{\mathbf k_{123}}\rangle|^2}
{\Delta_{i_{\mathbf k_1}j_{\mathbf k_2}}^{a_{\mathbf k_3}b_{\mathbf k_{123}}}},
\end{align}
where 
$\mathbf k_{123} = \mathbf k_1+\mathbf k_2-\mathbf k_3 \:\:\text{mod}\:\mathbf b$.
The computational cost of the MP2 correlation energy is then $\mathcal O(N_k^3 n^4)$.
The integral transformation is still the bottleneck, which costs
$\mathcal O(N_k^3 n^5)$.
Owing to the translational symmetry, we reduce the MP2 correlation energy cost by a factor of $1/N_k$ and the integral transformation cost by a factor of $1/N_k^2$.

The third-order MP (MP3) method
has seen relatively narrower applications compared to MP2
because of its higher computational cost.
The MP3 correlation energy reads
\begin{equation}
E_\text{MP3} = 
E_\text{vv}^{(3)}
+E_\text{oo}^{(3)}
+E_\text{ov}^{(3)},
\label{eq:mp3}
\end{equation}
where the spin-orbital expressions for each term are:
\begin{align}\label{eq:mp3vv}
E_\text{vv}^{(3)} &= \frac 18
\sum_{ijabcd} (t_{ij}^{ab})^* \langle ab || cd \rangle t_{ij}^{cd},\\ \label{eq:mp3oo}
E_\text{oo}^{(3)} &=
\frac 18
\sum_{ijklab} (t_{ij}^{ab})^* \langle ij || kl \rangle t_{kl}^{ab},\\ \label{eq:mp3ov}
E_\text{ov}^{(3)} &= -
\sum_{ijabkc} (t_{ij}^{ab})^* \langle ic || kb \rangle t_{kj}^{ac},
\end{align}
and the amplitude reads
\begin{equation}
t_{ij}^{ab}
=
\frac{\langle ij||ab\rangle}{\Delta_{ij}^{ab}}.
\end{equation}
Each term scales as $\mathcal O(n_\text{occ}^2n_\text{vir}^4)$, $\mathcal O(n_\text{occ}^4n_\text{vir}^2)$, and $\mathcal O(n_\text{occ}^3n_\text{vir}^3)$ with system size (i.e., sextic scaling), respectively.
The most expensive term is the particle-particle ladder term denoted as $E_\text{vv}^\text{(3)}$.
Similarly to MP2, converting, as an example, \cref{eq:mp3vv} to a momentum-explicit expression (per unit cell)
is straightforward,
\begin{align}\nonumber
\frac{E_\text{vv}^{(3)}}{N_k} =& \frac1{N_k}\frac {1}{8}
\sum_{ijabcd}
\sum_{\mathbf k_1
\mathbf k_2
\mathbf k_3
\mathbf k_4}
(t_{i_{\mathbf k_1}j_{\mathbf k_2}}^{a_{\mathbf k_3}b_{\mathbf k_{123}}})^* \\
&\langle a_{\mathbf k_3}b_{\mathbf k_{123}} || c_{\mathbf k_4}d_{\mathbf k_{124}} \rangle t_{i_{\mathbf k_1}j_{\mathbf k_2}}^{c_{\mathbf k_4}d_{\mathbf k_{124}}},
\end{align}
where $\mathbf k_{124}$ is defined completely analogously to $\mathbf k_{123}$.
Due to the momentum conservation,
instead of having $\mathcal O(N_k^6 n_\text{occ}^2
n_\text{vir}^4)$,
we have
$\mathcal O(N_k^4 n_\text{occ}^2
n_\text{vir}^4)$
cost 
highlighting
the $\mathcal O(N_k^2)$
computational cost saving.

There have been recent research activities on MP2 and MP3 for molecular systems,\cite{Bertels2019Aug,Loipersberger2021Sep} pertaining
to using different reference orbitals other than Hartree-Fock orbitals,\cite{Rettig2020Dec}
orbital-optimization,\cite{Lee2018Oct,Rettig2022Sep}
and 
regularizing MP2.\cite{Lee2018Oct,Loipersberger2021Sep,Rettig2022Sep}
As these are relatively recent developments in molecular quantum chemistry,
their utility in solid-state applications
is less understood.

\subsection{Coupled cluster theory}
Beyond MP methods,
more sophisticated diagrammatic methods
are available
under the umbrella of
coupled-cluster (CC) theory,\cite{Bartlett2007Feb}
\begin{equation}
|\Psi_\text{CC}\rangle
=
\exp(\hat{T})
|\Psi_\text{SD}\rangle,
\end{equation}
where the coupled-cluster wavefunction $|\Psi_\text{CC}\rangle$ is generated by acting a cluster operator $\exp(\hat{T})$ on a single Slater determinant, $|\Psi_\text{SD}\rangle$.
One of the most commonly used CC methods
is 
CC with singles and doubles (CCSD),
which has a computational cost
scaling sextically with system size.
In CCSD, we limit $\hat{T}$ to $\hat{T}_1 + \hat{T}_2$, incorporating all possible particle-hole type single and double excitations.

One of the reasons for the success of CCSD is
the role of single excitations. 
Owing to the Thouless theorem,\cite{thouless1960stability} the single excitations in CCSD can be understood as
orbital rotation.
Because of this orbital rotation in the presence of electron correlation effects (i.e., double excitations),
CCSD is often quite robust to 
different choices of $|\Psi_\text{SD}\rangle$.
This has been the key to the ``black-box'' nature of CCSD and higher-order CC methods.
Due to the steep scaling of CCSD, its application to sophisticated solid-state systems has been very limited.
Moreover, the addition of
triple excitations, either approximately~\cite{Gruber2018May,Neufeld2023Oct,Masios2023Oct} or fully,~\cite{Neufeld2023Oct} has been even more challenging as well.
Approaches to accelerate CC methods and broaden their applicability are ongoing research thrusts in both molecular and solid-state quantum chemistry.\cite{zhang_coupled_2019}

Efficient implementation of CC methods, in general, is not a straightforward task for molecules and more so for periodic systems.\cite{Irmler2023Jul}
Dealing with solid-state systems
adds another layer of complexity
from the translational invariance, which can provide a significant speedup.
Recent progress made by Irmler et al. applying the Cyclops Tensor Framework (CTF)\cite{Solomonik} is particularly noteworthy because it leveraged 
a production-level general sparse tensor library.\cite{Irmler2023Jul} The applications of coupled cluster theory to condensed phase quantum chemistry have been recently reviewed in Ref. \citenum{zhang_coupled_2019}, and we refer the readers there for an in-depth description.

\subsection{Random Phase Approximation}
The random phase approximation (RPA) is among the oldest methods of calculating correlation energy in many-body systems. 
The concept was first developed in the 1950s in three seminal papers by Bohm and Pines, 
\cite{Bohm_Pines_1951, Bohm_Pines_1952, Bohm_Pines_1953} as a method of solving the many-body electron problem for the uniform electron gas model. 
The next important development for chemists came in the 1970s when Langreth and Perdew \cite{Langreth_1975, Langreth_1977} 
as well as Gunnerson and Lundqvist \cite{Gunnarsson_1976} showed that one can obtain the RPA formalism by combining the adiabatic connection approach to Kohn-Sham DFT
with the fluctuation-dissipation theorem. 
It was as late as the start of this century when RPA became an electronic structure method applicable to molecules, as developed 
in works by Furche and others. \cite{Furche_DensityMatrixBased2001, Furche_MolecularTestsRandom2001, Fuchs_AccurateDensityFunctionals2002, Furche_FluctuationdissipationTheoremDensityfunctional2005, Furche_DevelopingRandomPhase2008, Eshuis_FastComputationMolecular2010} 
Furthermore, RPA can be viewed as an approximate coupled-cluster doubles (CCD) method where only ring diagrams are included.~\cite{Scuseria2008Dec}
Nowadays, RPA is an increasingly popular correlation method due to attractive features such as 
more accurate treatment of noncovalent interactions and applicability to small-gap systems. \cite{Eshuis_ElectronCorrelationMethods2012,Nguyen_DivergenceManyBodyPerturbation2020} 
We direct readers interested in the history of RPA development in quantum chemistry to excellent reviews \cite{Eshuis_ElectronCorrelationMethods2012, Chen_RPA2017, Ren_RandomPhaseApproximation2019} that cover this topic in greater detail.

RPA has been developed extensively for periodic first-principles calculations in a plane wave basis, \cite{Marini_2006, Deyu_2009, Nguyen_2009} with the PAW approach, \cite{Harl_CohesiveEnergyCurves2008, Harl_AccurateBulkProperties2009, Harl_AssessingQualityRandom2010, Olsen_RandomPhaseApproximation2013} and with the GPW approach. \cite{DelBen_ElectronCorrelationCondensed2013}. Generally, in a plane wave basis, the scaling is 
$\mathcal{O}(N^4)$ with respect to system size $N$ and $\mathcal{O}(N_k^2)$ with respect to number of k-points. \cite{Harl_CohesiveEnergyCurves2008, Yeh_LowScalingAlgorithmRandom2023} A cubic scaling algorithm has also been proposed by Kaltak and co-workers using an imaginary time method. \cite{Kaltak_CubicScalingAlgorithm2014, Kaltak_LowScalingAlgorithms2014}
This approach also reduces the scaling with respect to the number of k-points to $\mathcal O(N_k)$.

In a GTO basis, due to the locality of the basis functions, one could have a smaller prefactor compared to a plane-wave basis.
Standard Casida's equations are widely used in molecular time-dependent DFT~\cite{Casida_TimeDependentDensityFunctional1995} calculations, but are applicable only for the case of real orbitals. So, for periodic systems we need to evaluate the more general non-Hermitian eigenvalue equation \cite{Hirata_ConfigurationInteractionSingles1999, Dreuw_SingleReferenceInitioMethods2005a}
\begin{equation}
	\begin{pmatrix}
        \mathbf{A}  & \mathbf{B} \\
        \mathbf{B}^{*} & \mathbf{A}^{*}
    \end{pmatrix}
    \begin{pmatrix}
        \mathbf{X} \\
        \mathbf{Y}
    \end{pmatrix}
    =
    \begin{pmatrix}
        \Omega & 0 \\
        0 & -\Omega
    \end{pmatrix}
    \begin{pmatrix}
        \mathbf{X} \\
        \mathbf{Y}
    \end{pmatrix}.
\end{equation}
Here $\mathbf{A}_{iajb} = (\varepsilon_a - \varepsilon_i)\delta_{ij}\delta_{ab} + (ia|bj)$ and $\mathbf{B}_{iajb} = (ia|jb)$. Note that here we neglect the exchange part within the orbital rotation Hessians, which is commonly referred to as direct RPA (dRPA) in quantum chemistry literature. \cite{Eshuis_FastComputationMolecular2010} 
Because the linear dependency
problem could cause increased density fitting error for RI-RPA, in a recent study, Cholesky decomposition was carefully examined. \cite{Yeh_LowScalingAlgorithmRandom2023}  
Nonetheless, several groups have developed different schemes of RI-RPA for periodic systems using varying auxiliary basis sets. \cite{Grundei_RandomPhaseApproximation2017, Wilhelm_LargeScaleCubicScalingRandom2016a} 
Lastly, Yeh and Morales used ISDF as the method to decompose the ERIs for periodic RPA with GTOs, which gives an improved scaling with respect to system size and number of k-points.~\cite{Yeh_LowScalingAlgorithmRandom2023} Nonetheless, this does not offer any additional reduction in cost over the available low-scaling plane waves approach.~\cite{Kaltak_CubicScalingAlgorithm2014, Kaltak_LowScalingAlgorithms2014}

\subsection{Quantum Monte Carlo methods}
Quantum Monte Carlo (QMC) methods offer a new computational paradigm 
that differs from other methods discussed above.
In general, QMC methods rely on efficient stochastic sampling of the (approximate) ground state either through variational optimization (variational QMC) or through imaginary time evolution (projector QMC).
While real-space QMC methods have been used as a {\it de facto} many-body method for materials simulations for many years,~\cite{Foulkes2001Jan} second-quantized QMC methods are relatively new.~\cite{Blankenbecler1981Oct,Sorella2000Jan,Booth2009Aug,Motta2018Sep,Lee2022AFQMC}
These second-quantized QMC methods work in the space of determinants and can be directly compared with quantum chemistry methods.

The two notable second-quantized QMC methods relevant to the current review are full configuration interaction QMC (FCIQMC)\cite{Booth2009Aug} and auxiliary-field QMC (AFQMC).\cite{Blankenbecler1981Oct,Zhang2003Apr} FCIQMC is a formally exact approach, and its sample complexity, in general, is $\mathcal O(\exp(N))$ due to the fermionic sign problem. The initiator approximation\cite{Cleland2010Jan} was developed to reduce the sample complexity, but the asymptotic scaling stayed exponential. 
Despite its steep scaling, FCIQMC was used to solve some simple solid-state problems.\cite{Booth2013Jan}

AFQMC relies on a constraint set by a trial wavefunction. AFQMC controls the fermionic sign problem with the constraint, and its sample complexity remains polynomial-scaling.
Because of the constraint, the resulting imaginary-time evolution is not performed exactly, leading to an error in the final AFQMC energy.
In recent years, the quantum chemistry community has been
testing AFQMC in various molecular systems~\cite{Lee2022AFQMC,Sukurma2023Aug} and solid-state systems.\cite{SimonsCollaborationontheMany-ElectronProblem2017Sep,Motta2019Jul,Malone2020Jul,Malone2020Oct}
For a fixed statistical error per particle and with a single determinant trial, AFQMC offers an overall scaling of $\mathcal O(N^4)$, which is cheaper than most many-body methods.
Furthermore, a recent study showed that AFQMC was as accurate as CCSD(T) with a single determinant trial for problems that CCSD(T) is good for.\cite{Lee2022AFQMC}
One of the open questions in AFQMC is the design of efficient and accurate trial wavefunctions beyond HF theory.~\cite{Huggins2022Mar}
Another is to obtain excited state information (e.g., spectral functions) from AFQMC.\cite{Lee2021Jun} 
Finally, broadly assessing the utility of AFQMC in materials science would be beneficial in establishing an understanding of its potential.

\section{Lattice Dynamics}

The study of lattice dynamics is nearly a century old and has consistently remained at the forefront of condensed matter research. 
\cite{mahan_many-particle_2000, fetter_quantum_2003, landau_electron_1933, giustino_electron-phonon_2017}
In recent years, it has finally become practical to compute lattice dynamics approximately in 
real solids from first principles.\cite{baroni_phonons_2001, giustino_electron-phonon_2017}
Quantities such as phonon frequencies, \acp{epi}, and anharmonic contributions can be calculated in ever-larger systems and provide insight into experimental observables such as temperature-dependent band gaps, zero-point band gap renormalization, charge carrier transport, thermal expansion, heat capacity, superconductivity, polaron formation, to name just a few examples.\cite{stoffel_ab_2010,pokluda_ab_2015, giustino_electron-phonon_2017, lindsay_survey_2018}
Understanding the dynamics of nuclei (with or without \acp{epi}) is foundational to quantitative descriptions of solids.

It is practically impossible to obtain exact ``quantum'' dynamics of a lattice (i.e., nuclear degrees of freedom) in realistic solids.
Therefore, many available approaches assume that the nuclear degrees of freedom in solids are 
well-described by harmonic vibrational modes (i.e., phonons).
Subsequently, the coupling between lattice vibrations and electronic degrees of freedom is also approximated to
low-order in the phonon displacements.
Once the Hamiltonian for lattice dynamics is approximated in this form,
there are available tools to obtain the dynamics of phonons within these Hamiltonians.

Accounting only for the lowest order \acp{epi}, the simplest \textit{ab initio} Hamiltonian with electron-phonon coupling is given by
\begin{align}
    &H =  
    \sum_{\mathbf k}\sum_{p} \epsilon_{p_{\textbf{k}}} a_{p_{\textbf{k}}}^\dagger a_{p_{\textbf{k}}}  
    + \sum_{\mathbf{q}} \sum_{\nu} \hbar \omega_{\nu_{\mathbf{q}}} (b_{\nu_{\mathbf{q}}}^\dagger b_{\nu_{\mathbf{q}}}  + \frac12)\nonumber
    \\ &+  
    \frac{1}{\sqrt{N_k}}\sum_{\mathbf k \mathbf q}\sum_{pq\nu} 
    g_{p_{\mathbf k+\mathbf q}q_{\mathbf q}}^{\nu_{\mathbf q}}
    a^{\dagger}_{p_{\textbf{k+q}}} a_{q_{\textbf{k}}} (b_{\nu_{\mathbf q}} + b^\dagger_{\nu_{-\mathbf q}}),
    \label{eq:eph_ham}
\end{align}
where $\epsilon_{p_{\textbf{k}}}$ are the band energies (i.e., diagonal elements of the Fock operator), $\omega_{\nu_{\mathbf q}}$ are the phonon frequencies of the $\nu$-th phonon band at momentum $\mathbf q$ with corresponding creation/annihilation operator $b_{\mathbf{q}\nu}^\dagger / b_{\mathbf{q}\nu}$, and $g_{p_{\mathbf k+\mathbf q}q_{\mathbf q}}^{\nu_{\mathbf q}}$ is the electron-phonon interaction
matrix element.\cite{giustino_electron-phonon_2017}

\textit{Ab initio} approaches start by calculating the phonon frequencies and \ac{epi} matrix elements and then solving the interacting Hamiltonian (usually approximately).
While Eq. \ref{eq:eph_ham} contains much of the physics needed to treat \ac{epi} problems, further steps towards realism include additional phonon-phonon interaction (e.g., anharmonicity) terms and higher-order \acp{epi}.
Much work has been done on all aspects of this problem, and reviewing every progress of the field is beyond the scope of this article. We refer the reader to excellent reviews focused on this field.\cite{baroni_phonons_2001, giustino_electron-phonon_2017}
Numerous successful and well-used codes can compute phonon spectra and \acp{epi} in materials.
Much development, however, has occurred in the context of plane wave codes.\cite{giustino_electron-phonon_2007, giustino_electron-phonon_2017, zhou_perturbo_2021}
In this Section, we will focus on the current state of phonon and electron-phonon calculations specifically in GTO-based codes. 

Once mean-field calculations are performed to obtain band wavefunctions, the next step towards approximate lattice dynamics is the computation of harmonic phonon frequencies.
Completely analogous to molecular cases,~\cite{helgaker1988analytical} the vibrational modes are determined by computing the `interatomic force constants' between all atoms in the solid. 
We are interested in determining the eigenvalues of the following force matrix in momentum space,
\begin{align}
    D_{A \alpha, B \beta}^{\textbf{q}} = \frac{1}{\sqrt{M_A M_{B}}} \sum_{\mathbf R} \frac{\partial^2 E}{\partial X_{A \alpha}^{\mathbf{0}} \partial X_{B \beta}^{ \mathbf{R}}
    }
    e^{ i \mathbf{q \cdot R}},
\end{align}
where $E$ is the total electronic energy, $A$ and $B$ are nuclear indices, $\alpha$ and $\beta$ denote Cartesian components, $M_A$ is the mass of the $A$-th nucleus, 
and $X_{B \beta}^{\mathbf R}$ is the displacement of the $B$-th nucleus in the $\mathbf R$-th unit cell along $\beta$-direction.
This matrix is block-diagonal in $\mathbf q$, so diagonalizing the matrix for each $\mathbf q$ leads to
\begin{align}
\sum_{B\beta} D_{A \alpha, B\beta}^{\textbf{q}} v_{B\beta, \nu_{\mathbf q}} =  v_{A \alpha, \nu_{\textbf{q}}} \omega_{\nu_{\mathbf q}}^2,
\end{align}
 which provides the normal modes ($\mathbf v$) and the phonon frequencies/bands ($\omega_{\nu_{\mathbf q}}$).\cite{giustino_electron-phonon_2017}

The key quantity here is the computation of the second derivative of the total electronic energy, $E$, with respect to the nuclei positions, $X_{A \beta}^{\mathbf R}$. 
Computing this via finite differences in the so-called ``frozen phonon approximation''~\cite{Lloyd-Williams2015Nov} is conceptually straightforward but requires large supercells to sufficiently converge phonon bands. 
Nonetheless, the relative lack of complexity makes it a common method and an often-used reference implementation. 
The supercell scaling problem is ameliorated by \ac{dfpt} \cite{baroni2001phonons}, which corresponds to the analytic Hessian computation in molecular quantum chemistry.
\ac{dfpt} is a self-consistent set of equations that compute the linear response of the Kohn-Sham electronic potential due to phonons $\partial_{\nu_{\mathbf{q}}} v^\textnormal{KS}$.
This is equivalent to the coupled-perturbed self-consistent field equations.

The phonon calculation simultaneously contains all the information needed for an \ac{epi} calculation. Assuming plane waves and the Hellmann-Feynman theorem, the \ac{epi} matrix element is given by
\begin{equation}
    g_{p_{\mathbf k+\mathbf q}q_{\mathbf k}}^{\nu_{\mathbf q}} 
    = 
\sum_{A\alpha}
\sqrt{\frac1{2M_A\omega_{\nu_{\mathbf q}}}}
v_{A\alpha,\nu_{\mathbf q}}
    \langle {p_{\mathbf{k+q}}} | \frac{\mathrm d\hat{V}_\text{KS}}{\mathrm dX_{A\alpha}} | {q_{\mathbf{k}}} \rangle,
\label{eq:gkq}
\end{equation}
where $\hat{V}_\text{KS}$ is the Kohn-Sham potential operator used to obtain quasiparticle (i.e., band) energies.
It is important to note that unlike in plane wave calculations, Pulay terms arising from the incompleteness of the basis set will not vanish and must be included.\cite{baroni_phonons_2001}

In planewave codes, the \acp{epi} are usually transformed to a basis of \acp{wf} for the subsequent Wannier interpolation. This allows one to obtain the \ac{epi} tensor over a much more dense k-grid than can be directly computed with \ac{dfpt}.~\cite{Giustino2007Oct}
For more complicated systems, the process of obtaining the Wannier representation becomes increasingly complicated, especially for conduction bands.~\cite{Subotnik2005Sep}
While approaches to computing the \ac{epi} via atomic orbital bases are newer and less widespread in their application, they have been recognized as providing a pathway to bringing black-box style electron-phonon calculations to arbitrarily complicated materials where the \acp{wf} approach requires a user intensive trial-and-error process.~\cite{agapito_ab_2018}
Recognizing this potential, numeric atomic orbitals were used in Ref.~\citenum{agapito_ab_2018} to interpolate \acp{epi} with results comparable to Wannier interpolation.
To the best of our knowledge, there has not yet been an implementation of \ac{epi} computations specifically for \acp{gto}, and this remains an open area for development.

There are quite a few open questions left in the field. One example is that the evaluation of \cref{eq:gkq} has been done mostly using local density functionals.
A natural question is to assess the improvement of the accuracy using more sophisticated approaches such as hybrid density functionals (\cref{sec:hybrid}) and wavefunction methods (\cref{sec:wfn}).
There have been related works using plane waves codes, ~\cite{Marini2015Jun,Li2019May,Yang2022Sep} but GTO-based codes may have more potential for more efficient calculations due to their compact representation.
Another interesting question is to develop systematically improvable wavefunction methods to solve 
\cref{eq:eph_ham}.
Currently, there are methods based on Green's function approaches, ~\cite{sio_ab_2019, sio_polarons_2019, lafuente-bartolome_ab_2022, lafuente-bartolome_unified_2022} and these have been applied to realistic {\it ab initio} set-ups already.
On the wavefunction method side, there are coupled-cluster methods~\cite{White2020Dec} and auxiliary-field quantum Monte Carlo, ~\cite{lee2021constrained} but their applications have been limited to lattice models and small {\it ab initio} model systems.

\section{Conclusions and Outlook}
In this Focus Article, we reviewed some useful formalisms in translating molecular quantum chemistry tools to solid-state quantum chemistry ones.
We also focused on challenges and opportunities in basis set development, hybrid density functional development and applications, wavefunction methods development,
and method development for lattice dynamics.
These only reflect a small subset of topics that represent active research areas in condensed-phase quantum chemistry.
We hope that the readers of this Focus Article will feel the excitement for 
developing quantum chemistry methods for solid-state applications, 
problems that are considered less common for quantum chemistry methods.
We also hope that
major quantum chemistry packages that are known for their molecular capability will
start to expand in this research area.
In recent years, we have seen such development in quite a few quantum chemistry packages
such as CRYSTAL,~\cite{Erba2022Dec}, FHI-AIMS,~\cite{Blum2009Nov} SIESTA,~\cite{Garcia2020May} TURBOMOLE,~\cite{Balasubramani2020May} PySCF,~\cite{Sun2020Jul} and Q-Chem.~\cite{Epifanovsky2021Aug}
We expect that the enthusiasm for condensed-phase quantum chemistry will
only grow with time, and many of the machineries of molecular quantum chemistry will 
make an impact in simulating larger and more complex systems for which
current state-of-the-art methods cannot produce quantitative results.
\bibliography{references}

\begin{thebibliography}{246}%
\makeatletter
\providecommand \@ifxundefined [1]{%
 \@ifx{#1\undefined}
}%
\providecommand \@ifnum [1]{%
 \ifnum #1\expandafter \@firstoftwo
 \else \expandafter \@secondoftwo
 \fi
}%
\providecommand \@ifx [1]{%
 \ifx #1\expandafter \@firstoftwo
 \else \expandafter \@secondoftwo
 \fi
}%
\providecommand \natexlab [1]{#1}%
\providecommand \enquote  [1]{``#1''}%
\providecommand \bibnamefont  [1]{#1}%
\providecommand \bibfnamefont [1]{#1}%
\providecommand \citenamefont [1]{#1}%
\providecommand \href@noop [0]{\@secondoftwo}%
\providecommand \href [0]{\begingroup \@sanitize@url \@href}%
\providecommand \@href[1]{\@@startlink{#1}\@@href}%
\providecommand \@@href[1]{\endgroup#1\@@endlink}%
\providecommand \@sanitize@url [0]{\catcode `\\12\catcode `\$12\catcode
  `\&12\catcode `\#12\catcode `\^12\catcode `\_12\catcode `\%12\relax}%
\providecommand \@@startlink[1]{}%
\providecommand \@@endlink[0]{}%
\providecommand \url  [0]{\begingroup\@sanitize@url \@url }%
\providecommand \@url [1]{\endgroup\@href {#1}{\urlprefix }}%
\providecommand \urlprefix  [0]{URL }%
\providecommand \Eprint [0]{\href }%
\providecommand \doibase [0]{http://dx.doi.org/}%
\providecommand \selectlanguage [0]{\@gobble}%
\providecommand \bibinfo  [0]{\@secondoftwo}%
\providecommand \bibfield  [0]{\@secondoftwo}%
\providecommand \translation [1]{[#1]}%
\providecommand \BibitemOpen [0]{}%
\providecommand \bibitemStop [0]{}%
\providecommand \bibitemNoStop [0]{.\EOS\space}%
\providecommand \EOS [0]{\spacefactor3000\relax}%
\providecommand \BibitemShut  [1]{\csname bibitem#1\endcsname}%
\let\auto@bib@innerbib\@empty
\bibitem [{\citenamefont {Head-Gordon}\ and\ \citenamefont
  {Artacho}(2008)}]{Head-Gordon2008Apr}%
  \BibitemOpen
  \bibfield  {author} {\bibinfo {author} {\bibfnamefont {M.}~\bibnamefont
  {Head-Gordon}}\ and\ \bibinfo {author} {\bibfnamefont {E.}~\bibnamefont
  {Artacho}},\ }\href {\doibase 10.1063/1.2911179} {\bibfield  {journal}
  {\bibinfo  {journal} {Phys. Today}\ }\textbf {\bibinfo {volume} {61}},\
  \bibinfo {pages} {58} (\bibinfo {year} {2008})}\BibitemShut {NoStop}%
\bibitem [{\citenamefont {Pople}(1999)}]{Pople1999Jul}%
  \BibitemOpen
  \bibfield  {author} {\bibinfo {author} {\bibfnamefont {J.~A.}\ \bibnamefont
  {Pople}},\ }\href {\doibase
  10.1002/(SICI)1521-3773(19990712)38:13/14<1894::AID-ANIE1894>3.0.CO;2-H}
  {\bibfield  {journal} {\bibinfo  {journal} {Angew. Chem. Int. Ed.}\ }\textbf
  {\bibinfo {volume} {38}},\ \bibinfo {pages} {1894} (\bibinfo {year}
  {1999})}\BibitemShut {NoStop}%
\bibitem [{\citenamefont {Gro{\ss}}\ and\ \citenamefont
  {Sakong}(2022)}]{Gross2022Jan}%
  \BibitemOpen
  \bibfield  {author} {\bibinfo {author} {\bibfnamefont {A.}~\bibnamefont
  {Gro{\ss}}}\ and\ \bibinfo {author} {\bibfnamefont {S.}~\bibnamefont
  {Sakong}},\ }\href {\doibase 10.1021/acs.chemrev.1c00679} {\bibfield
  {journal} {\bibinfo  {journal} {Chem. Rev.}\ }\textbf {\bibinfo {volume}
  {122}},\ \bibinfo {pages} {10746} (\bibinfo {year} {2022})}\BibitemShut
  {NoStop}%
\bibitem [{\citenamefont {Dovesi}\ \emph {et~al.}(2005)\citenamefont {Dovesi},
  \citenamefont {Civalleri}, \citenamefont {Roetti}, \citenamefont {Saunders},\
  and\ \citenamefont {Orlando}}]{Dovesi2005Jan}%
  \BibitemOpen
  \bibfield  {author} {\bibinfo {author} {\bibfnamefont {R.}~\bibnamefont
  {Dovesi}}, \bibinfo {author} {\bibfnamefont {B.}~\bibnamefont {Civalleri}},
  \bibinfo {author} {\bibfnamefont {C.}~\bibnamefont {Roetti}}, \bibinfo
  {author} {\bibfnamefont {V.~R.}\ \bibnamefont {Saunders}}, \ and\ \bibinfo
  {author} {\bibfnamefont {R.}~\bibnamefont {Orlando}},\ }\enquote {\bibinfo
  {title} {Ab initio quantum simulation in solid state chemistry},}\ in\ \href
  {\doibase https://doi.org/10.1002/0471720895.ch1} {\emph {\bibinfo
  {booktitle} {Reviews in Computational Chemistry}}}\ (\bibinfo  {publisher}
  {John Wiley {\&} Sons, Ltd},\ \bibinfo {year} {2005})\ Chap.~\bibinfo
  {chapter} {1}, pp.\ \bibinfo {pages} {1--125}\BibitemShut {NoStop}%
\bibitem [{\citenamefont {Hirata}(2009)}]{Hirata2009}%
  \BibitemOpen
  \bibfield  {author} {\bibinfo {author} {\bibfnamefont {S.}~\bibnamefont
  {Hirata}},\ }\href {\doibase 10.1039/B905812P} {\bibfield  {journal}
  {\bibinfo  {journal} {Phys. Chem. Chem. Phys.}\ }\textbf {\bibinfo {volume}
  {11}},\ \bibinfo {pages} {8397} (\bibinfo {year} {2009})}\BibitemShut
  {NoStop}%
\bibitem [{\citenamefont {Albuquerque}\ \emph {et~al.}(2021)\citenamefont
  {Albuquerque}, \citenamefont {Fulco}, \citenamefont {Caetano},\ and\
  \citenamefont {Freire}}]{albuquerque2021quantum}%
  \BibitemOpen
  \bibfield  {author} {\bibinfo {author} {\bibfnamefont {E.~L.}\ \bibnamefont
  {Albuquerque}}, \bibinfo {author} {\bibfnamefont {U.~L.}\ \bibnamefont
  {Fulco}}, \bibinfo {author} {\bibfnamefont {E.~W.}\ \bibnamefont {Caetano}},
  \ and\ \bibinfo {author} {\bibfnamefont {V.~N.}\ \bibnamefont {Freire}},\
  }\href@noop {} {\emph {\bibinfo {title} {Quantum chemistry simulation of
  biological molecules}}}\ (\bibinfo  {publisher} {Cambridge University
  Press},\ \bibinfo {year} {2021})\BibitemShut {NoStop}%
\bibitem [{\citenamefont {Keith}\ \emph {et~al.}(2021)\citenamefont {Keith},
  \citenamefont {Vassilev-Galindo}, \citenamefont {Cheng}, \citenamefont
  {Chmiela}, \citenamefont {Gastegger}, \citenamefont {Müller},\ and\
  \citenamefont {Tkatchenko}}]{Keith2021Jul}%
  \BibitemOpen
  \bibfield  {author} {\bibinfo {author} {\bibfnamefont {J.~A.}\ \bibnamefont
  {Keith}}, \bibinfo {author} {\bibfnamefont {V.}~\bibnamefont
  {Vassilev-Galindo}}, \bibinfo {author} {\bibfnamefont {B.}~\bibnamefont
  {Cheng}}, \bibinfo {author} {\bibfnamefont {S.}~\bibnamefont {Chmiela}},
  \bibinfo {author} {\bibfnamefont {M.}~\bibnamefont {Gastegger}}, \bibinfo
  {author} {\bibfnamefont {K.-R.}\ \bibnamefont {Müller}}, \ and\ \bibinfo
  {author} {\bibfnamefont {A.}~\bibnamefont {Tkatchenko}},\ }\href {\doibase
  10.1021/acs.chemrev.1c00107} {\bibfield  {journal} {\bibinfo  {journal}
  {Chem. Rev.}\ }\textbf {\bibinfo {volume} {121}},\ \bibinfo {pages} {9816}
  (\bibinfo {year} {2021})}\BibitemShut {NoStop}%
\bibitem [{\citenamefont {Cao}\ \emph {et~al.}(2019)\citenamefont {Cao},
  \citenamefont {Romero}, \citenamefont {Olson}, \citenamefont {Degroote},
  \citenamefont {Johnson}, \citenamefont {Kieferová}, \citenamefont
  {Kivlichan}, \citenamefont {Menke}, \citenamefont {Peropadre}, \citenamefont
  {Sawaya}, \citenamefont {Sim}, \citenamefont {Veis},\ and\ \citenamefont
  {Aspuru-Guzik}}]{cao2019quantum}%
  \BibitemOpen
  \bibfield  {author} {\bibinfo {author} {\bibfnamefont {Y.}~\bibnamefont
  {Cao}}, \bibinfo {author} {\bibfnamefont {J.}~\bibnamefont {Romero}},
  \bibinfo {author} {\bibfnamefont {J.~P.}\ \bibnamefont {Olson}}, \bibinfo
  {author} {\bibfnamefont {M.}~\bibnamefont {Degroote}}, \bibinfo {author}
  {\bibfnamefont {P.~D.}\ \bibnamefont {Johnson}}, \bibinfo {author}
  {\bibfnamefont {M.}~\bibnamefont {Kieferová}}, \bibinfo {author}
  {\bibfnamefont {I.~D.}\ \bibnamefont {Kivlichan}}, \bibinfo {author}
  {\bibfnamefont {T.}~\bibnamefont {Menke}}, \bibinfo {author} {\bibfnamefont
  {B.}~\bibnamefont {Peropadre}}, \bibinfo {author} {\bibfnamefont {N.~P.~D.}\
  \bibnamefont {Sawaya}}, \bibinfo {author} {\bibfnamefont {S.}~\bibnamefont
  {Sim}}, \bibinfo {author} {\bibfnamefont {L.}~\bibnamefont {Veis}}, \ and\
  \bibinfo {author} {\bibfnamefont {A.}~\bibnamefont {Aspuru-Guzik}},\ }\href
  {\doibase 10.1021/acs.chemrev.8b00803} {\bibfield  {journal} {\bibinfo
  {journal} {Chem. Rev.}\ }\textbf {\bibinfo {volume} {119}},\ \bibinfo {pages}
  {10856} (\bibinfo {year} {2019})}\BibitemShut {NoStop}%
\bibitem [{\citenamefont {Westermayr}\ \emph {et~al.}(2021)\citenamefont
  {Westermayr}, \citenamefont {Gastegger}, \citenamefont {Sch{\"u}tt},\ and\
  \citenamefont {Maurer}}]{westermayr2021perspective}%
  \BibitemOpen
  \bibfield  {author} {\bibinfo {author} {\bibfnamefont {J.}~\bibnamefont
  {Westermayr}}, \bibinfo {author} {\bibfnamefont {M.}~\bibnamefont
  {Gastegger}}, \bibinfo {author} {\bibfnamefont {K.~T.}\ \bibnamefont
  {Sch{\"u}tt}}, \ and\ \bibinfo {author} {\bibfnamefont {R.~J.}\ \bibnamefont
  {Maurer}},\ }\href@noop {} {\bibfield  {journal} {\bibinfo  {journal} {J.
  Chem. Phys.}\ }\textbf {\bibinfo {volume} {154}} (\bibinfo {year}
  {2021})}\BibitemShut {NoStop}%
\bibitem [{\citenamefont {N{\o}rskov}\ \emph {et~al.}(2009)\citenamefont
  {N{\o}rskov}, \citenamefont {Bligaard}, \citenamefont {Rossmeisl},\ and\
  \citenamefont {Christensen}}]{norskov2009towards}%
  \BibitemOpen
  \bibfield  {author} {\bibinfo {author} {\bibfnamefont {J.~K.}\ \bibnamefont
  {N{\o}rskov}}, \bibinfo {author} {\bibfnamefont {T.}~\bibnamefont
  {Bligaard}}, \bibinfo {author} {\bibfnamefont {J.}~\bibnamefont {Rossmeisl}},
  \ and\ \bibinfo {author} {\bibfnamefont {C.~H.}\ \bibnamefont
  {Christensen}},\ }\href {\doibase 10.1038/nchem.121} {\bibfield  {journal}
  {\bibinfo  {journal} {Nat. Chem.}\ }\textbf {\bibinfo {volume} {1}},\
  \bibinfo {pages} {37} (\bibinfo {year} {2009})}\BibitemShut {NoStop}%
\bibitem [{\citenamefont {Qin}\ \emph {et~al.}(2020)\citenamefont {Qin},
  \citenamefont {Liu}, \citenamefont {Hu},\ and\ \citenamefont
  {Yang}}]{Qin2020Jun}%
  \BibitemOpen
  \bibfield  {author} {\bibinfo {author} {\bibfnamefont {X.}~\bibnamefont
  {Qin}}, \bibinfo {author} {\bibfnamefont {J.}~\bibnamefont {Liu}}, \bibinfo
  {author} {\bibfnamefont {W.}~\bibnamefont {Hu}}, \ and\ \bibinfo {author}
  {\bibfnamefont {J.}~\bibnamefont {Yang}},\ }\href {\doibase
  10.1021/acs.jpca.0c02826} {\bibfield  {journal} {\bibinfo  {journal} {The
  Journal of Physical Chemistry A}\ }\textbf {\bibinfo {volume} {124}},\
  \bibinfo {pages} {5664} (\bibinfo {year} {2020})}\BibitemShut {NoStop}%
\bibitem [{\citenamefont {Lee}\ \emph {et~al.}(2022{\natexlab{a}})\citenamefont
  {Lee}, \citenamefont {Rettig}, \citenamefont {Feng}, \citenamefont
  {Epifanovsky},\ and\ \citenamefont {Head-Gordon}}]{Lee2022Dec}%
  \BibitemOpen
  \bibfield  {author} {\bibinfo {author} {\bibfnamefont {J.}~\bibnamefont
  {Lee}}, \bibinfo {author} {\bibfnamefont {A.}~\bibnamefont {Rettig}},
  \bibinfo {author} {\bibfnamefont {X.}~\bibnamefont {Feng}}, \bibinfo {author}
  {\bibfnamefont {E.}~\bibnamefont {Epifanovsky}}, \ and\ \bibinfo {author}
  {\bibfnamefont {M.}~\bibnamefont {Head-Gordon}},\ }\href {\doibase
  10.1021/acs.jctc.2c00742} {\bibfield  {journal} {\bibinfo  {journal} {J.
  Chem. Theory Comput.}\ }\textbf {\bibinfo {volume} {18}},\ \bibinfo {pages}
  {7336} (\bibinfo {year} {2022}{\natexlab{a}})}\BibitemShut {NoStop}%
\bibitem [{\citenamefont {Sharma}\ \emph {et~al.}(2022)\citenamefont {Sharma},
  \citenamefont {White},\ and\ \citenamefont {Beylkin}}]{Sharma2022Nov}%
  \BibitemOpen
  \bibfield  {author} {\bibinfo {author} {\bibfnamefont {S.}~\bibnamefont
  {Sharma}}, \bibinfo {author} {\bibfnamefont {A.~F.}\ \bibnamefont {White}}, \
  and\ \bibinfo {author} {\bibfnamefont {G.}~\bibnamefont {Beylkin}},\ }\href
  {\doibase 10.1021/acs.jctc.2c00720} {\bibfield  {journal} {\bibinfo
  {journal} {J. Chem. Theory Comput.}\ }\textbf {\bibinfo {volume} {18}},\
  \bibinfo {pages} {7306} (\bibinfo {year} {2022})}\BibitemShut {NoStop}%
\bibitem [{\citenamefont {Rettig}\ \emph {et~al.}(2023)\citenamefont {Rettig},
  \citenamefont {Lee},\ and\ \citenamefont {Head-Gordon}}]{rettig_even_2023}%
  \BibitemOpen
  \bibfield  {author} {\bibinfo {author} {\bibfnamefont {A.}~\bibnamefont
  {Rettig}}, \bibinfo {author} {\bibfnamefont {J.}~\bibnamefont {Lee}}, \ and\
  \bibinfo {author} {\bibfnamefont {M.}~\bibnamefont {Head-Gordon}},\ }\href
  {\doibase 10.1021/acs.jctc.3c00407} {\bibfield  {journal} {\bibinfo
  {journal} {J. Chem. Theory Comput.}\ }\textbf {\bibinfo {volume} {19}},\
  \bibinfo {pages} {5773} (\bibinfo {year} {2023})}\BibitemShut {NoStop}%
\bibitem [{\citenamefont {Bussy}\ and\ \citenamefont
  {Hutter}(2024)}]{Bussy2024Feb}%
  \BibitemOpen
  \bibfield  {author} {\bibinfo {author} {\bibfnamefont {A.}~\bibnamefont
  {Bussy}}\ and\ \bibinfo {author} {\bibfnamefont {J.}~\bibnamefont {Hutter}},\
  }\href {\doibase 10.1063/5.0189659} {\bibfield  {journal} {\bibinfo
  {journal} {J. Chem. Phys.}\ }\textbf {\bibinfo {volume} {160}},\ \bibinfo
  {pages} {064116} (\bibinfo {year} {2024})}\BibitemShut {NoStop}%
\bibitem [{\citenamefont {Ren}\ \emph {et~al.}(2012)\citenamefont {Ren},
  \citenamefont {Rinke}, \citenamefont {Joas},\ and\ \citenamefont
  {Scheffler}}]{Ren2012Nov}%
  \BibitemOpen
  \bibfield  {author} {\bibinfo {author} {\bibfnamefont {X.}~\bibnamefont
  {Ren}}, \bibinfo {author} {\bibfnamefont {P.}~\bibnamefont {Rinke}}, \bibinfo
  {author} {\bibfnamefont {C.}~\bibnamefont {Joas}}, \ and\ \bibinfo {author}
  {\bibfnamefont {M.}~\bibnamefont {Scheffler}},\ }\href {\doibase
  10.1007/s10853-012-6570-4} {\bibfield  {journal} {\bibinfo  {journal} {J.
  Mater. Sci.}\ }\textbf {\bibinfo {volume} {47}},\ \bibinfo {pages} {7447}
  (\bibinfo {year} {2012})}\BibitemShut {NoStop}%
\bibitem [{\citenamefont {Grundei}\ and\ \citenamefont
  {Burow}(2017{\natexlab{a}})}]{Grundei2017Feb}%
  \BibitemOpen
  \bibfield  {author} {\bibinfo {author} {\bibfnamefont {M.~M.~J.}\
  \bibnamefont {Grundei}}\ and\ \bibinfo {author} {\bibfnamefont {A.~M.}\
  \bibnamefont {Burow}},\ }\href {\doibase 10.1021/acs.jctc.6b01146} {\bibfield
   {journal} {\bibinfo  {journal} {J. Chem. Theory Comput.}\ }\textbf {\bibinfo
  {volume} {13}},\ \bibinfo {pages} {1159} (\bibinfo {year}
  {2017}{\natexlab{a}})}\BibitemShut {NoStop}%
\bibitem [{\citenamefont {Yeh}\ and\ \citenamefont
  {Morales}(2023{\natexlab{a}})}]{Yeh2023Sep}%
  \BibitemOpen
  \bibfield  {author} {\bibinfo {author} {\bibfnamefont {C.-N.}\ \bibnamefont
  {Yeh}}\ and\ \bibinfo {author} {\bibfnamefont {M.~A.}\ \bibnamefont
  {Morales}},\ }\href {\doibase 10.1021/acs.jctc.3c00615} {\bibfield  {journal}
  {\bibinfo  {journal} {J. Chem. Theory Comput.}\ }\textbf {\bibinfo {volume}
  {19}},\ \bibinfo {pages} {6197} (\bibinfo {year}
  {2023}{\natexlab{a}})}\BibitemShut {NoStop}%
\bibitem [{\citenamefont {Pisani}\ \emph {et~al.}(2008)\citenamefont {Pisani},
  \citenamefont {Maschio}, \citenamefont {Casassa}, \citenamefont {Halo},
  \citenamefont {Sch{\ifmmode\ddot{u}\else\"{u}\fi}tz},\ and\ \citenamefont
  {Usvyat}}]{Pisani2008Oct}%
  \BibitemOpen
  \bibfield  {author} {\bibinfo {author} {\bibfnamefont {C.}~\bibnamefont
  {Pisani}}, \bibinfo {author} {\bibfnamefont {L.}~\bibnamefont {Maschio}},
  \bibinfo {author} {\bibfnamefont {S.}~\bibnamefont {Casassa}}, \bibinfo
  {author} {\bibfnamefont {M.}~\bibnamefont {Halo}}, \bibinfo {author}
  {\bibfnamefont {M.}~\bibnamefont {Sch{\ifmmode\ddot{u}\else\"{u}\fi}tz}}, \
  and\ \bibinfo {author} {\bibfnamefont {D.}~\bibnamefont {Usvyat}},\ }\href
  {\doibase 10.1002/jcc.20975} {\bibfield  {journal} {\bibinfo  {journal} {J.
  Comput. Chem.}\ }\textbf {\bibinfo {volume} {29}},\ \bibinfo {pages} {2113}
  (\bibinfo {year} {2008})}\BibitemShut {NoStop}%
\bibitem [{\citenamefont {Del~Ben}\ \emph
  {et~al.}(2012{\natexlab{a}})\citenamefont {Del~Ben}, \citenamefont {Hutter},\
  and\ \citenamefont {VandeVondele}}]{del2012second}%
  \BibitemOpen
  \bibfield  {author} {\bibinfo {author} {\bibfnamefont {M.}~\bibnamefont
  {Del~Ben}}, \bibinfo {author} {\bibfnamefont {J.}~\bibnamefont {Hutter}}, \
  and\ \bibinfo {author} {\bibfnamefont {J.}~\bibnamefont {VandeVondele}},\
  }\href@noop {} {\bibfield  {journal} {\bibinfo  {journal} {J. Chem. Theory
  Comput.}\ }\textbf {\bibinfo {volume} {8}},\ \bibinfo {pages} {4177}
  (\bibinfo {year} {2012}{\natexlab{a}})}\BibitemShut {NoStop}%
\bibitem [{\citenamefont {Foulkes}\ \emph {et~al.}(2001)\citenamefont
  {Foulkes}, \citenamefont {Mitas}, \citenamefont {Needs},\ and\ \citenamefont
  {Rajagopal}}]{Foulkes2001Jan}%
  \BibitemOpen
  \bibfield  {author} {\bibinfo {author} {\bibfnamefont {W.~M.~C.}\
  \bibnamefont {Foulkes}}, \bibinfo {author} {\bibfnamefont {L.}~\bibnamefont
  {Mitas}}, \bibinfo {author} {\bibfnamefont {R.~J.}\ \bibnamefont {Needs}}, \
  and\ \bibinfo {author} {\bibfnamefont {G.}~\bibnamefont {Rajagopal}},\ }\href
  {\doibase 10.1103/RevModPhys.73.33} {\bibfield  {journal} {\bibinfo
  {journal} {Rev. Mod. Phys.}\ }\textbf {\bibinfo {volume} {73}},\ \bibinfo
  {pages} {33} (\bibinfo {year} {2001})}\BibitemShut {NoStop}%
\bibitem [{\citenamefont {Booth}\ \emph {et~al.}(2013)\citenamefont {Booth},
  \citenamefont {Gr{\"u}neis}, \citenamefont {Kresse},\ and\ \citenamefont
  {Alavi}}]{Booth2013Jan}%
  \BibitemOpen
  \bibfield  {author} {\bibinfo {author} {\bibfnamefont {G.~H.}\ \bibnamefont
  {Booth}}, \bibinfo {author} {\bibfnamefont {A.}~\bibnamefont {Gr{\"u}neis}},
  \bibinfo {author} {\bibfnamefont {G.}~\bibnamefont {Kresse}}, \ and\ \bibinfo
  {author} {\bibfnamefont {A.}~\bibnamefont {Alavi}},\ }\href {\doibase
  10.1038/nature11770} {\bibfield  {journal} {\bibinfo  {journal} {Nature}\
  }\textbf {\bibinfo {volume} {493}},\ \bibinfo {pages} {365} (\bibinfo {year}
  {2013})}\BibitemShut {NoStop}%
\bibitem [{\citenamefont {Motta}\ \emph {et~al.}(2019)\citenamefont {Motta},
  \citenamefont {Zhang},\ and\ \citenamefont {Chan}}]{Motta2019Jul}%
  \BibitemOpen
  \bibfield  {author} {\bibinfo {author} {\bibfnamefont {M.}~\bibnamefont
  {Motta}}, \bibinfo {author} {\bibfnamefont {S.}~\bibnamefont {Zhang}}, \ and\
  \bibinfo {author} {\bibfnamefont {G.~K.-L.}\ \bibnamefont {Chan}},\ }\href
  {\doibase 10.1103/PhysRevB.100.045127} {\bibfield  {journal} {\bibinfo
  {journal} {Phys. Rev. B}\ }\textbf {\bibinfo {volume} {100}},\ \bibinfo
  {pages} {045127} (\bibinfo {year} {2019})}\BibitemShut {NoStop}%
\bibitem [{\citenamefont {Malone}\ \emph
  {et~al.}(2020{\natexlab{a}})\citenamefont {Malone}, \citenamefont {Zhang},\
  and\ \citenamefont {Morales}}]{Malone2020Jul}%
  \BibitemOpen
  \bibfield  {author} {\bibinfo {author} {\bibfnamefont {F.~D.}\ \bibnamefont
  {Malone}}, \bibinfo {author} {\bibfnamefont {S.}~\bibnamefont {Zhang}}, \
  and\ \bibinfo {author} {\bibfnamefont {M.~A.}\ \bibnamefont {Morales}},\
  }\href {\doibase 10.1021/acs.jctc.0c00262} {\bibfield  {journal} {\bibinfo
  {journal} {J. Chem. Theory Comput.}\ }\textbf {\bibinfo {volume} {16}},\
  \bibinfo {pages} {4286} (\bibinfo {year} {2020}{\natexlab{a}})}\BibitemShut
  {NoStop}%
\bibitem [{\citenamefont {McClain}\ \emph {et~al.}(2017)\citenamefont
  {McClain}, \citenamefont {Sun}, \citenamefont {Chan},\ and\ \citenamefont
  {Berkelbach}}]{McClain2017Mar}%
  \BibitemOpen
  \bibfield  {author} {\bibinfo {author} {\bibfnamefont {J.}~\bibnamefont
  {McClain}}, \bibinfo {author} {\bibfnamefont {Q.}~\bibnamefont {Sun}},
  \bibinfo {author} {\bibfnamefont {G.~K.-L.}\ \bibnamefont {Chan}}, \ and\
  \bibinfo {author} {\bibfnamefont {T.~C.}\ \bibnamefont {Berkelbach}},\ }\href
  {\doibase 10.1021/acs.jctc.7b00049} {\bibfield  {journal} {\bibinfo
  {journal} {J. Chem. Theory Comput.}\ }\textbf {\bibinfo {volume} {13}},\
  \bibinfo {pages} {1209} (\bibinfo {year} {2017})}\BibitemShut {NoStop}%
\bibitem [{\citenamefont {Gruber}\ \emph {et~al.}(2018)\citenamefont {Gruber},
  \citenamefont {Liao}, \citenamefont {Tsatsoulis}, \citenamefont {Hummel},\
  and\ \citenamefont {Gr{\ifmmode\ddot{u}\else\"{u}\fi}neis}}]{Gruber2018May}%
  \BibitemOpen
  \bibfield  {author} {\bibinfo {author} {\bibfnamefont {T.}~\bibnamefont
  {Gruber}}, \bibinfo {author} {\bibfnamefont {K.}~\bibnamefont {Liao}},
  \bibinfo {author} {\bibfnamefont {T.}~\bibnamefont {Tsatsoulis}}, \bibinfo
  {author} {\bibfnamefont {F.}~\bibnamefont {Hummel}}, \ and\ \bibinfo {author}
  {\bibfnamefont {A.}~\bibnamefont {Gr{\ifmmode\ddot{u}\else\"{u}\fi}neis}},\
  }\href {\doibase 10.1103/PhysRevX.8.021043} {\bibfield  {journal} {\bibinfo
  {journal} {Phys. Rev. X}\ }\textbf {\bibinfo {volume} {8}},\ \bibinfo {pages}
  {021043} (\bibinfo {year} {2018})}\BibitemShut {NoStop}%
\bibitem [{\citenamefont {Zhang}\ and\ \citenamefont
  {Gr{\ifmmode\ddot{u}\else\"{u}\fi}neis}(2019)}]{Zhang2019Jun}%
  \BibitemOpen
  \bibfield  {author} {\bibinfo {author} {\bibfnamefont {I.~Y.}\ \bibnamefont
  {Zhang}}\ and\ \bibinfo {author} {\bibfnamefont {A.}~\bibnamefont
  {Gr{\ifmmode\ddot{u}\else\"{u}\fi}neis}},\ }\href {\doibase
  10.3389/fmats.2019.00123} {\bibfield  {journal} {\bibinfo  {journal} {Front.
  Mater.}\ }\textbf {\bibinfo {volume} {6}},\ \bibinfo {pages} {432749}
  (\bibinfo {year} {2019})}\BibitemShut {NoStop}%
\bibitem [{\citenamefont {Hoffmann}(1987)}]{Hoffmann1987Sep}%
  \BibitemOpen
  \bibfield  {author} {\bibinfo {author} {\bibfnamefont {R.}~\bibnamefont
  {Hoffmann}},\ }\href {\doibase 10.1002/anie.198708461} {\bibfield  {journal}
  {\bibinfo  {journal} {Angew. Chem., Int. Ed. Engl.}\ }\textbf {\bibinfo
  {volume} {26}},\ \bibinfo {pages} {846} (\bibinfo {year} {1987})}\BibitemShut
  {NoStop}%
\bibitem [{\citenamefont {Monkhorst}\ and\ \citenamefont
  {Pack}(1976)}]{monkhorst1976special}%
  \BibitemOpen
  \bibfield  {author} {\bibinfo {author} {\bibfnamefont {H.~J.}\ \bibnamefont
  {Monkhorst}}\ and\ \bibinfo {author} {\bibfnamefont {J.~D.}\ \bibnamefont
  {Pack}},\ }\href@noop {} {\bibfield  {journal} {\bibinfo  {journal} {Phys.
  Rev. B}\ }\textbf {\bibinfo {volume} {13}},\ \bibinfo {pages} {5188}
  (\bibinfo {year} {1976})}\BibitemShut {NoStop}%
\bibitem [{\citenamefont {Drummond}\ \emph {et~al.}(2008)\citenamefont
  {Drummond}, \citenamefont {Needs}, \citenamefont {Sorouri},\ and\
  \citenamefont {Foulkes}}]{Drummond2008Jun}%
  \BibitemOpen
  \bibfield  {author} {\bibinfo {author} {\bibfnamefont {N.~D.}\ \bibnamefont
  {Drummond}}, \bibinfo {author} {\bibfnamefont {R.~J.}\ \bibnamefont {Needs}},
  \bibinfo {author} {\bibfnamefont {A.}~\bibnamefont {Sorouri}}, \ and\
  \bibinfo {author} {\bibfnamefont {W.~M.~C.}\ \bibnamefont {Foulkes}},\ }\href
  {\doibase 10.1103/PhysRevB.78.125106} {\bibfield  {journal} {\bibinfo
  {journal} {Phys. Rev. B}\ }\textbf {\bibinfo {volume} {78}},\ \bibinfo
  {pages} {125106} (\bibinfo {year} {2008})}\BibitemShut {NoStop}%
\bibitem [{\citenamefont {Xing}\ \emph {et~al.}(2024)\citenamefont {Xing},
  \citenamefont {Li},\ and\ \citenamefont {Lin}}]{xing2024unified}%
  \BibitemOpen
  \bibfield  {author} {\bibinfo {author} {\bibfnamefont {X.}~\bibnamefont
  {Xing}}, \bibinfo {author} {\bibfnamefont {X.}~\bibnamefont {Li}}, \ and\
  \bibinfo {author} {\bibfnamefont {L.}~\bibnamefont {Lin}},\ }\href@noop {}
  {\bibfield  {journal} {\bibinfo  {journal} {Math. Comput.}\ }\textbf
  {\bibinfo {volume} {93}},\ \bibinfo {pages} {679} (\bibinfo {year}
  {2024})}\BibitemShut {NoStop}%
\bibitem [{\citenamefont {Janesko}\ \emph {et~al.}(2009)\citenamefont
  {Janesko}, \citenamefont {Henderson},\ and\ \citenamefont
  {Scuseria}}]{Janesko2009}%
  \BibitemOpen
  \bibfield  {author} {\bibinfo {author} {\bibfnamefont {B.~G.}\ \bibnamefont
  {Janesko}}, \bibinfo {author} {\bibfnamefont {T.~M.}\ \bibnamefont
  {Henderson}}, \ and\ \bibinfo {author} {\bibfnamefont {G.~E.}\ \bibnamefont
  {Scuseria}},\ }\href {\doibase 10.1039/B812838C} {\bibfield  {journal}
  {\bibinfo  {journal} {Phys. Chem. Chem. Phys.}\ }\textbf {\bibinfo {volume}
  {11}},\ \bibinfo {pages} {443} (\bibinfo {year} {2009})}\BibitemShut
  {NoStop}%
\bibitem [{\citenamefont {Evarestov}(2007)}]{evarestov2007quantum}%
  \BibitemOpen
  \bibfield  {author} {\bibinfo {author} {\bibfnamefont {R.~A.}\ \bibnamefont
  {Evarestov}},\ }\href@noop {} {\emph {\bibinfo {title} {Quantum chemistry of
  solids: the LCAO first principles treatment of crystals}}},\ Vol.\ \bibinfo
  {volume} {153}\ (\bibinfo  {publisher} {Springer Science \& Business Media},\
  \bibinfo {year} {2007})\BibitemShut {NoStop}%
\bibitem [{\citenamefont {Huzinaga}\ \emph {et~al.}(1984)\citenamefont
  {Huzinaga}, \citenamefont {Andzelm}, \citenamefont {Kłobukowski},
  \citenamefont {Radzio-Andzelm}, \citenamefont {Sakai},\ and\ \citenamefont
  {Tatewaki}}]{huzinaga2012gaussian}%
  \BibitemOpen
  \bibfield  {author} {\bibinfo {author} {\bibfnamefont {S.}~\bibnamefont
  {Huzinaga}}, \bibinfo {author} {\bibfnamefont {J.}~\bibnamefont {Andzelm}},
  \bibinfo {author} {\bibfnamefont {M.}~\bibnamefont {Kłobukowski}}, \bibinfo
  {author} {\bibfnamefont {E.}~\bibnamefont {Radzio-Andzelm}}, \bibinfo
  {author} {\bibfnamefont {Y.}~\bibnamefont {Sakai}}, \ and\ \bibinfo {author}
  {\bibfnamefont {H.}~\bibnamefont {Tatewaki}},\ }\href@noop {} {\emph
  {\bibinfo {title} {Gaussian Basis Sets for Molecular Calculations}}},\
  \bibinfo {series} {Gaussian basis sets for molecular calculations}\ No.\
  \bibinfo {number} {v. 16}\ (\bibinfo  {publisher} {Elsevier Science},\
  \bibinfo {year} {1984})\BibitemShut {NoStop}%
\bibitem [{\citenamefont {Dunning}\ and\ \citenamefont
  {Hay}(1977)}]{dunning1977gaussian}%
  \BibitemOpen
  \bibfield  {author} {\bibinfo {author} {\bibfnamefont {T.~H.}\ \bibnamefont
  {Dunning}}\ and\ \bibinfo {author} {\bibfnamefont {P.~J.}\ \bibnamefont
  {Hay}},\ }\enquote {\bibinfo {title} {Gaussian basis sets for molecular
  calculations},}\ in\ \href {\doibase 10.1007/978-1-4757-0887-5_1} {\emph
  {\bibinfo {booktitle} {Methods of Electronic Structure Theory}}},\ \bibinfo
  {editor} {edited by\ \bibinfo {editor} {\bibfnamefont {H.~F.}\ \bibnamefont
  {Schaefer}}}\ (\bibinfo  {publisher} {Springer US},\ \bibinfo {address}
  {Boston, MA},\ \bibinfo {year} {1977})\ pp.\ \bibinfo {pages}
  {1--27}\BibitemShut {NoStop}%
\bibitem [{\citenamefont {Kato}(1957)}]{kato1957eigenfunctions}%
  \BibitemOpen
  \bibfield  {author} {\bibinfo {author} {\bibfnamefont {T.}~\bibnamefont
  {Kato}},\ }\href@noop {} {\bibfield  {journal} {\bibinfo  {journal} {Commun.
  Pure Appl. Math.}\ }\textbf {\bibinfo {volume} {10}},\ \bibinfo {pages} {151}
  (\bibinfo {year} {1957})}\BibitemShut {NoStop}%
\bibitem [{\citenamefont {Pritchard}\ \emph {et~al.}(2019)\citenamefont
  {Pritchard}, \citenamefont {Altarawy}, \citenamefont {Didier}, \citenamefont
  {Gibson},\ and\ \citenamefont {Windus}}]{pritchard2019new}%
  \BibitemOpen
  \bibfield  {author} {\bibinfo {author} {\bibfnamefont {B.~P.}\ \bibnamefont
  {Pritchard}}, \bibinfo {author} {\bibfnamefont {D.}~\bibnamefont {Altarawy}},
  \bibinfo {author} {\bibfnamefont {B.}~\bibnamefont {Didier}}, \bibinfo
  {author} {\bibfnamefont {T.~D.}\ \bibnamefont {Gibson}}, \ and\ \bibinfo
  {author} {\bibfnamefont {T.~L.}\ \bibnamefont {Windus}},\ }\href@noop {}
  {\bibfield  {journal} {\bibinfo  {journal} {J. Chem. Inf. Model.}\ }\textbf
  {\bibinfo {volume} {59}},\ \bibinfo {pages} {4814} (\bibinfo {year}
  {2019})}\BibitemShut {NoStop}%
\bibitem [{\citenamefont {Sch{\"a}fer}\ \emph {et~al.}(1992)\citenamefont
  {Sch{\"a}fer}, \citenamefont {Horn},\ and\ \citenamefont
  {Ahlrichs}}]{schafer1992fully}%
  \BibitemOpen
  \bibfield  {author} {\bibinfo {author} {\bibfnamefont {A.}~\bibnamefont
  {Sch{\"a}fer}}, \bibinfo {author} {\bibfnamefont {H.}~\bibnamefont {Horn}}, \
  and\ \bibinfo {author} {\bibfnamefont {R.}~\bibnamefont {Ahlrichs}},\
  }\href@noop {} {\bibfield  {journal} {\bibinfo  {journal} {J. Chem. Phys.}\
  }\textbf {\bibinfo {volume} {97}},\ \bibinfo {pages} {2571} (\bibinfo {year}
  {1992})}\BibitemShut {NoStop}%
\bibitem [{\citenamefont {Sch{\"a}fer}\ \emph {et~al.}(1994)\citenamefont
  {Sch{\"a}fer}, \citenamefont {Huber},\ and\ \citenamefont
  {Ahlrichs}}]{schafer1994fully}%
  \BibitemOpen
  \bibfield  {author} {\bibinfo {author} {\bibfnamefont {A.}~\bibnamefont
  {Sch{\"a}fer}}, \bibinfo {author} {\bibfnamefont {C.}~\bibnamefont {Huber}},
  \ and\ \bibinfo {author} {\bibfnamefont {R.}~\bibnamefont {Ahlrichs}},\
  }\href@noop {} {\bibfield  {journal} {\bibinfo  {journal} {J. Chem. Phys.}\
  }\textbf {\bibinfo {volume} {100}},\ \bibinfo {pages} {5829} (\bibinfo {year}
  {1994})}\BibitemShut {NoStop}%
\bibitem [{\citenamefont {Dunning~Jr}(1989)}]{dunning1989gaussian}%
  \BibitemOpen
  \bibfield  {author} {\bibinfo {author} {\bibfnamefont {T.~H.}\ \bibnamefont
  {Dunning~Jr}},\ }\href@noop {} {\bibfield  {journal} {\bibinfo  {journal} {J.
  Chem. Phys.}\ }\textbf {\bibinfo {volume} {90}},\ \bibinfo {pages} {1007}
  (\bibinfo {year} {1989})}\BibitemShut {NoStop}%
\bibitem [{\citenamefont {Woon}\ and\ \citenamefont
  {Dunning~Jr}(1993)}]{woon1993gaussian}%
  \BibitemOpen
  \bibfield  {author} {\bibinfo {author} {\bibfnamefont {D.~E.}\ \bibnamefont
  {Woon}}\ and\ \bibinfo {author} {\bibfnamefont {T.~H.}\ \bibnamefont
  {Dunning~Jr}},\ }\href@noop {} {\bibfield  {journal} {\bibinfo  {journal} {J.
  Chem. Phys.}\ }\textbf {\bibinfo {volume} {98}},\ \bibinfo {pages} {1358}
  (\bibinfo {year} {1993})}\BibitemShut {NoStop}%
\bibitem [{\citenamefont {Jensen}(2001)}]{jensen2001polarization}%
  \BibitemOpen
  \bibfield  {author} {\bibinfo {author} {\bibfnamefont {F.}~\bibnamefont
  {Jensen}},\ }\href@noop {} {\bibfield  {journal} {\bibinfo  {journal} {J.
  Chem. Phys.}\ }\textbf {\bibinfo {volume} {115}},\ \bibinfo {pages} {9113}
  (\bibinfo {year} {2001})}\BibitemShut {NoStop}%
\bibitem [{\citenamefont {Almlöf}\ and\ \citenamefont
  {Taylor}(1991)}]{almlof1991atomic}%
  \BibitemOpen
  \bibfield  {author} {\bibinfo {author} {\bibfnamefont {J.}~\bibnamefont
  {Almlöf}}\ and\ \bibinfo {author} {\bibfnamefont {P.~R.}\ \bibnamefont
  {Taylor}}\ }(\bibinfo  {publisher} {Academic Press},\ \bibinfo {year}
  {1991})\ pp.\ \bibinfo {pages} {301--373}\BibitemShut {NoStop}%
\bibitem [{\citenamefont {Neese}\ and\ \citenamefont
  {Valeev}(2011)}]{neese2011revisiting}%
  \BibitemOpen
  \bibfield  {author} {\bibinfo {author} {\bibfnamefont {F.}~\bibnamefont
  {Neese}}\ and\ \bibinfo {author} {\bibfnamefont {E.~F.}\ \bibnamefont
  {Valeev}},\ }\href@noop {} {\bibfield  {journal} {\bibinfo  {journal} {J.
  Chem. Theory Comput.}\ }\textbf {\bibinfo {volume} {7}},\ \bibinfo {pages}
  {33} (\bibinfo {year} {2011})}\BibitemShut {NoStop}%
\bibitem [{\citenamefont {Kirschner}\ \emph {et~al.}(2020)\citenamefont
  {Kirschner}, \citenamefont {Reith},\ and\ \citenamefont
  {Heiden}}]{kirschner2020performance}%
  \BibitemOpen
  \bibfield  {author} {\bibinfo {author} {\bibfnamefont {K.~N.}\ \bibnamefont
  {Kirschner}}, \bibinfo {author} {\bibfnamefont {D.}~\bibnamefont {Reith}}, \
  and\ \bibinfo {author} {\bibfnamefont {W.}~\bibnamefont {Heiden}},\
  }\href@noop {} {\bibfield  {journal} {\bibinfo  {journal} {Soft Matter.}\
  }\textbf {\bibinfo {volume} {18}},\ \bibinfo {pages} {200} (\bibinfo {year}
  {2020})}\BibitemShut {NoStop}%
\bibitem [{\citenamefont {Klahn}\ and\ \citenamefont
  {Bingel}(1977)}]{klahn1977completeness}%
  \BibitemOpen
  \bibfield  {author} {\bibinfo {author} {\bibfnamefont {B.}~\bibnamefont
  {Klahn}}\ and\ \bibinfo {author} {\bibfnamefont {W.~A.}\ \bibnamefont
  {Bingel}},\ }\href@noop {} {\bibfield  {journal} {\bibinfo  {journal} {Int.
  J. Quantum Chem.}\ }\textbf {\bibinfo {volume} {11}},\ \bibinfo {pages} {943}
  (\bibinfo {year} {1977})}\BibitemShut {NoStop}%
\bibitem [{\citenamefont {Peintinger}\ \emph {et~al.}(2013)\citenamefont
  {Peintinger}, \citenamefont {Oliveira},\ and\ \citenamefont
  {Bredow}}]{peintinger2013consistent}%
  \BibitemOpen
  \bibfield  {author} {\bibinfo {author} {\bibfnamefont {M.~F.}\ \bibnamefont
  {Peintinger}}, \bibinfo {author} {\bibfnamefont {D.~V.}\ \bibnamefont
  {Oliveira}}, \ and\ \bibinfo {author} {\bibfnamefont {T.}~\bibnamefont
  {Bredow}},\ }\href@noop {} {\bibfield  {journal} {\bibinfo  {journal} {J.
  Comput. Chem.}\ }\textbf {\bibinfo {volume} {34}},\ \bibinfo {pages} {451}
  (\bibinfo {year} {2013})}\BibitemShut {NoStop}%
\bibitem [{\citenamefont {Lee}\ \emph {et~al.}(2021{\natexlab{a}})\citenamefont
  {Lee}, \citenamefont {Feng}, \citenamefont {Cunha}, \citenamefont {Gonthier},
  \citenamefont {Epifanovsky},\ and\ \citenamefont
  {Head-Gordon}}]{lee2021approaching}%
  \BibitemOpen
  \bibfield  {author} {\bibinfo {author} {\bibfnamefont {J.}~\bibnamefont
  {Lee}}, \bibinfo {author} {\bibfnamefont {X.}~\bibnamefont {Feng}}, \bibinfo
  {author} {\bibfnamefont {L.~A.}\ \bibnamefont {Cunha}}, \bibinfo {author}
  {\bibfnamefont {J.~F.}\ \bibnamefont {Gonthier}}, \bibinfo {author}
  {\bibfnamefont {E.}~\bibnamefont {Epifanovsky}}, \ and\ \bibinfo {author}
  {\bibfnamefont {M.}~\bibnamefont {Head-Gordon}},\ }\href@noop {} {\bibfield
  {journal} {\bibinfo  {journal} {J. Chem. Phys.}\ }\textbf {\bibinfo {volume}
  {155}} (\bibinfo {year} {2021}{\natexlab{a}})}\BibitemShut {NoStop}%
\bibitem [{\citenamefont {Li}\ \emph {et~al.}(2021)\citenamefont {Li},
  \citenamefont {Chen}, \citenamefont {Rossomme}, \citenamefont {Head-Gordon},\
  and\ \citenamefont {Head-Gordon}}]{li2021optimized}%
  \BibitemOpen
  \bibfield  {author} {\bibinfo {author} {\bibfnamefont {W.-L.}\ \bibnamefont
  {Li}}, \bibinfo {author} {\bibfnamefont {K.}~\bibnamefont {Chen}}, \bibinfo
  {author} {\bibfnamefont {E.}~\bibnamefont {Rossomme}}, \bibinfo {author}
  {\bibfnamefont {M.}~\bibnamefont {Head-Gordon}}, \ and\ \bibinfo {author}
  {\bibfnamefont {T.}~\bibnamefont {Head-Gordon}},\ }\href@noop {} {\bibfield
  {journal} {\bibinfo  {journal} {J. Phys. Chem. Lett.}\ }\textbf {\bibinfo
  {volume} {12}},\ \bibinfo {pages} {10304} (\bibinfo {year}
  {2021})}\BibitemShut {NoStop}%
\bibitem [{\citenamefont {VandeVondele}\ and\ \citenamefont
  {Hutter}(2007)}]{vandevondele2007gaussian}%
  \BibitemOpen
  \bibfield  {author} {\bibinfo {author} {\bibfnamefont {J.}~\bibnamefont
  {VandeVondele}}\ and\ \bibinfo {author} {\bibfnamefont {J.}~\bibnamefont
  {Hutter}},\ }\href@noop {} {\bibfield  {journal} {\bibinfo  {journal} {J.
  Chem. Phys.}\ }\textbf {\bibinfo {volume} {127}} (\bibinfo {year}
  {2007})}\BibitemShut {NoStop}%
\bibitem [{\citenamefont {Daga}\ \emph {et~al.}(2020)\citenamefont {Daga},
  \citenamefont {Civalleri},\ and\ \citenamefont {Maschio}}]{daga2020gaussian}%
  \BibitemOpen
  \bibfield  {author} {\bibinfo {author} {\bibfnamefont {L.~E.}\ \bibnamefont
  {Daga}}, \bibinfo {author} {\bibfnamefont {B.}~\bibnamefont {Civalleri}}, \
  and\ \bibinfo {author} {\bibfnamefont {L.}~\bibnamefont {Maschio}},\
  }\href@noop {} {\bibfield  {journal} {\bibinfo  {journal} {J. Chem. Theory
  Comput.}\ }\textbf {\bibinfo {volume} {16}},\ \bibinfo {pages} {2192}
  (\bibinfo {year} {2020})}\BibitemShut {NoStop}%
\bibitem [{\citenamefont {Laun}\ \emph {et~al.}(2018)\citenamefont {Laun},
  \citenamefont {Vilela~Oliveira},\ and\ \citenamefont
  {Bredow}}]{laun2018consistent}%
  \BibitemOpen
  \bibfield  {author} {\bibinfo {author} {\bibfnamefont {J.}~\bibnamefont
  {Laun}}, \bibinfo {author} {\bibfnamefont {D.}~\bibnamefont
  {Vilela~Oliveira}}, \ and\ \bibinfo {author} {\bibfnamefont {T.}~\bibnamefont
  {Bredow}},\ }\href@noop {} {\bibfield  {journal} {\bibinfo  {journal} {J.
  Comput. Chem.}\ }\textbf {\bibinfo {volume} {39}},\ \bibinfo {pages} {1285}
  (\bibinfo {year} {2018})}\BibitemShut {NoStop}%
\bibitem [{\citenamefont {Ye}\ and\ \citenamefont
  {Berkelbach}(2022)}]{ye2022correlation}%
  \BibitemOpen
  \bibfield  {author} {\bibinfo {author} {\bibfnamefont {H.-Z.}\ \bibnamefont
  {Ye}}\ and\ \bibinfo {author} {\bibfnamefont {T.~C.}\ \bibnamefont
  {Berkelbach}},\ }\href@noop {} {\bibfield  {journal} {\bibinfo  {journal} {J.
  Chem. Theory Comput.}\ }\textbf {\bibinfo {volume} {18}},\ \bibinfo {pages}
  {1595} (\bibinfo {year} {2022})}\BibitemShut {NoStop}%
\bibitem [{\citenamefont {Szabo}\ and\ \citenamefont
  {Ostlund}(1996)}]{szabo1996modern}%
  \BibitemOpen
  \bibfield  {author} {\bibinfo {author} {\bibfnamefont {A.}~\bibnamefont
  {Szabo}}\ and\ \bibinfo {author} {\bibfnamefont {N.}~\bibnamefont
  {Ostlund}},\ }\href@noop {} {\emph {\bibinfo {title} {Modern Quantum
  Chemistry: Introduction to Advanced Electronic Structure Theory}}},\ Dover
  Books on Chemistry\ (\bibinfo  {publisher} {Dover Publications},\ \bibinfo
  {year} {1996})\BibitemShut {NoStop}%
\bibitem [{\citenamefont {Kresse}\ and\ \citenamefont
  {Furthm{\"u}ller}(1996)}]{kresse1996efficient}%
  \BibitemOpen
  \bibfield  {author} {\bibinfo {author} {\bibfnamefont {G.}~\bibnamefont
  {Kresse}}\ and\ \bibinfo {author} {\bibfnamefont {J.}~\bibnamefont
  {Furthm{\"u}ller}},\ }\href@noop {} {\bibfield  {journal} {\bibinfo
  {journal} {Phys. Rev. B}\ }\textbf {\bibinfo {volume} {54}},\ \bibinfo
  {pages} {11169} (\bibinfo {year} {1996})}\BibitemShut {NoStop}%
\bibitem [{\citenamefont {Hafner}(2008)}]{hafner2008ab}%
  \BibitemOpen
  \bibfield  {author} {\bibinfo {author} {\bibfnamefont {J.}~\bibnamefont
  {Hafner}},\ }\href@noop {} {\bibfield  {journal} {\bibinfo  {journal} {J.
  Comput. Chem.}\ }\textbf {\bibinfo {volume} {29}},\ \bibinfo {pages} {2044}
  (\bibinfo {year} {2008})}\BibitemShut {NoStop}%
\bibitem [{\citenamefont {Booth}\ \emph {et~al.}(2016)\citenamefont {Booth},
  \citenamefont {Tsatsoulis}, \citenamefont {Chan},\ and\ \citenamefont
  {Gr{\ifmmode\ddot{u}\else\"{u}\fi}neis}}]{Booth2016Aug}%
  \BibitemOpen
  \bibfield  {author} {\bibinfo {author} {\bibfnamefont {G.~H.}\ \bibnamefont
  {Booth}}, \bibinfo {author} {\bibfnamefont {T.}~\bibnamefont {Tsatsoulis}},
  \bibinfo {author} {\bibfnamefont {G.~K.-L.}\ \bibnamefont {Chan}}, \ and\
  \bibinfo {author} {\bibfnamefont {A.}~\bibnamefont
  {Gr{\ifmmode\ddot{u}\else\"{u}\fi}neis}},\ }\href {\doibase
  10.1063/1.4961301} {\bibfield  {journal} {\bibinfo  {journal} {J. Chem.
  Phys.}\ }\textbf {\bibinfo {volume} {145}},\ \bibinfo {pages} {084111}
  (\bibinfo {year} {2016})}\BibitemShut {NoStop}%
\bibitem [{\citenamefont {Helgaker}\ \emph {et~al.}(2013)\citenamefont
  {Helgaker}, \citenamefont {Jorgensen},\ and\ \citenamefont
  {Olsen}}]{helgaker2013molecular}%
  \BibitemOpen
  \bibfield  {author} {\bibinfo {author} {\bibfnamefont {T.}~\bibnamefont
  {Helgaker}}, \bibinfo {author} {\bibfnamefont {P.}~\bibnamefont {Jorgensen}},
  \ and\ \bibinfo {author} {\bibfnamefont {J.}~\bibnamefont {Olsen}},\
  }\href@noop {} {\emph {\bibinfo {title} {Molecular electronic-structure
  theory}}}\ (\bibinfo  {publisher} {John Wiley \& Sons},\ \bibinfo {year}
  {2013})\BibitemShut {NoStop}%
\bibitem [{\citenamefont {Ihm}\ \emph {et~al.}(1979)\citenamefont {Ihm},
  \citenamefont {Zunger},\ and\ \citenamefont {Cohen}}]{ihm1979momentum}%
  \BibitemOpen
  \bibfield  {author} {\bibinfo {author} {\bibfnamefont {J.}~\bibnamefont
  {Ihm}}, \bibinfo {author} {\bibfnamefont {A.}~\bibnamefont {Zunger}}, \ and\
  \bibinfo {author} {\bibfnamefont {M.~L.}\ \bibnamefont {Cohen}},\ }\href@noop
  {} {\bibfield  {journal} {\bibinfo  {journal} {J. Phys. C: Solid State Phys}\
  }\textbf {\bibinfo {volume} {12}},\ \bibinfo {pages} {4409} (\bibinfo {year}
  {1979})}\BibitemShut {NoStop}%
\bibitem [{\citenamefont {Lippert}\ \emph {et~al.}(1997)\citenamefont
  {Lippert}, \citenamefont {Hutter},\ and\ \citenamefont
  {Parrinello}}]{lippert_hybrid_1997}%
  \BibitemOpen
  \bibfield  {author} {\bibinfo {author} {\bibfnamefont {G.}~\bibnamefont
  {Lippert}}, \bibinfo {author} {\bibfnamefont {J.}~\bibnamefont {Hutter}}, \
  and\ \bibinfo {author} {\bibfnamefont {M.}~\bibnamefont {Parrinello}},\
  }\href {\doibase 10.1080/002689797170220} {\bibfield  {journal} {\bibinfo
  {journal} {Mol. Phys.}\ }\textbf {\bibinfo {volume} {92}},\ \bibinfo {pages}
  {477} (\bibinfo {year} {1997})}\BibitemShut {NoStop}%
\bibitem [{\citenamefont {VandeVondele}\ \emph {et~al.}(2005)\citenamefont
  {VandeVondele}, \citenamefont {Krack}, \citenamefont {Mohamed}, \citenamefont
  {Parrinello}, \citenamefont {Chassaing},\ and\ \citenamefont
  {Hutter}}]{vandevondele_quickstep_2005}%
  \BibitemOpen
  \bibfield  {author} {\bibinfo {author} {\bibfnamefont {J.}~\bibnamefont
  {VandeVondele}}, \bibinfo {author} {\bibfnamefont {M.}~\bibnamefont {Krack}},
  \bibinfo {author} {\bibfnamefont {F.}~\bibnamefont {Mohamed}}, \bibinfo
  {author} {\bibfnamefont {M.}~\bibnamefont {Parrinello}}, \bibinfo {author}
  {\bibfnamefont {T.}~\bibnamefont {Chassaing}}, \ and\ \bibinfo {author}
  {\bibfnamefont {J.}~\bibnamefont {Hutter}},\ }\href {\doibase
  10.1016/j.cpc.2004.12.014} {\bibfield  {journal} {\bibinfo  {journal}
  {Comput. Phys. Commun.}\ }\textbf {\bibinfo {volume} {167}},\ \bibinfo
  {pages} {103} (\bibinfo {year} {2005})}\BibitemShut {NoStop}%
\bibitem [{\citenamefont {Kühne}\ \emph {et~al.}(2020)\citenamefont {Kühne},
  \citenamefont {Iannuzzi}, \citenamefont {Del~Ben}, \citenamefont {Rybkin},
  \citenamefont {Seewald}, \citenamefont {Stein}, \citenamefont {Laino},
  \citenamefont {Khaliullin}, \citenamefont {Schütt}, \citenamefont
  {Schiffmann}, \citenamefont {Golze}, \citenamefont {Wilhelm}, \citenamefont
  {Chulkov}, \citenamefont {Bani-Hashemian}, \citenamefont {Weber},
  \citenamefont {Borštnik}, \citenamefont {Taillefumier}, \citenamefont
  {Jakobovits}, \citenamefont {Lazzaro}, \citenamefont {Pabst}, \citenamefont
  {Müller}, \citenamefont {Schade}, \citenamefont {Guidon}, \citenamefont
  {Andermatt}, \citenamefont {Holmberg}, \citenamefont {Schenter},
  \citenamefont {Hehn}, \citenamefont {Bussy}, \citenamefont {Belleflamme},
  \citenamefont {Tabacchi}, \citenamefont {Glöß}, \citenamefont {Lass},
  \citenamefont {Bethune}, \citenamefont {Mundy}, \citenamefont {Plessl},
  \citenamefont {Watkins}, \citenamefont {VandeVondele}, \citenamefont
  {Krack},\ and\ \citenamefont {Hutter}}]{kuhne_cp2k_2020}%
  \BibitemOpen
  \bibfield  {author} {\bibinfo {author} {\bibfnamefont {T.~D.}\ \bibnamefont
  {Kühne}}, \bibinfo {author} {\bibfnamefont {M.}~\bibnamefont {Iannuzzi}},
  \bibinfo {author} {\bibfnamefont {M.}~\bibnamefont {Del~Ben}}, \bibinfo
  {author} {\bibfnamefont {V.~V.}\ \bibnamefont {Rybkin}}, \bibinfo {author}
  {\bibfnamefont {P.}~\bibnamefont {Seewald}}, \bibinfo {author} {\bibfnamefont
  {F.}~\bibnamefont {Stein}}, \bibinfo {author} {\bibfnamefont
  {T.}~\bibnamefont {Laino}}, \bibinfo {author} {\bibfnamefont {R.~Z.}\
  \bibnamefont {Khaliullin}}, \bibinfo {author} {\bibfnamefont
  {O.}~\bibnamefont {Schütt}}, \bibinfo {author} {\bibfnamefont
  {F.}~\bibnamefont {Schiffmann}}, \bibinfo {author} {\bibfnamefont
  {D.}~\bibnamefont {Golze}}, \bibinfo {author} {\bibfnamefont
  {J.}~\bibnamefont {Wilhelm}}, \bibinfo {author} {\bibfnamefont
  {S.}~\bibnamefont {Chulkov}}, \bibinfo {author} {\bibfnamefont {M.~H.}\
  \bibnamefont {Bani-Hashemian}}, \bibinfo {author} {\bibfnamefont
  {V.}~\bibnamefont {Weber}}, \bibinfo {author} {\bibfnamefont
  {U.}~\bibnamefont {Borštnik}}, \bibinfo {author} {\bibfnamefont
  {M.}~\bibnamefont {Taillefumier}}, \bibinfo {author} {\bibfnamefont {A.~S.}\
  \bibnamefont {Jakobovits}}, \bibinfo {author} {\bibfnamefont
  {A.}~\bibnamefont {Lazzaro}}, \bibinfo {author} {\bibfnamefont
  {H.}~\bibnamefont {Pabst}}, \bibinfo {author} {\bibfnamefont
  {T.}~\bibnamefont {Müller}}, \bibinfo {author} {\bibfnamefont
  {R.}~\bibnamefont {Schade}}, \bibinfo {author} {\bibfnamefont
  {M.}~\bibnamefont {Guidon}}, \bibinfo {author} {\bibfnamefont
  {S.}~\bibnamefont {Andermatt}}, \bibinfo {author} {\bibfnamefont
  {N.}~\bibnamefont {Holmberg}}, \bibinfo {author} {\bibfnamefont {G.~K.}\
  \bibnamefont {Schenter}}, \bibinfo {author} {\bibfnamefont {A.}~\bibnamefont
  {Hehn}}, \bibinfo {author} {\bibfnamefont {A.}~\bibnamefont {Bussy}},
  \bibinfo {author} {\bibfnamefont {F.}~\bibnamefont {Belleflamme}}, \bibinfo
  {author} {\bibfnamefont {G.}~\bibnamefont {Tabacchi}}, \bibinfo {author}
  {\bibfnamefont {A.}~\bibnamefont {Glöß}}, \bibinfo {author} {\bibfnamefont
  {M.}~\bibnamefont {Lass}}, \bibinfo {author} {\bibfnamefont {I.}~\bibnamefont
  {Bethune}}, \bibinfo {author} {\bibfnamefont {C.~J.}\ \bibnamefont {Mundy}},
  \bibinfo {author} {\bibfnamefont {C.}~\bibnamefont {Plessl}}, \bibinfo
  {author} {\bibfnamefont {M.}~\bibnamefont {Watkins}}, \bibinfo {author}
  {\bibfnamefont {J.}~\bibnamefont {VandeVondele}}, \bibinfo {author}
  {\bibfnamefont {M.}~\bibnamefont {Krack}}, \ and\ \bibinfo {author}
  {\bibfnamefont {J.}~\bibnamefont {Hutter}},\ }\href {\doibase
  10.1063/5.0007045} {\bibfield  {journal} {\bibinfo  {journal} {J. Chem.
  Phys.}\ }\textbf {\bibinfo {volume} {152}},\ \bibinfo {pages} {194103}
  (\bibinfo {year} {2020})}\BibitemShut {NoStop}%
\bibitem [{\citenamefont {Sun}\ \emph {et~al.}(2020{\natexlab{a}})\citenamefont
  {Sun}, \citenamefont {Zhang}, \citenamefont {Banerjee}, \citenamefont {Bao},
  \citenamefont {Barbry}, \citenamefont {Blunt}, \citenamefont {Bogdanov},
  \citenamefont {Booth}, \citenamefont {Chen}, \citenamefont {Cui},
  \citenamefont {Eriksen}, \citenamefont {Gao}, \citenamefont {Guo},
  \citenamefont {Hermann}, \citenamefont {Hermes}, \citenamefont {Koh},
  \citenamefont {Koval}, \citenamefont {Lehtola}, \citenamefont {Li},
  \citenamefont {Liu}, \citenamefont {Mardirossian}, \citenamefont {McClain},
  \citenamefont {Motta}, \citenamefont {Mussard}, \citenamefont {Pham},
  \citenamefont {Pulkin}, \citenamefont {Purwanto}, \citenamefont {Robinson},
  \citenamefont {Ronca}, \citenamefont {Sayfutyarova}, \citenamefont
  {Scheurer}, \citenamefont {Schurkus}, \citenamefont {Smith}, \citenamefont
  {Sun}, \citenamefont {Sun}, \citenamefont {Upadhyay}, \citenamefont {Wagner},
  \citenamefont {Wang}, \citenamefont {White}, \citenamefont {Whitfield},
  \citenamefont {Williamson}, \citenamefont {Wouters}, \citenamefont {Yang},
  \citenamefont {Yu}, \citenamefont {Zhu}, \citenamefont {Berkelbach},
  \citenamefont {Sharma}, \citenamefont {Sokolov},\ and\ \citenamefont
  {Chan}}]{sun_recent_2020}%
  \BibitemOpen
  \bibfield  {author} {\bibinfo {author} {\bibfnamefont {Q.}~\bibnamefont
  {Sun}}, \bibinfo {author} {\bibfnamefont {X.}~\bibnamefont {Zhang}}, \bibinfo
  {author} {\bibfnamefont {S.}~\bibnamefont {Banerjee}}, \bibinfo {author}
  {\bibfnamefont {P.}~\bibnamefont {Bao}}, \bibinfo {author} {\bibfnamefont
  {M.}~\bibnamefont {Barbry}}, \bibinfo {author} {\bibfnamefont {N.~S.}\
  \bibnamefont {Blunt}}, \bibinfo {author} {\bibfnamefont {N.~A.}\ \bibnamefont
  {Bogdanov}}, \bibinfo {author} {\bibfnamefont {G.~H.}\ \bibnamefont {Booth}},
  \bibinfo {author} {\bibfnamefont {J.}~\bibnamefont {Chen}}, \bibinfo {author}
  {\bibfnamefont {Z.-H.}\ \bibnamefont {Cui}}, \bibinfo {author} {\bibfnamefont
  {J.~J.}\ \bibnamefont {Eriksen}}, \bibinfo {author} {\bibfnamefont
  {Y.}~\bibnamefont {Gao}}, \bibinfo {author} {\bibfnamefont {S.}~\bibnamefont
  {Guo}}, \bibinfo {author} {\bibfnamefont {J.}~\bibnamefont {Hermann}},
  \bibinfo {author} {\bibfnamefont {M.~R.}\ \bibnamefont {Hermes}}, \bibinfo
  {author} {\bibfnamefont {K.}~\bibnamefont {Koh}}, \bibinfo {author}
  {\bibfnamefont {P.}~\bibnamefont {Koval}}, \bibinfo {author} {\bibfnamefont
  {S.}~\bibnamefont {Lehtola}}, \bibinfo {author} {\bibfnamefont
  {Z.}~\bibnamefont {Li}}, \bibinfo {author} {\bibfnamefont {J.}~\bibnamefont
  {Liu}}, \bibinfo {author} {\bibfnamefont {N.}~\bibnamefont {Mardirossian}},
  \bibinfo {author} {\bibfnamefont {J.~D.}\ \bibnamefont {McClain}}, \bibinfo
  {author} {\bibfnamefont {M.}~\bibnamefont {Motta}}, \bibinfo {author}
  {\bibfnamefont {B.}~\bibnamefont {Mussard}}, \bibinfo {author} {\bibfnamefont
  {H.~Q.}\ \bibnamefont {Pham}}, \bibinfo {author} {\bibfnamefont
  {A.}~\bibnamefont {Pulkin}}, \bibinfo {author} {\bibfnamefont
  {W.}~\bibnamefont {Purwanto}}, \bibinfo {author} {\bibfnamefont {P.~J.}\
  \bibnamefont {Robinson}}, \bibinfo {author} {\bibfnamefont {E.}~\bibnamefont
  {Ronca}}, \bibinfo {author} {\bibfnamefont {E.~R.}\ \bibnamefont
  {Sayfutyarova}}, \bibinfo {author} {\bibfnamefont {M.}~\bibnamefont
  {Scheurer}}, \bibinfo {author} {\bibfnamefont {H.~F.}\ \bibnamefont
  {Schurkus}}, \bibinfo {author} {\bibfnamefont {J.~E.~T.}\ \bibnamefont
  {Smith}}, \bibinfo {author} {\bibfnamefont {C.}~\bibnamefont {Sun}}, \bibinfo
  {author} {\bibfnamefont {S.-N.}\ \bibnamefont {Sun}}, \bibinfo {author}
  {\bibfnamefont {S.}~\bibnamefont {Upadhyay}}, \bibinfo {author}
  {\bibfnamefont {L.~K.}\ \bibnamefont {Wagner}}, \bibinfo {author}
  {\bibfnamefont {X.}~\bibnamefont {Wang}}, \bibinfo {author} {\bibfnamefont
  {A.}~\bibnamefont {White}}, \bibinfo {author} {\bibfnamefont {J.~D.}\
  \bibnamefont {Whitfield}}, \bibinfo {author} {\bibfnamefont {M.~J.}\
  \bibnamefont {Williamson}}, \bibinfo {author} {\bibfnamefont
  {S.}~\bibnamefont {Wouters}}, \bibinfo {author} {\bibfnamefont
  {J.}~\bibnamefont {Yang}}, \bibinfo {author} {\bibfnamefont {J.~M.}\
  \bibnamefont {Yu}}, \bibinfo {author} {\bibfnamefont {T.}~\bibnamefont
  {Zhu}}, \bibinfo {author} {\bibfnamefont {T.~C.}\ \bibnamefont {Berkelbach}},
  \bibinfo {author} {\bibfnamefont {S.}~\bibnamefont {Sharma}}, \bibinfo
  {author} {\bibfnamefont {A.~Y.}\ \bibnamefont {Sokolov}}, \ and\ \bibinfo
  {author} {\bibfnamefont {G.~K.-L.}\ \bibnamefont {Chan}},\ }\href {\doibase
  10.1063/5.0006074} {\bibfield  {journal} {\bibinfo  {journal} {J. Chem.
  Phys.}\ }\textbf {\bibinfo {volume} {153}},\ \bibinfo {pages} {024109}
  (\bibinfo {year} {2020}{\natexlab{a}})}\BibitemShut {NoStop}%
\bibitem [{\citenamefont {Frigo}\ and\ \citenamefont
  {Johnson}(2005)}]{frigo2005design}%
  \BibitemOpen
  \bibfield  {author} {\bibinfo {author} {\bibfnamefont {M.}~\bibnamefont
  {Frigo}}\ and\ \bibinfo {author} {\bibfnamefont {S.~G.}\ \bibnamefont
  {Johnson}},\ }\href@noop {} {\bibfield  {journal} {\bibinfo  {journal} {Proc.
  IEEE}\ }\textbf {\bibinfo {volume} {93}},\ \bibinfo {pages} {216} (\bibinfo
  {year} {2005})}\BibitemShut {NoStop}%
\bibitem [{\citenamefont {Goedecker}\ \emph {et~al.}(1996)\citenamefont
  {Goedecker}, \citenamefont {Teter},\ and\ \citenamefont
  {Hutter}}]{goedecker1996separable}%
  \BibitemOpen
  \bibfield  {author} {\bibinfo {author} {\bibfnamefont {S.}~\bibnamefont
  {Goedecker}}, \bibinfo {author} {\bibfnamefont {M.}~\bibnamefont {Teter}}, \
  and\ \bibinfo {author} {\bibfnamefont {J.}~\bibnamefont {Hutter}},\
  }\href@noop {} {\bibfield  {journal} {\bibinfo  {journal} {Phys. Rev. B}\
  }\textbf {\bibinfo {volume} {54}},\ \bibinfo {pages} {1703} (\bibinfo {year}
  {1996})}\BibitemShut {NoStop}%
\bibitem [{\citenamefont {Vanderbilt}(1990)}]{vanderbilt1990soft}%
  \BibitemOpen
  \bibfield  {author} {\bibinfo {author} {\bibfnamefont {D.}~\bibnamefont
  {Vanderbilt}},\ }\href@noop {} {\bibfield  {journal} {\bibinfo  {journal}
  {Phys. Rev. B}\ }\textbf {\bibinfo {volume} {41}},\ \bibinfo {pages} {7892}
  (\bibinfo {year} {1990})}\BibitemShut {NoStop}%
\bibitem [{\citenamefont {{Del Ben}}\ \emph {et~al.}(2015)\citenamefont {{Del
  Ben}}, \citenamefont {{Sch{\"u}tt}}, \citenamefont {{Wentz}}, \citenamefont
  {{Messmer}}, \citenamefont {{Hutter}},\ and\ \citenamefont
  {{VandeVondele}}}]{BibEntry2015Feb}%
  \BibitemOpen
  \bibfield  {author} {\bibinfo {author} {\bibfnamefont {M.}~\bibnamefont {{Del
  Ben}}}, \bibinfo {author} {\bibfnamefont {O.}~\bibnamefont {{Sch{\"u}tt}}},
  \bibinfo {author} {\bibfnamefont {T.}~\bibnamefont {{Wentz}}}, \bibinfo
  {author} {\bibfnamefont {P.}~\bibnamefont {{Messmer}}}, \bibinfo {author}
  {\bibfnamefont {J.}~\bibnamefont {{Hutter}}}, \ and\ \bibinfo {author}
  {\bibfnamefont {J.}~\bibnamefont {{VandeVondele}}},\ }\href {\doibase
  10.1016/j.cpc.2014.10.021} {\bibfield  {journal} {\bibinfo  {journal}
  {Comput. Phys. Commun.}\ }\textbf {\bibinfo {volume} {187}},\ \bibinfo
  {pages} {120} (\bibinfo {year} {2015})}\BibitemShut {NoStop}%
\bibitem [{\citenamefont {Wilhelm}\ and\ \citenamefont
  {Hutter}(2017)}]{Wilhelm2017Jun}%
  \BibitemOpen
  \bibfield  {author} {\bibinfo {author} {\bibfnamefont {J.}~\bibnamefont
  {Wilhelm}}\ and\ \bibinfo {author} {\bibfnamefont {J.}~\bibnamefont
  {Hutter}},\ }\href {\doibase 10.1103/PhysRevB.95.235123} {\bibfield
  {journal} {\bibinfo  {journal} {Phys. Rev. B}\ }\textbf {\bibinfo {volume}
  {95}},\ \bibinfo {pages} {235123} (\bibinfo {year} {2017})}\BibitemShut
  {NoStop}%
\bibitem [{\citenamefont {Del~Ben}\ \emph
  {et~al.}(2012{\natexlab{b}})\citenamefont {Del~Ben}, \citenamefont {Hutter},\
  and\ \citenamefont {VandeVondele}}]{DelBen2012Sep}%
  \BibitemOpen
  \bibfield  {author} {\bibinfo {author} {\bibfnamefont {M.}~\bibnamefont
  {Del~Ben}}, \bibinfo {author} {\bibfnamefont {J.}~\bibnamefont {Hutter}}, \
  and\ \bibinfo {author} {\bibfnamefont {J.}~\bibnamefont {VandeVondele}},\
  }\href {\doibase 10.1021/ct300531w} {\bibfield  {journal} {\bibinfo
  {journal} {J. Chem. Theory Comput.}\ }\textbf {\bibinfo {volume} {8}},\
  \bibinfo {pages} {4177} (\bibinfo {year} {2012}{\natexlab{b}})}\BibitemShut
  {NoStop}%
\bibitem [{\citenamefont {Del~Ben}\ \emph
  {et~al.}(2013{\natexlab{a}})\citenamefont {Del~Ben}, \citenamefont {Hutter},\
  and\ \citenamefont {VandeVondele}}]{DelBen2013May}%
  \BibitemOpen
  \bibfield  {author} {\bibinfo {author} {\bibfnamefont {M.}~\bibnamefont
  {Del~Ben}}, \bibinfo {author} {\bibfnamefont {J.}~\bibnamefont {Hutter}}, \
  and\ \bibinfo {author} {\bibfnamefont {J.}~\bibnamefont {VandeVondele}},\
  }\href {\doibase 10.1021/ct4002202} {\bibfield  {journal} {\bibinfo
  {journal} {J. Chem. Theory Comput.}\ }\textbf {\bibinfo {volume} {9}},\
  \bibinfo {pages} {2654} (\bibinfo {year} {2013}{\natexlab{a}})}\BibitemShut
  {NoStop}%
\bibitem [{\citenamefont {Saunders}\ \emph {et~al.}(1992)\citenamefont
  {Saunders}, \citenamefont {Freyria-Fava}, \citenamefont {Dovesi},
  \citenamefont {Salasco},\ and\ \citenamefont
  {Roetti}}]{saunders_electrostatic_1992}%
  \BibitemOpen
  \bibfield  {author} {\bibinfo {author} {\bibfnamefont {V.}~\bibnamefont
  {Saunders}}, \bibinfo {author} {\bibfnamefont {C.}~\bibnamefont
  {Freyria-Fava}}, \bibinfo {author} {\bibfnamefont {R.}~\bibnamefont
  {Dovesi}}, \bibinfo {author} {\bibfnamefont {L.}~\bibnamefont {Salasco}}, \
  and\ \bibinfo {author} {\bibfnamefont {C.}~\bibnamefont {Roetti}},\ }\href
  {\doibase 10.1080/00268979200102671} {\bibfield  {journal} {\bibinfo
  {journal} {Mol. Phys.}\ }\textbf {\bibinfo {volume} {77}},\ \bibinfo {pages}
  {629} (\bibinfo {year} {1992})}\BibitemShut {NoStop}%
\bibitem [{\citenamefont {Sun}(2023{\natexlab{a}})}]{sun_exact_2023}%
  \BibitemOpen
  \bibfield  {author} {\bibinfo {author} {\bibfnamefont {Q.}~\bibnamefont
  {Sun}},\ }\href {\doibase 10.1063/5.0155815} {\bibfield  {journal} {\bibinfo
  {journal} {J. Chem. Phys.}\ }\textbf {\bibinfo {volume} {159}},\ \bibinfo
  {pages} {024108} (\bibinfo {year} {2023}{\natexlab{a}})}\BibitemShut
  {NoStop}%
\bibitem [{\citenamefont {Stenberg}\ and\ \citenamefont
  {Stenqvist}(2020)}]{stenberg2020exact}%
  \BibitemOpen
  \bibfield  {author} {\bibinfo {author} {\bibfnamefont {S.}~\bibnamefont
  {Stenberg}}\ and\ \bibinfo {author} {\bibfnamefont {B.}~\bibnamefont
  {Stenqvist}},\ }\href@noop {} {\bibfield  {journal} {\bibinfo  {journal} {J.
  Phys. Chem. A}\ }\textbf {\bibinfo {volume} {124}},\ \bibinfo {pages} {3943}
  (\bibinfo {year} {2020})}\BibitemShut {NoStop}%
\bibitem [{\citenamefont {Gill}\ \emph {et~al.}(1989)\citenamefont {Gill},
  \citenamefont {Head-Gordon},\ and\ \citenamefont
  {Pople}}]{gill1989efficient}%
  \BibitemOpen
  \bibfield  {author} {\bibinfo {author} {\bibfnamefont {P.~M.}\ \bibnamefont
  {Gill}}, \bibinfo {author} {\bibfnamefont {M.}~\bibnamefont {Head-Gordon}}, \
  and\ \bibinfo {author} {\bibfnamefont {J.~A.}\ \bibnamefont {Pople}},\
  }\href@noop {} {\bibfield  {journal} {\bibinfo  {journal} {Int. J. Quantum
  Chem.}\ }\textbf {\bibinfo {volume} {36}},\ \bibinfo {pages} {269} (\bibinfo
  {year} {1989})}\BibitemShut {NoStop}%
\bibitem [{\citenamefont {Head-Gordon}\ and\ \citenamefont
  {Pople}(1988)}]{head1988method}%
  \BibitemOpen
  \bibfield  {author} {\bibinfo {author} {\bibfnamefont {M.}~\bibnamefont
  {Head-Gordon}}\ and\ \bibinfo {author} {\bibfnamefont {J.~A.}\ \bibnamefont
  {Pople}},\ }\href@noop {} {\bibfield  {journal} {\bibinfo  {journal} {J.
  Chem. Phys.}\ }\textbf {\bibinfo {volume} {89}},\ \bibinfo {pages} {5777}
  (\bibinfo {year} {1988})}\BibitemShut {NoStop}%
\bibitem [{\citenamefont {Guidon}\ \emph {et~al.}(2008)\citenamefont {Guidon},
  \citenamefont {Schiffmann}, \citenamefont {Hutter},\ and\ \citenamefont
  {VandeVondele}}]{guidon2008ab}%
  \BibitemOpen
  \bibfield  {author} {\bibinfo {author} {\bibfnamefont {M.}~\bibnamefont
  {Guidon}}, \bibinfo {author} {\bibfnamefont {F.}~\bibnamefont {Schiffmann}},
  \bibinfo {author} {\bibfnamefont {J.}~\bibnamefont {Hutter}}, \ and\ \bibinfo
  {author} {\bibfnamefont {J.}~\bibnamefont {VandeVondele}},\ }\href@noop {}
  {\bibfield  {journal} {\bibinfo  {journal} {J. Chem. Phys.}\ }\textbf
  {\bibinfo {volume} {128}} (\bibinfo {year} {2008})}\BibitemShut {NoStop}%
\bibitem [{\citenamefont {Heyd}\ \emph {et~al.}(2003)\citenamefont {Heyd},
  \citenamefont {Scuseria},\ and\ \citenamefont {Ernzerhof}}]{heyd2003hybrid}%
  \BibitemOpen
  \bibfield  {author} {\bibinfo {author} {\bibfnamefont {J.}~\bibnamefont
  {Heyd}}, \bibinfo {author} {\bibfnamefont {G.~E.}\ \bibnamefont {Scuseria}},
  \ and\ \bibinfo {author} {\bibfnamefont {M.}~\bibnamefont {Ernzerhof}},\
  }\href@noop {} {\bibfield  {journal} {\bibinfo  {journal} {J. Chem. Phys.}\
  }\textbf {\bibinfo {volume} {118}},\ \bibinfo {pages} {8207} (\bibinfo {year}
  {2003})}\BibitemShut {NoStop}%
\bibitem [{\citenamefont {Izmaylov}\ \emph {et~al.}(2006)\citenamefont
  {Izmaylov}, \citenamefont {Scuseria},\ and\ \citenamefont
  {Frisch}}]{izmaylov2006efficient}%
  \BibitemOpen
  \bibfield  {author} {\bibinfo {author} {\bibfnamefont {A.~F.}\ \bibnamefont
  {Izmaylov}}, \bibinfo {author} {\bibfnamefont {G.~E.}\ \bibnamefont
  {Scuseria}}, \ and\ \bibinfo {author} {\bibfnamefont {M.~J.}\ \bibnamefont
  {Frisch}},\ }\href@noop {} {\bibfield  {journal} {\bibinfo  {journal} {J.
  Chem. Phys.}\ }\textbf {\bibinfo {volume} {125}} (\bibinfo {year}
  {2006})}\BibitemShut {NoStop}%
\bibitem [{\citenamefont {Ye}\ and\ \citenamefont
  {Berkelbach}(2021{\natexlab{a}})}]{ye2021tight}%
  \BibitemOpen
  \bibfield  {author} {\bibinfo {author} {\bibfnamefont {H.-Z.}\ \bibnamefont
  {Ye}}\ and\ \bibinfo {author} {\bibfnamefont {T.~C.}\ \bibnamefont
  {Berkelbach}},\ }\href@noop {} {\bibfield  {journal} {\bibinfo  {journal} {J.
  Chem. Phys.}\ }\textbf {\bibinfo {volume} {155}} (\bibinfo {year}
  {2021}{\natexlab{a}})}\BibitemShut {NoStop}%
\bibitem [{\citenamefont {Łazarski}\ \emph {et~al.}(2015)\citenamefont
  {Łazarski}, \citenamefont {Burow},\ and\ \citenamefont
  {Sierka}}]{Lazarski2015Jun}%
  \BibitemOpen
  \bibfield  {author} {\bibinfo {author} {\bibfnamefont {R.}~\bibnamefont
  {Łazarski}}, \bibinfo {author} {\bibfnamefont {A.~M.}\ \bibnamefont
  {Burow}}, \ and\ \bibinfo {author} {\bibfnamefont {M.}~\bibnamefont
  {Sierka}},\ }\href {\doibase 10.1021/acs.jctc.5b00252} {\bibfield  {journal}
  {\bibinfo  {journal} {J. Chem. Theory Comput.}\ }\textbf {\bibinfo {volume}
  {11}},\ \bibinfo {pages} {3029} (\bibinfo {year} {2015})}\BibitemShut
  {NoStop}%
\bibitem [{\citenamefont {Ye}\ and\ \citenamefont
  {Berkelbach}(2021{\natexlab{b}})}]{ye_fast_2021}%
  \BibitemOpen
  \bibfield  {author} {\bibinfo {author} {\bibfnamefont {H.-Z.}\ \bibnamefont
  {Ye}}\ and\ \bibinfo {author} {\bibfnamefont {T.~C.}\ \bibnamefont
  {Berkelbach}},\ }\href {\doibase 10.1063/5.0046617} {\bibfield  {journal}
  {\bibinfo  {journal} {J. Chem. Phys.}\ }\textbf {\bibinfo {volume} {154}},\
  \bibinfo {pages} {131104} (\bibinfo {year} {2021}{\natexlab{b}})}\BibitemShut
  {NoStop}%
\bibitem [{\citenamefont {Sun}(2023{\natexlab{b}})}]{sun2023efficient}%
  \BibitemOpen
  \bibfield  {author} {\bibinfo {author} {\bibfnamefont {Q.}~\bibnamefont
  {Sun}},\ }\href@noop {} {\bibfield  {journal} {\bibinfo  {journal} {J. Chem.
  Phys.}\ }\textbf {\bibinfo {volume} {159}} (\bibinfo {year}
  {2023}{\natexlab{b}})}\BibitemShut {NoStop}%
\bibitem [{\citenamefont {F{\"u}sti-Moln{\'a}r}\ and\ \citenamefont
  {Pulay}(2002)}]{fusti2002fourier}%
  \BibitemOpen
  \bibfield  {author} {\bibinfo {author} {\bibfnamefont {L.}~\bibnamefont
  {F{\"u}sti-Moln{\'a}r}}\ and\ \bibinfo {author} {\bibfnamefont
  {P.}~\bibnamefont {Pulay}},\ }\href@noop {} {\bibfield  {journal} {\bibinfo
  {journal} {J. Chem. Phys.}\ }\textbf {\bibinfo {volume} {117}},\ \bibinfo
  {pages} {7827} (\bibinfo {year} {2002})}\BibitemShut {NoStop}%
\bibitem [{\citenamefont {Weigend}(2002)}]{weigend2002fully}%
  \BibitemOpen
  \bibfield  {author} {\bibinfo {author} {\bibfnamefont {F.}~\bibnamefont
  {Weigend}},\ }\href@noop {} {\bibfield  {journal} {\bibinfo  {journal} {Phys.
  Chem. Chem. Phys.}\ }\textbf {\bibinfo {volume} {4}},\ \bibinfo {pages}
  {4285} (\bibinfo {year} {2002})}\BibitemShut {NoStop}%
\bibitem [{\citenamefont {Hesselmann}\ \emph {et~al.}(2005)\citenamefont
  {Hesselmann}, \citenamefont {Jansen},\ and\ \citenamefont
  {Sch{\"u}tz}}]{hesselmann2005density}%
  \BibitemOpen
  \bibfield  {author} {\bibinfo {author} {\bibfnamefont {A.}~\bibnamefont
  {Hesselmann}}, \bibinfo {author} {\bibfnamefont {G.}~\bibnamefont {Jansen}},
  \ and\ \bibinfo {author} {\bibfnamefont {M.}~\bibnamefont {Sch{\"u}tz}},\
  }\href@noop {} {\bibfield  {journal} {\bibinfo  {journal} {J. Chem. Phys.}\
  }\textbf {\bibinfo {volume} {122}} (\bibinfo {year} {2005})}\BibitemShut
  {NoStop}%
\bibitem [{\citenamefont {Weigend}\ \emph {et~al.}(1998)\citenamefont
  {Weigend}, \citenamefont {H{\"a}ser}, \citenamefont {Patzelt},\ and\
  \citenamefont {Ahlrichs}}]{weigend1998ri}%
  \BibitemOpen
  \bibfield  {author} {\bibinfo {author} {\bibfnamefont {F.}~\bibnamefont
  {Weigend}}, \bibinfo {author} {\bibfnamefont {M.}~\bibnamefont {H{\"a}ser}},
  \bibinfo {author} {\bibfnamefont {H.}~\bibnamefont {Patzelt}}, \ and\
  \bibinfo {author} {\bibfnamefont {R.}~\bibnamefont {Ahlrichs}},\ }\href@noop
  {} {\bibfield  {journal} {\bibinfo  {journal} {Chem. Phys. Lett.}\ }\textbf
  {\bibinfo {volume} {294}},\ \bibinfo {pages} {143} (\bibinfo {year}
  {1998})}\BibitemShut {NoStop}%
\bibitem [{\citenamefont {H{\"a}ttig}\ and\ \citenamefont
  {Weigend}(2000)}]{hattig2000cc2}%
  \BibitemOpen
  \bibfield  {author} {\bibinfo {author} {\bibfnamefont {C.}~\bibnamefont
  {H{\"a}ttig}}\ and\ \bibinfo {author} {\bibfnamefont {F.}~\bibnamefont
  {Weigend}},\ }\href@noop {} {\bibfield  {journal} {\bibinfo  {journal} {J.
  Chem. Phys.}\ }\textbf {\bibinfo {volume} {113}},\ \bibinfo {pages} {5154}
  (\bibinfo {year} {2000})}\BibitemShut {NoStop}%
\bibitem [{\citenamefont {Eichkorn}\ \emph {et~al.}(1995)\citenamefont
  {Eichkorn}, \citenamefont {Treutler}, \citenamefont {{\"O}hm}, \citenamefont
  {H{\"a}ser},\ and\ \citenamefont {Ahlrichs}}]{eichkorn1995auxiliary}%
  \BibitemOpen
  \bibfield  {author} {\bibinfo {author} {\bibfnamefont {K.}~\bibnamefont
  {Eichkorn}}, \bibinfo {author} {\bibfnamefont {O.}~\bibnamefont {Treutler}},
  \bibinfo {author} {\bibfnamefont {H.}~\bibnamefont {{\"O}hm}}, \bibinfo
  {author} {\bibfnamefont {M.}~\bibnamefont {H{\"a}ser}}, \ and\ \bibinfo
  {author} {\bibfnamefont {R.}~\bibnamefont {Ahlrichs}},\ }\href@noop {}
  {\bibfield  {journal} {\bibinfo  {journal} {Chem. Phys. Lett.}\ }\textbf
  {\bibinfo {volume} {240}},\ \bibinfo {pages} {283} (\bibinfo {year}
  {1995})}\BibitemShut {NoStop}%
\bibitem [{\citenamefont {Weigend}(2008)}]{weigend2008hartree}%
  \BibitemOpen
  \bibfield  {author} {\bibinfo {author} {\bibfnamefont {F.}~\bibnamefont
  {Weigend}},\ }\href@noop {} {\bibfield  {journal} {\bibinfo  {journal} {J.
  Comput. Chem.}\ }\textbf {\bibinfo {volume} {29}},\ \bibinfo {pages} {167}
  (\bibinfo {year} {2008})}\BibitemShut {NoStop}%
\bibitem [{\citenamefont {Stoychev}\ \emph {et~al.}(2017)\citenamefont
  {Stoychev}, \citenamefont {Auer},\ and\ \citenamefont
  {Neese}}]{stoychev2017automatic}%
  \BibitemOpen
  \bibfield  {author} {\bibinfo {author} {\bibfnamefont {G.~L.}\ \bibnamefont
  {Stoychev}}, \bibinfo {author} {\bibfnamefont {A.~A.}\ \bibnamefont {Auer}},
  \ and\ \bibinfo {author} {\bibfnamefont {F.}~\bibnamefont {Neese}},\
  }\href@noop {} {\bibfield  {journal} {\bibinfo  {journal} {J. Chem. Theory
  Comput.}\ }\textbf {\bibinfo {volume} {13}},\ \bibinfo {pages} {554}
  (\bibinfo {year} {2017})}\BibitemShut {NoStop}%
\bibitem [{\citenamefont {Sodt}\ \emph {et~al.}(2006)\citenamefont {Sodt},
  \citenamefont {Subotnik},\ and\ \citenamefont
  {Head-Gordon}}]{sodt2006linear}%
  \BibitemOpen
  \bibfield  {author} {\bibinfo {author} {\bibfnamefont {A.}~\bibnamefont
  {Sodt}}, \bibinfo {author} {\bibfnamefont {J.~E.}\ \bibnamefont {Subotnik}},
  \ and\ \bibinfo {author} {\bibfnamefont {M.}~\bibnamefont {Head-Gordon}},\
  }\href@noop {} {\bibfield  {journal} {\bibinfo  {journal} {J. Chem. Phys.}\
  }\textbf {\bibinfo {volume} {125}} (\bibinfo {year} {2006})}\BibitemShut
  {NoStop}%
\bibitem [{\citenamefont {Van~Alsenoy}(1988)}]{van1988ab}%
  \BibitemOpen
  \bibfield  {author} {\bibinfo {author} {\bibfnamefont {C.}~\bibnamefont
  {Van~Alsenoy}},\ }\href@noop {} {\bibfield  {journal} {\bibinfo  {journal}
  {J. Comput. Chem.}\ }\textbf {\bibinfo {volume} {9}},\ \bibinfo {pages} {620}
  (\bibinfo {year} {1988})}\BibitemShut {NoStop}%
\bibitem [{\citenamefont {Maschio}\ and\ \citenamefont
  {Usvyat}(2008)}]{maschio2008fitting}%
  \BibitemOpen
  \bibfield  {author} {\bibinfo {author} {\bibfnamefont {L.}~\bibnamefont
  {Maschio}}\ and\ \bibinfo {author} {\bibfnamefont {D.}~\bibnamefont
  {Usvyat}},\ }\href@noop {} {\bibfield  {journal} {\bibinfo  {journal} {Phys.
  Rev. B}\ }\textbf {\bibinfo {volume} {78}},\ \bibinfo {pages} {073102}
  (\bibinfo {year} {2008})}\BibitemShut {NoStop}%
\bibitem [{\citenamefont {Sun}\ \emph {et~al.}(2017{\natexlab{a}})\citenamefont
  {Sun}, \citenamefont {Berkelbach}, \citenamefont {McClain},\ and\
  \citenamefont {Chan}}]{sun2017gaussian}%
  \BibitemOpen
  \bibfield  {author} {\bibinfo {author} {\bibfnamefont {Q.}~\bibnamefont
  {Sun}}, \bibinfo {author} {\bibfnamefont {T.~C.}\ \bibnamefont {Berkelbach}},
  \bibinfo {author} {\bibfnamefont {J.~D.}\ \bibnamefont {McClain}}, \ and\
  \bibinfo {author} {\bibfnamefont {G.~K.}\ \bibnamefont {Chan}},\ }\href@noop
  {} {\bibfield  {journal} {\bibinfo  {journal} {J. Chem. Phys.}\ }\textbf
  {\bibinfo {volume} {147}} (\bibinfo {year} {2017}{\natexlab{a}})}\BibitemShut
  {NoStop}%
\bibitem [{\citenamefont {Bintrim}\ \emph {et~al.}(2022)\citenamefont
  {Bintrim}, \citenamefont {Berkelbach},\ and\ \citenamefont
  {Ye}}]{bintrim2022integral}%
  \BibitemOpen
  \bibfield  {author} {\bibinfo {author} {\bibfnamefont {S.~J.}\ \bibnamefont
  {Bintrim}}, \bibinfo {author} {\bibfnamefont {T.~C.}\ \bibnamefont
  {Berkelbach}}, \ and\ \bibinfo {author} {\bibfnamefont {H.-Z.}\ \bibnamefont
  {Ye}},\ }\href@noop {} {\bibfield  {journal} {\bibinfo  {journal} {J. Chem.
  Theory Comput.}\ }\textbf {\bibinfo {volume} {18}},\ \bibinfo {pages} {5374}
  (\bibinfo {year} {2022})}\BibitemShut {NoStop}%
\bibitem [{\citenamefont {Jung}\ \emph {et~al.}(2005)\citenamefont {Jung},
  \citenamefont {Sodt}, \citenamefont {Gill},\ and\ \citenamefont
  {Head-Gordon}}]{jung2005auxiliary}%
  \BibitemOpen
  \bibfield  {author} {\bibinfo {author} {\bibfnamefont {Y.}~\bibnamefont
  {Jung}}, \bibinfo {author} {\bibfnamefont {A.}~\bibnamefont {Sodt}}, \bibinfo
  {author} {\bibfnamefont {P.~M.}\ \bibnamefont {Gill}}, \ and\ \bibinfo
  {author} {\bibfnamefont {M.}~\bibnamefont {Head-Gordon}},\ }\href@noop {}
  {\bibfield  {journal} {\bibinfo  {journal} {Proc. Natl. Acad. Sci. U.S.A.}\
  }\textbf {\bibinfo {volume} {102}},\ \bibinfo {pages} {6692} (\bibinfo {year}
  {2005})}\BibitemShut {NoStop}%
\bibitem [{\citenamefont {Baerends}\ \emph {et~al.}(1973)\citenamefont
  {Baerends}, \citenamefont {Ellis},\ and\ \citenamefont
  {Ros}}]{baerends1973self}%
  \BibitemOpen
  \bibfield  {author} {\bibinfo {author} {\bibfnamefont {E.}~\bibnamefont
  {Baerends}}, \bibinfo {author} {\bibfnamefont {D.}~\bibnamefont {Ellis}}, \
  and\ \bibinfo {author} {\bibfnamefont {P.}~\bibnamefont {Ros}},\ }\href@noop
  {} {\bibfield  {journal} {\bibinfo  {journal} {Chem. Phys.}\ }\textbf
  {\bibinfo {volume} {2}},\ \bibinfo {pages} {41} (\bibinfo {year}
  {1973})}\BibitemShut {NoStop}%
\bibitem [{\citenamefont {Gill}\ \emph {et~al.}(2005)\citenamefont {Gill},
  \citenamefont {Gilbert}, \citenamefont {Taylor}, \citenamefont {Friesecke},\
  and\ \citenamefont {Head-Gordon}}]{gill2005decay}%
  \BibitemOpen
  \bibfield  {author} {\bibinfo {author} {\bibfnamefont {P.~M.}\ \bibnamefont
  {Gill}}, \bibinfo {author} {\bibfnamefont {A.~T.}\ \bibnamefont {Gilbert}},
  \bibinfo {author} {\bibfnamefont {S.}~\bibnamefont {Taylor}}, \bibinfo
  {author} {\bibfnamefont {G.}~\bibnamefont {Friesecke}}, \ and\ \bibinfo
  {author} {\bibfnamefont {M.}~\bibnamefont {Head-Gordon}},\ }\href@noop {}
  {\bibfield  {journal} {\bibinfo  {journal} {J. Chem. Phys.}\ }\textbf
  {\bibinfo {volume} {123}} (\bibinfo {year} {2005})}\BibitemShut {NoStop}%
\bibitem [{\citenamefont {Reine}\ \emph {et~al.}(2008)\citenamefont {Reine},
  \citenamefont {Tellgren}, \citenamefont {Krapp}, \citenamefont
  {Kj{\ae}rgaard}, \citenamefont {Helgaker}, \citenamefont {Jansik},
  \citenamefont {H{\o}st},\ and\ \citenamefont {Salek}}]{reine2008variational}%
  \BibitemOpen
  \bibfield  {author} {\bibinfo {author} {\bibfnamefont {S.}~\bibnamefont
  {Reine}}, \bibinfo {author} {\bibfnamefont {E.}~\bibnamefont {Tellgren}},
  \bibinfo {author} {\bibfnamefont {A.}~\bibnamefont {Krapp}}, \bibinfo
  {author} {\bibfnamefont {T.}~\bibnamefont {Kj{\ae}rgaard}}, \bibinfo {author}
  {\bibfnamefont {T.}~\bibnamefont {Helgaker}}, \bibinfo {author}
  {\bibfnamefont {B.}~\bibnamefont {Jansik}}, \bibinfo {author} {\bibfnamefont
  {S.}~\bibnamefont {H{\o}st}}, \ and\ \bibinfo {author} {\bibfnamefont
  {P.}~\bibnamefont {Salek}},\ }\href@noop {} {\bibfield  {journal} {\bibinfo
  {journal} {J. Chem. Phys.}\ }\textbf {\bibinfo {volume} {129}} (\bibinfo
  {year} {2008})}\BibitemShut {NoStop}%
\bibitem [{\citenamefont {Yang}\ and\ \citenamefont
  {Lee}(1995)}]{yang1995density}%
  \BibitemOpen
  \bibfield  {author} {\bibinfo {author} {\bibfnamefont {W.}~\bibnamefont
  {Yang}}\ and\ \bibinfo {author} {\bibfnamefont {T.-S.}\ \bibnamefont {Lee}},\
  }\href@noop {} {\bibfield  {journal} {\bibinfo  {journal} {J. Chem. Phys.}\
  }\textbf {\bibinfo {volume} {103}},\ \bibinfo {pages} {5674} (\bibinfo {year}
  {1995})}\BibitemShut {NoStop}%
\bibitem [{\citenamefont {Gallant}\ and\ \citenamefont
  {St-Amant}(1996)}]{gallant1996linear}%
  \BibitemOpen
  \bibfield  {author} {\bibinfo {author} {\bibfnamefont {R.~T.}\ \bibnamefont
  {Gallant}}\ and\ \bibinfo {author} {\bibfnamefont {A.}~\bibnamefont
  {St-Amant}},\ }\href@noop {} {\bibfield  {journal} {\bibinfo  {journal}
  {Chem. Phys. Lett.}\ }\textbf {\bibinfo {volume} {256}},\ \bibinfo {pages}
  {569} (\bibinfo {year} {1996})}\BibitemShut {NoStop}%
\bibitem [{\citenamefont {Sodt}\ and\ \citenamefont
  {Head-Gordon}(2008)}]{sodt2008hartree}%
  \BibitemOpen
  \bibfield  {author} {\bibinfo {author} {\bibfnamefont {A.}~\bibnamefont
  {Sodt}}\ and\ \bibinfo {author} {\bibfnamefont {M.}~\bibnamefont
  {Head-Gordon}},\ }\href@noop {} {\bibfield  {journal} {\bibinfo  {journal}
  {J. Chem. Phys.}\ }\textbf {\bibinfo {volume} {128}} (\bibinfo {year}
  {2008})}\BibitemShut {NoStop}%
\bibitem [{\citenamefont {Fonseca~Guerra}\ \emph {et~al.}(1998)\citenamefont
  {Fonseca~Guerra}, \citenamefont {Snijders}, \citenamefont {Te~Velde},\ and\
  \citenamefont {Baerends}}]{fonseca1998towards}%
  \BibitemOpen
  \bibfield  {author} {\bibinfo {author} {\bibfnamefont {C.}~\bibnamefont
  {Fonseca~Guerra}}, \bibinfo {author} {\bibfnamefont {J.}~\bibnamefont
  {Snijders}}, \bibinfo {author} {\bibfnamefont {G.~t.}\ \bibnamefont
  {Te~Velde}}, \ and\ \bibinfo {author} {\bibfnamefont {E.~J.}\ \bibnamefont
  {Baerends}},\ }\href@noop {} {\bibfield  {journal} {\bibinfo  {journal}
  {Theor. Chem. Acc.}\ }\textbf {\bibinfo {volume} {99}},\ \bibinfo {pages}
  {391} (\bibinfo {year} {1998})}\BibitemShut {NoStop}%
\bibitem [{\citenamefont {Hollman}\ \emph {et~al.}(2017)\citenamefont
  {Hollman}, \citenamefont {Schaefer},\ and\ \citenamefont
  {Valeev}}]{hollman2017fast}%
  \BibitemOpen
  \bibfield  {author} {\bibinfo {author} {\bibfnamefont {D.~S.}\ \bibnamefont
  {Hollman}}, \bibinfo {author} {\bibfnamefont {H.~F.}\ \bibnamefont
  {Schaefer}}, \ and\ \bibinfo {author} {\bibfnamefont {E.~F.}\ \bibnamefont
  {Valeev}},\ }\href@noop {} {\bibfield  {journal} {\bibinfo  {journal} {Mol.
  Phys.}\ }\textbf {\bibinfo {volume} {115}},\ \bibinfo {pages} {2065}
  (\bibinfo {year} {2017})}\BibitemShut {NoStop}%
\bibitem [{\citenamefont {Manzer}\ \emph {et~al.}(2015)\citenamefont {Manzer},
  \citenamefont {Epifanovsky},\ and\ \citenamefont
  {Head-Gordon}}]{manzer2015efficient}%
  \BibitemOpen
  \bibfield  {author} {\bibinfo {author} {\bibfnamefont {S.~F.}\ \bibnamefont
  {Manzer}}, \bibinfo {author} {\bibfnamefont {E.}~\bibnamefont {Epifanovsky}},
  \ and\ \bibinfo {author} {\bibfnamefont {M.}~\bibnamefont {Head-Gordon}},\
  }\href@noop {} {\bibfield  {journal} {\bibinfo  {journal} {J. Chem. Theory
  Comput.}\ }\textbf {\bibinfo {volume} {11}},\ \bibinfo {pages} {518}
  (\bibinfo {year} {2015})}\BibitemShut {NoStop}%
\bibitem [{\citenamefont {Rebolini}\ \emph {et~al.}(2016)\citenamefont
  {Rebolini}, \citenamefont {Izsak}, \citenamefont {Reine}, \citenamefont
  {Helgaker},\ and\ \citenamefont {Pedersen}}]{rebolini2016comparison}%
  \BibitemOpen
  \bibfield  {author} {\bibinfo {author} {\bibfnamefont {E.}~\bibnamefont
  {Rebolini}}, \bibinfo {author} {\bibfnamefont {R.}~\bibnamefont {Izsak}},
  \bibinfo {author} {\bibfnamefont {S.~S.}\ \bibnamefont {Reine}}, \bibinfo
  {author} {\bibfnamefont {T.}~\bibnamefont {Helgaker}}, \ and\ \bibinfo
  {author} {\bibfnamefont {T.~B.}\ \bibnamefont {Pedersen}},\ }\href@noop {}
  {\bibfield  {journal} {\bibinfo  {journal} {J. Chem. Theory Comput.}\
  }\textbf {\bibinfo {volume} {12}},\ \bibinfo {pages} {3514} (\bibinfo {year}
  {2016})}\BibitemShut {NoStop}%
\bibitem [{\citenamefont {Mintmire}\ and\ \citenamefont
  {Dunlap}(1982)}]{Mintmire1982Jan}%
  \BibitemOpen
  \bibfield  {author} {\bibinfo {author} {\bibfnamefont {J.~W.}\ \bibnamefont
  {Mintmire}}\ and\ \bibinfo {author} {\bibfnamefont {B.~I.}\ \bibnamefont
  {Dunlap}},\ }\href {\doibase 10.1103/PhysRevA.25.88} {\bibfield  {journal}
  {\bibinfo  {journal} {Phys. Rev. A}\ }\textbf {\bibinfo {volume} {25}},\
  \bibinfo {pages} {88} (\bibinfo {year} {1982})}\BibitemShut {NoStop}%
\bibitem [{\citenamefont {Manby}\ \emph {et~al.}(2001)\citenamefont {Manby},
  \citenamefont {Knowles},\ and\ \citenamefont {Lloyd}}]{manby2001poisson}%
  \BibitemOpen
  \bibfield  {author} {\bibinfo {author} {\bibfnamefont {F.}~\bibnamefont
  {Manby}}, \bibinfo {author} {\bibfnamefont {P.~J.}\ \bibnamefont {Knowles}},
  \ and\ \bibinfo {author} {\bibfnamefont {A.}~\bibnamefont {Lloyd}},\
  }\href@noop {} {\bibfield  {journal} {\bibinfo  {journal} {J. Chem. Phys.}\
  }\textbf {\bibinfo {volume} {115}},\ \bibinfo {pages} {9144} (\bibinfo {year}
  {2001})}\BibitemShut {NoStop}%
\bibitem [{\citenamefont {McCurdy}\ \emph {et~al.}(2001)\citenamefont
  {McCurdy}, \citenamefont {Horner},\ and\ \citenamefont
  {Rescigno}}]{McCurdy2001Jan}%
  \BibitemOpen
  \bibfield  {author} {\bibinfo {author} {\bibfnamefont {C.~W.}\ \bibnamefont
  {McCurdy}}, \bibinfo {author} {\bibfnamefont {D.~A.}\ \bibnamefont {Horner}},
  \ and\ \bibinfo {author} {\bibfnamefont {T.~N.}\ \bibnamefont {Rescigno}},\
  }\href {\doibase 10.1103/PhysRevA.63.022711} {\bibfield  {journal} {\bibinfo
  {journal} {Phys. Rev. A}\ }\textbf {\bibinfo {volume} {63}},\ \bibinfo
  {pages} {022711} (\bibinfo {year} {2001})}\BibitemShut {NoStop}%
\bibitem [{\citenamefont {Lambrecht}\ \emph {et~al.}(2011)\citenamefont
  {Lambrecht}, \citenamefont {Brandhorst}, \citenamefont {Miller},
  \citenamefont {McCurdy},\ and\ \citenamefont
  {Head-Gordon}}]{Lambrecht2011Mar}%
  \BibitemOpen
  \bibfield  {author} {\bibinfo {author} {\bibfnamefont {D.~S.}\ \bibnamefont
  {Lambrecht}}, \bibinfo {author} {\bibfnamefont {K.}~\bibnamefont
  {Brandhorst}}, \bibinfo {author} {\bibfnamefont {W.~H.}\ \bibnamefont
  {Miller}}, \bibinfo {author} {\bibfnamefont {C.~W.}\ \bibnamefont {McCurdy}},
  \ and\ \bibinfo {author} {\bibfnamefont {M.}~\bibnamefont {Head-Gordon}},\
  }\href {\doibase 10.1021/jp108218w} {\bibfield  {journal} {\bibinfo
  {journal} {J. Phys. Chem. A}\ }\textbf {\bibinfo {volume} {115}},\ \bibinfo
  {pages} {2794} (\bibinfo {year} {2011})}\BibitemShut {NoStop}%
\bibitem [{\citenamefont {Wang}\ \emph {et~al.}(2020)\citenamefont {Wang},
  \citenamefont {Lewis},\ and\ \citenamefont {Valeev}}]{wang2020efficient}%
  \BibitemOpen
  \bibfield  {author} {\bibinfo {author} {\bibfnamefont {X.}~\bibnamefont
  {Wang}}, \bibinfo {author} {\bibfnamefont {C.~A.}\ \bibnamefont {Lewis}}, \
  and\ \bibinfo {author} {\bibfnamefont {E.~F.}\ \bibnamefont {Valeev}},\
  }\href@noop {} {\bibfield  {journal} {\bibinfo  {journal} {J. Chem. Phys.}\
  }\textbf {\bibinfo {volume} {153}} (\bibinfo {year} {2020})}\BibitemShut
  {NoStop}%
\bibitem [{\citenamefont {Sun}\ \emph {et~al.}(2017{\natexlab{b}})\citenamefont
  {Sun}, \citenamefont {Berkelbach}, \citenamefont {McClain},\ and\
  \citenamefont {Chan}}]{sun_gaussian_2017}%
  \BibitemOpen
  \bibfield  {author} {\bibinfo {author} {\bibfnamefont {Q.}~\bibnamefont
  {Sun}}, \bibinfo {author} {\bibfnamefont {T.~C.}\ \bibnamefont {Berkelbach}},
  \bibinfo {author} {\bibfnamefont {J.~D.}\ \bibnamefont {McClain}}, \ and\
  \bibinfo {author} {\bibfnamefont {G.~K.-L.}\ \bibnamefont {Chan}},\ }\href
  {\doibase 10.1063/1.4998644} {\bibfield  {journal} {\bibinfo  {journal} {J.
  Chem. Phys.}\ }\textbf {\bibinfo {volume} {147}},\ \bibinfo {pages} {164119}
  (\bibinfo {year} {2017}{\natexlab{b}})}\BibitemShut {NoStop}%
\bibitem [{\citenamefont {Benoit}(1924)}]{benoit1924note}%
  \BibitemOpen
  \bibfield  {author} {\bibinfo {author} {\bibnamefont {Benoit}},\ }\href
  {\doibase 10.1007/BF03031308} {\bibfield  {journal} {\bibinfo  {journal}
  {Bulletin Géodésique}\ }\textbf {\bibinfo {volume} {2}},\ \bibinfo {pages}
  {67} (\bibinfo {year} {1924})}\BibitemShut {NoStop}%
\bibitem [{\citenamefont {Beebe}\ and\ \citenamefont
  {Linderberg}(1977)}]{Beebe1977}%
  \BibitemOpen
  \bibfield  {author} {\bibinfo {author} {\bibfnamefont {N.~H.~F.}\
  \bibnamefont {Beebe}}\ and\ \bibinfo {author} {\bibfnamefont
  {J.}~\bibnamefont {Linderberg}},\ }\href@noop {} {\bibfield  {journal}
  {\bibinfo  {journal} {Int. J. Quantum Chem.}\ }\textbf {\bibinfo {volume}
  {12}},\ \bibinfo {pages} {683} (\bibinfo {year} {1977})}\BibitemShut
  {NoStop}%
\bibitem [{\citenamefont {Aquilante}\ \emph {et~al.}(2011)\citenamefont
  {Aquilante}, \citenamefont {Boman}, \citenamefont {Bostr{\"o}m},
  \citenamefont {Koch}, \citenamefont {Lindh}, \citenamefont {de~Mer{\'a}s},\
  and\ \citenamefont {Pedersen}}]{Aquilante2011}%
  \BibitemOpen
  \bibfield  {author} {\bibinfo {author} {\bibfnamefont {F.}~\bibnamefont
  {Aquilante}}, \bibinfo {author} {\bibfnamefont {L.}~\bibnamefont {Boman}},
  \bibinfo {author} {\bibfnamefont {J.}~\bibnamefont {Bostr{\"o}m}}, \bibinfo
  {author} {\bibfnamefont {H.}~\bibnamefont {Koch}}, \bibinfo {author}
  {\bibfnamefont {R.}~\bibnamefont {Lindh}}, \bibinfo {author} {\bibfnamefont
  {A.~S.}\ \bibnamefont {de~Mer{\'a}s}}, \ and\ \bibinfo {author}
  {\bibfnamefont {T.~B.}\ \bibnamefont {Pedersen}},\ }\enquote {\bibinfo
  {title} {Cholesky decomposition techniques in electronic structure theory},}\
  in\ \href {\doibase 10.1007/978-90-481-2853-2_13} {\emph {\bibinfo
  {booktitle} {Linear-Scaling Techniques in Computational Chemistry and
  Physics: Methods and Applications}}},\ \bibinfo {editor} {edited by\ \bibinfo
  {editor} {\bibfnamefont {R.}~\bibnamefont {Zalesny}}, \bibinfo {editor}
  {\bibfnamefont {M.~G.}\ \bibnamefont {Papadopoulos}}, \bibinfo {editor}
  {\bibfnamefont {P.~G.}\ \bibnamefont {Mezey}}, \ and\ \bibinfo {editor}
  {\bibfnamefont {J.}~\bibnamefont {Leszczynski}}}\ (\bibinfo  {publisher}
  {Springer Netherlands},\ \bibinfo {address} {Dordrecht},\ \bibinfo {year}
  {2011})\ pp.\ \bibinfo {pages} {301--343}\BibitemShut {NoStop}%
\bibitem [{\citenamefont {Hohenstein}\ \emph
  {et~al.}(2012{\natexlab{a}})\citenamefont {Hohenstein}, \citenamefont
  {Parrish},\ and\ \citenamefont {Mart{\'{i}}nez}}]{Hohenstein2012}%
  \BibitemOpen
  \bibfield  {author} {\bibinfo {author} {\bibfnamefont {E.~G.}\ \bibnamefont
  {Hohenstein}}, \bibinfo {author} {\bibfnamefont {R.~M.}\ \bibnamefont
  {Parrish}}, \ and\ \bibinfo {author} {\bibfnamefont {T.~J.}\ \bibnamefont
  {Mart{\'{i}}nez}},\ }\href {\doibase 10.1063/1.4732310} {\bibfield  {journal}
  {\bibinfo  {journal} {J. Chem. Phys.}\ }\textbf {\bibinfo {volume} {137}},\
  \bibinfo {pages} {1085} (\bibinfo {year} {2012}{\natexlab{a}})}\BibitemShut
  {NoStop}%
\bibitem [{\citenamefont {Parrish}\ \emph {et~al.}(2012)\citenamefont
  {Parrish}, \citenamefont {Hohenstein}, \citenamefont {Mart{\'{i}}nez},\ and\
  \citenamefont {Sherrill}}]{Parrish2012}%
  \BibitemOpen
  \bibfield  {author} {\bibinfo {author} {\bibfnamefont {R.~M.}\ \bibnamefont
  {Parrish}}, \bibinfo {author} {\bibfnamefont {E.~G.}\ \bibnamefont
  {Hohenstein}}, \bibinfo {author} {\bibfnamefont {T.~J.}\ \bibnamefont
  {Mart{\'{i}}nez}}, \ and\ \bibinfo {author} {\bibfnamefont {C.~D.}\
  \bibnamefont {Sherrill}},\ }\href {\doibase 10.1063/1.4768233} {\bibfield
  {journal} {\bibinfo  {journal} {J. Chem. Phys.}\ }\textbf {\bibinfo {volume}
  {137}},\ \bibinfo {pages} {224106} (\bibinfo {year} {2012})}\BibitemShut
  {NoStop}%
\bibitem [{\citenamefont {Hohenstein}\ \emph
  {et~al.}(2012{\natexlab{b}})\citenamefont {Hohenstein}, \citenamefont
  {Parrish}, \citenamefont {Sherrill},\ and\ \citenamefont
  {Mart{\'{i}}nez}}]{Hohenstein2012a}%
  \BibitemOpen
  \bibfield  {author} {\bibinfo {author} {\bibfnamefont {E.~G.}\ \bibnamefont
  {Hohenstein}}, \bibinfo {author} {\bibfnamefont {R.~M.}\ \bibnamefont
  {Parrish}}, \bibinfo {author} {\bibfnamefont {C.~D.}\ \bibnamefont
  {Sherrill}}, \ and\ \bibinfo {author} {\bibfnamefont {T.~J.}\ \bibnamefont
  {Mart{\'{i}}nez}},\ }\bibfield  {booktitle} {\emph {\bibinfo {booktitle} {J.
  Chem. Phys.}},\ }\href {\doibase 10.1063/1.4768241} {\bibfield  {journal}
  {\bibinfo  {journal} {J. Chem. Phys.}\ }\textbf {\bibinfo {volume} {137}},\
  \bibinfo {pages} {221101} (\bibinfo {year} {2012}{\natexlab{b}})}\BibitemShut
  {NoStop}%
\bibitem [{\citenamefont {Parrish}\ \emph {et~al.}(2013)\citenamefont
  {Parrish}, \citenamefont {Hohenstein}, \citenamefont {Mart{\'{i}}nez},\ and\
  \citenamefont {Sherrill}}]{Parrish2013a}%
  \BibitemOpen
  \bibfield  {author} {\bibinfo {author} {\bibfnamefont {R.~M.}\ \bibnamefont
  {Parrish}}, \bibinfo {author} {\bibfnamefont {E.~G.}\ \bibnamefont
  {Hohenstein}}, \bibinfo {author} {\bibfnamefont {T.~J.}\ \bibnamefont
  {Mart{\'{i}}nez}}, \ and\ \bibinfo {author} {\bibfnamefont {C.~D.}\
  \bibnamefont {Sherrill}},\ }\href {\doibase 10.1063/1.4802773} {\bibfield
  {journal} {\bibinfo  {journal} {J. Chem. Phys.}\ }\textbf {\bibinfo {volume}
  {138}},\ \bibinfo {pages} {194107} (\bibinfo {year} {2013})}\BibitemShut
  {NoStop}%
\bibitem [{\citenamefont {Hohenstein}\ \emph
  {et~al.}(2013{\natexlab{a}})\citenamefont {Hohenstein}, \citenamefont
  {Kokkila}, \citenamefont {Parrish},\ and\ \citenamefont
  {Mart{\'{i}}nez}}]{Hohenstein2013a}%
  \BibitemOpen
  \bibfield  {author} {\bibinfo {author} {\bibfnamefont {E.~G.}\ \bibnamefont
  {Hohenstein}}, \bibinfo {author} {\bibfnamefont {S.~I.}\ \bibnamefont
  {Kokkila}}, \bibinfo {author} {\bibfnamefont {R.~M.}\ \bibnamefont
  {Parrish}}, \ and\ \bibinfo {author} {\bibfnamefont {T.~J.}\ \bibnamefont
  {Mart{\'{i}}nez}},\ }\href {\doibase 10.1063/1.4795514} {\bibfield  {journal}
  {\bibinfo  {journal} {J. Chem. Phys.}\ }\textbf {\bibinfo {volume} {138}},\
  \bibinfo {pages} {124111} (\bibinfo {year} {2013}{\natexlab{a}})}\BibitemShut
  {NoStop}%
\bibitem [{\citenamefont {Hohenstein}\ \emph
  {et~al.}(2013{\natexlab{b}})\citenamefont {Hohenstein}, \citenamefont
  {Kokkila}, \citenamefont {Parrish},\ and\ \citenamefont
  {Mart{\'{i}}nez}}]{Hohenstein2013}%
  \BibitemOpen
  \bibfield  {author} {\bibinfo {author} {\bibfnamefont {E.~G.}\ \bibnamefont
  {Hohenstein}}, \bibinfo {author} {\bibfnamefont {S.~I.}\ \bibnamefont
  {Kokkila}}, \bibinfo {author} {\bibfnamefont {R.~M.}\ \bibnamefont
  {Parrish}}, \ and\ \bibinfo {author} {\bibfnamefont {T.~J.}\ \bibnamefont
  {Mart{\'{i}}nez}},\ }\href {\doibase 10.1021/jp4021905} {\bibfield  {journal}
  {\bibinfo  {journal} {J. Phys. Chem. B}\ }\textbf {\bibinfo {volume} {117}},\
  \bibinfo {pages} {12972} (\bibinfo {year} {2013}{\natexlab{b}})}\BibitemShut
  {NoStop}%
\bibitem [{\citenamefont {Parrish}\ \emph {et~al.}(2014)\citenamefont
  {Parrish}, \citenamefont {Sherrill}, \citenamefont {Hohenstein},
  \citenamefont {Kokkila},\ and\ \citenamefont {Mart{\'{i}}nez}}]{Parrish2014}%
  \BibitemOpen
  \bibfield  {author} {\bibinfo {author} {\bibfnamefont {R.~M.}\ \bibnamefont
  {Parrish}}, \bibinfo {author} {\bibfnamefont {C.~D.}\ \bibnamefont
  {Sherrill}}, \bibinfo {author} {\bibfnamefont {E.~G.}\ \bibnamefont
  {Hohenstein}}, \bibinfo {author} {\bibfnamefont {S.~I.}\ \bibnamefont
  {Kokkila}}, \ and\ \bibinfo {author} {\bibfnamefont {T.~J.}\ \bibnamefont
  {Mart{\'{i}}nez}},\ }\href {\doibase 10.1063/1.4876016} {\bibfield  {journal}
  {\bibinfo  {journal} {J. Chem. Phys.}\ }\textbf {\bibinfo {volume} {140}},\
  \bibinfo {pages} {181102} (\bibinfo {year} {2014})}\BibitemShut {NoStop}%
\bibitem [{\citenamefont {Hitchcock}(1927)}]{Hitchcock1927Apr}%
  \BibitemOpen
  \bibfield  {author} {\bibinfo {author} {\bibfnamefont {F.~L.}\ \bibnamefont
  {Hitchcock}},\ }\href {\doibase 10.1002/sapm192761164} {\bibfield  {journal}
  {\bibinfo  {journal} {J. Math. Phys.}\ }\textbf {\bibinfo {volume} {6}},\
  \bibinfo {pages} {164} (\bibinfo {year} {1927})}\BibitemShut {NoStop}%
\bibitem [{\citenamefont {Carroll}\ and\ \citenamefont
  {Chang}(1970)}]{Carroll1970Sep}%
  \BibitemOpen
  \bibfield  {author} {\bibinfo {author} {\bibfnamefont {J.~D.}\ \bibnamefont
  {Carroll}}\ and\ \bibinfo {author} {\bibfnamefont {J.-J.}\ \bibnamefont
  {Chang}},\ }\href {\doibase 10.1007/BF02310791} {\bibfield  {journal}
  {\bibinfo  {journal} {Psychometrika}\ }\textbf {\bibinfo {volume} {35}},\
  \bibinfo {pages} {283} (\bibinfo {year} {1970})}\BibitemShut {NoStop}%
\bibitem [{\citenamefont {Malone}\ \emph {et~al.}(2019)\citenamefont {Malone},
  \citenamefont {Zhang},\ and\ \citenamefont {Morales}}]{Malone2018Dec}%
  \BibitemOpen
  \bibfield  {author} {\bibinfo {author} {\bibfnamefont {F.~D.}\ \bibnamefont
  {Malone}}, \bibinfo {author} {\bibfnamefont {S.}~\bibnamefont {Zhang}}, \
  and\ \bibinfo {author} {\bibfnamefont {M.~A.}\ \bibnamefont {Morales}},\
  }\href {\doibase 10.1021/acs.jctc.8b00944} {\bibfield  {journal} {\bibinfo
  {journal} {J. Chem. Theory Comput.}\ }\textbf {\bibinfo {volume} {15}},\
  \bibinfo {pages} {256} (\bibinfo {year} {2019})}\BibitemShut {NoStop}%
\bibitem [{\citenamefont {Matthews}(2020)}]{Matthews2020Jan}%
  \BibitemOpen
  \bibfield  {author} {\bibinfo {author} {\bibfnamefont {D.~A.}\ \bibnamefont
  {Matthews}},\ }\href {\doibase 10.1021/acs.jctc.9b01205} {\bibfield
  {journal} {\bibinfo  {journal} {J. Chem. Theory Comput.}\ }\textbf {\bibinfo
  {volume} {16}},\ \bibinfo {pages} {1382} (\bibinfo {year}
  {2020})}\BibitemShut {NoStop}%
\bibitem [{\citenamefont {Lee}\ \emph {et~al.}(2020)\citenamefont {Lee},
  \citenamefont {Lin},\ and\ \citenamefont
  {Head-Gordon}}]{lee_systematically_2020}%
  \BibitemOpen
  \bibfield  {author} {\bibinfo {author} {\bibfnamefont {J.}~\bibnamefont
  {Lee}}, \bibinfo {author} {\bibfnamefont {L.}~\bibnamefont {Lin}}, \ and\
  \bibinfo {author} {\bibfnamefont {M.}~\bibnamefont {Head-Gordon}},\ }\href
  {\doibase 10.1021/acs.jctc.9b00820} {\bibfield  {journal} {\bibinfo
  {journal} {J. Chem. Theory Comput.}\ }\textbf {\bibinfo {volume} {16}},\
  \bibinfo {pages} {243} (\bibinfo {year} {2020})}\BibitemShut {NoStop}%
\bibitem [{\citenamefont {Matthews}(2021)}]{Matthews2021Apr}%
  \BibitemOpen
  \bibfield  {author} {\bibinfo {author} {\bibfnamefont {D.~A.}\ \bibnamefont
  {Matthews}},\ }\href {\doibase 10.1063/5.0038764} {\bibfield  {journal}
  {\bibinfo  {journal} {J. Chem. Phys.}\ }\textbf {\bibinfo {volume} {154}},\
  \bibinfo {pages} {134102} (\bibinfo {year} {2021})}\BibitemShut {NoStop}%
\bibitem [{\citenamefont {Zhao}\ \emph {et~al.}(2023)\citenamefont {Zhao},
  \citenamefont {Simons},\ and\ \citenamefont {Matthews}}]{Zhao2023Jun}%
  \BibitemOpen
  \bibfield  {author} {\bibinfo {author} {\bibfnamefont {T.}~\bibnamefont
  {Zhao}}, \bibinfo {author} {\bibfnamefont {M.}~\bibnamefont {Simons}}, \ and\
  \bibinfo {author} {\bibfnamefont {D.~A.}\ \bibnamefont {Matthews}},\ }\href
  {\doibase 10.1021/acs.jctc.3c00392} {\bibfield  {journal} {\bibinfo
  {journal} {J. Chem. Theory Comput.}\ }\textbf {\bibinfo {volume} {19}},\
  \bibinfo {pages} {3996} (\bibinfo {year} {2023})}\BibitemShut {NoStop}%
\bibitem [{\citenamefont {Lu}\ and\ \citenamefont {Ying}(2015)}]{Lu2015}%
  \BibitemOpen
  \bibfield  {author} {\bibinfo {author} {\bibfnamefont {J.}~\bibnamefont
  {Lu}}\ and\ \bibinfo {author} {\bibfnamefont {L.}~\bibnamefont {Ying}},\
  }\href {\doibase 10.1016/j.jcp.2015.09.014} {\bibfield  {journal} {\bibinfo
  {journal} {J. Comput. Phys}\ }\textbf {\bibinfo {volume} {302}},\ \bibinfo
  {pages} {329} (\bibinfo {year} {2015})}\BibitemShut {NoStop}%
\bibitem [{\citenamefont {Dong}\ \emph {et~al.}(2018)\citenamefont {Dong},
  \citenamefont {Hu},\ and\ \citenamefont {Lin}}]{Dong2018Jan}%
  \BibitemOpen
  \bibfield  {author} {\bibinfo {author} {\bibfnamefont {K.}~\bibnamefont
  {Dong}}, \bibinfo {author} {\bibfnamefont {W.}~\bibnamefont {Hu}}, \ and\
  \bibinfo {author} {\bibfnamefont {L.}~\bibnamefont {Lin}},\ }\href {\doibase
  10.1021/acs.jctc.7b01113} {\bibfield  {journal} {\bibinfo  {journal} {J.
  Chem. Theory Comput.}\ }\textbf {\bibinfo {volume} {14}},\ \bibinfo {pages}
  {1311} (\bibinfo {year} {2018})}\BibitemShut {NoStop}%
\bibitem [{\citenamefont {Hu}\ \emph {et~al.}(2017)\citenamefont {Hu},
  \citenamefont {Lin},\ and\ \citenamefont {Yang}}]{hu_interpolative_2017}%
  \BibitemOpen
  \bibfield  {author} {\bibinfo {author} {\bibfnamefont {W.}~\bibnamefont
  {Hu}}, \bibinfo {author} {\bibfnamefont {L.}~\bibnamefont {Lin}}, \ and\
  \bibinfo {author} {\bibfnamefont {C.}~\bibnamefont {Yang}},\ }\href {\doibase
  10.1021/acs.jctc.7b00807} {\bibfield  {journal} {\bibinfo  {journal} {J.
  Chem. Theory Comput.}\ }\textbf {\bibinfo {volume} {13}},\ \bibinfo {pages}
  {5420} (\bibinfo {year} {2017})}\BibitemShut {NoStop}%
\bibitem [{\citenamefont {Wu}\ \emph {et~al.}(2022)\citenamefont {Wu},
  \citenamefont {Qin}, \citenamefont {Hu},\ and\ \citenamefont
  {Yang}}]{Wu2022Jan}%
  \BibitemOpen
  \bibfield  {author} {\bibinfo {author} {\bibfnamefont {K.}~\bibnamefont
  {Wu}}, \bibinfo {author} {\bibfnamefont {X.}~\bibnamefont {Qin}}, \bibinfo
  {author} {\bibfnamefont {W.}~\bibnamefont {Hu}}, \ and\ \bibinfo {author}
  {\bibfnamefont {J.}~\bibnamefont {Yang}},\ }\href {\doibase
  10.1021/acs.jctc.1c00874} {\bibfield  {journal} {\bibinfo  {journal} {J.
  Chem. Theory Comput.}\ }\textbf {\bibinfo {volume} {18}},\ \bibinfo {pages}
  {206} (\bibinfo {year} {2022})}\BibitemShut {NoStop}%
\bibitem [{\citenamefont {Qin}\ \emph {et~al.}(2023)\citenamefont {Qin},
  \citenamefont {Hu},\ and\ \citenamefont {Yang}}]{qin_interpolative_2023}%
  \BibitemOpen
  \bibfield  {author} {\bibinfo {author} {\bibfnamefont {X.}~\bibnamefont
  {Qin}}, \bibinfo {author} {\bibfnamefont {W.}~\bibnamefont {Hu}}, \ and\
  \bibinfo {author} {\bibfnamefont {J.}~\bibnamefont {Yang}},\ }\href {\doibase
  10.1021/acs.jctc.2c00927} {\bibfield  {journal} {\bibinfo  {journal} {J.
  Chem. Theory Comput.}\ }\textbf {\bibinfo {volume} {19}},\ \bibinfo {pages}
  {679} (\bibinfo {year} {2023})}\BibitemShut {NoStop}%
\bibitem [{\citenamefont {Greengard}\ and\ \citenamefont
  {Rokhlin}(1987)}]{greengard1987fast}%
  \BibitemOpen
  \bibfield  {author} {\bibinfo {author} {\bibfnamefont {L.}~\bibnamefont
  {Greengard}}\ and\ \bibinfo {author} {\bibfnamefont {V.}~\bibnamefont
  {Rokhlin}},\ }\href@noop {} {\bibfield  {journal} {\bibinfo  {journal} {J.
  Comput. Phys}\ }\textbf {\bibinfo {volume} {73}},\ \bibinfo {pages} {325}
  (\bibinfo {year} {1987})}\BibitemShut {NoStop}%
\bibitem [{\citenamefont {White}\ and\ \citenamefont
  {Head-Gordon}(1994)}]{White1994Oct}%
  \BibitemOpen
  \bibfield  {author} {\bibinfo {author} {\bibfnamefont {C.~A.}\ \bibnamefont
  {White}}\ and\ \bibinfo {author} {\bibfnamefont {M.}~\bibnamefont
  {Head-Gordon}},\ }\href {\doibase 10.1063/1.468354} {\bibfield  {journal}
  {\bibinfo  {journal} {J. Chem. Phys.}\ }\textbf {\bibinfo {volume} {101}},\
  \bibinfo {pages} {6593} (\bibinfo {year} {1994})}\BibitemShut {NoStop}%
\bibitem [{\citenamefont {White}\ \emph {et~al.}(1994)\citenamefont {White},
  \citenamefont {Johnson}, \citenamefont {Gill},\ and\ \citenamefont
  {Head-Gordon}}]{white_continuous_1994}%
  \BibitemOpen
  \bibfield  {author} {\bibinfo {author} {\bibfnamefont {C.~A.}\ \bibnamefont
  {White}}, \bibinfo {author} {\bibfnamefont {B.~G.}\ \bibnamefont {Johnson}},
  \bibinfo {author} {\bibfnamefont {P.~M.~W.}\ \bibnamefont {Gill}}, \ and\
  \bibinfo {author} {\bibfnamefont {M.}~\bibnamefont {Head-Gordon}},\ }\href
  {\doibase 10.1016/0009-2614(94)01128-1} {\bibfield  {journal} {\bibinfo
  {journal} {Chem. Phys. Lett.}\ }\textbf {\bibinfo {volume} {230}},\ \bibinfo
  {pages} {8} (\bibinfo {year} {1994})}\BibitemShut {NoStop}%
\bibitem [{\citenamefont {Strain}\ \emph {et~al.}(1996)\citenamefont {Strain},
  \citenamefont {Scuseria},\ and\ \citenamefont {Frisch}}]{Strain1996Jan}%
  \BibitemOpen
  \bibfield  {author} {\bibinfo {author} {\bibfnamefont {M.~C.}\ \bibnamefont
  {Strain}}, \bibinfo {author} {\bibfnamefont {G.~E.}\ \bibnamefont
  {Scuseria}}, \ and\ \bibinfo {author} {\bibfnamefont {M.~J.}\ \bibnamefont
  {Frisch}},\ }\href {\doibase 10.1126/science.271.5245.51} {\bibfield
  {journal} {\bibinfo  {journal} {Science}\ }\textbf {\bibinfo {volume}
  {271}},\ \bibinfo {pages} {51} (\bibinfo {year} {1996})}\BibitemShut
  {NoStop}%
\bibitem [{\citenamefont {Challacombe}\ \emph {et~al.}(1997)\citenamefont
  {Challacombe}, \citenamefont {White},\ and\ \citenamefont
  {Head-Gordon}}]{challacombe_periodic_1997}%
  \BibitemOpen
  \bibfield  {author} {\bibinfo {author} {\bibfnamefont {M.}~\bibnamefont
  {Challacombe}}, \bibinfo {author} {\bibfnamefont {C.}~\bibnamefont {White}},
  \ and\ \bibinfo {author} {\bibfnamefont {M.}~\bibnamefont {Head-Gordon}},\
  }\href {\doibase 10.1063/1.474150} {\bibfield  {journal} {\bibinfo  {journal}
  {J. Chem. Phys.}\ }\textbf {\bibinfo {volume} {107}},\ \bibinfo {pages}
  {10131} (\bibinfo {year} {1997})}\BibitemShut {NoStop}%
\bibitem [{\citenamefont {Kudin}\ and\ \citenamefont
  {Scuseria}(1998)}]{kudin1998fast}%
  \BibitemOpen
  \bibfield  {author} {\bibinfo {author} {\bibfnamefont {K.~N.}\ \bibnamefont
  {Kudin}}\ and\ \bibinfo {author} {\bibfnamefont {G.~E.}\ \bibnamefont
  {Scuseria}},\ }\href@noop {} {\bibfield  {journal} {\bibinfo  {journal}
  {Chem. Phys. Lett.}\ }\textbf {\bibinfo {volume} {289}},\ \bibinfo {pages}
  {611} (\bibinfo {year} {1998})}\BibitemShut {NoStop}%
\bibitem [{\citenamefont {Tymczak}\ and\ \citenamefont
  {Challacombe}(2005)}]{tymczak2005linear}%
  \BibitemOpen
  \bibfield  {author} {\bibinfo {author} {\bibfnamefont {C.}~\bibnamefont
  {Tymczak}}\ and\ \bibinfo {author} {\bibfnamefont {M.}~\bibnamefont
  {Challacombe}},\ }\href@noop {} {\bibfield  {journal} {\bibinfo  {journal}
  {J. Chem. Phys.}\ }\textbf {\bibinfo {volume} {122}} (\bibinfo {year}
  {2005})}\BibitemShut {NoStop}%
\bibitem [{\citenamefont {Burow}\ \emph {et~al.}(2009)\citenamefont {Burow},
  \citenamefont {Sierka},\ and\ \citenamefont
  {Mohamed}}]{burow_resolution_2009}%
  \BibitemOpen
  \bibfield  {author} {\bibinfo {author} {\bibfnamefont {A.~M.}\ \bibnamefont
  {Burow}}, \bibinfo {author} {\bibfnamefont {M.}~\bibnamefont {Sierka}}, \
  and\ \bibinfo {author} {\bibfnamefont {F.}~\bibnamefont {Mohamed}},\ }\href
  {\doibase 10.1063/1.3267858} {\bibfield  {journal} {\bibinfo  {journal} {J.
  Chem. Phys.}\ }\textbf {\bibinfo {volume} {131}},\ \bibinfo {pages} {214101}
  (\bibinfo {year} {2009})}\BibitemShut {NoStop}%
\bibitem [{\citenamefont {White}\ and\ \citenamefont
  {Head-Gordon}(1996)}]{White1996Feb}%
  \BibitemOpen
  \bibfield  {author} {\bibinfo {author} {\bibfnamefont {C.~A.}\ \bibnamefont
  {White}}\ and\ \bibinfo {author} {\bibfnamefont {M.}~\bibnamefont
  {Head-Gordon}},\ }\href {\doibase 10.1063/1.470986} {\bibfield  {journal}
  {\bibinfo  {journal} {J. Chem. Phys.}\ }\textbf {\bibinfo {volume} {104}},\
  \bibinfo {pages} {2620} (\bibinfo {year} {1996})}\BibitemShut {NoStop}%
\bibitem [{\citenamefont
  {F{\ifmmode\ddot{u}\else\"{u}\fi}sti-Moln{\ifmmode\acute{a}\else\'{a}\fi}r}\
  and\ \citenamefont {Kong}(2005)}]{Fusti-Molnar2005Feb}%
  \BibitemOpen
  \bibfield  {author} {\bibinfo {author} {\bibfnamefont {L.}~\bibnamefont
  {F{\ifmmode\ddot{u}\else\"{u}\fi}sti-Moln{\ifmmode\acute{a}\else\'{a}\fi}r}}\
  and\ \bibinfo {author} {\bibfnamefont {J.}~\bibnamefont {Kong}},\ }\href
  {\doibase 10.1063/1.1849168} {\bibfield  {journal} {\bibinfo  {journal} {J.
  Chem. Phys.}\ }\textbf {\bibinfo {volume} {122}},\ \bibinfo {pages} {074108}
  (\bibinfo {year} {2005})}\BibitemShut {NoStop}%
\bibitem [{\citenamefont {Mardirossian}\ and\ \citenamefont
  {Head-Gordon}(2017)}]{Mardirossian2017Oct}%
  \BibitemOpen
  \bibfield  {author} {\bibinfo {author} {\bibfnamefont {N.}~\bibnamefont
  {Mardirossian}}\ and\ \bibinfo {author} {\bibfnamefont {M.}~\bibnamefont
  {Head-Gordon}},\ }\href
  {https://www.tandfonline.com/doi/full/10.1080/00268976.2017.1333644}
  {\bibfield  {journal} {\bibinfo  {journal} {Mol. Phys.}\ } (\bibinfo {year}
  {2017})}\BibitemShut {NoStop}%
\bibitem [{\citenamefont {Becke}(1993)}]{Becke1993Jan}%
  \BibitemOpen
  \bibfield  {author} {\bibinfo {author} {\bibfnamefont {A.~D.}\ \bibnamefont
  {Becke}},\ }\href {\doibase 10.1063/1.464304} {\bibfield  {journal} {\bibinfo
   {journal} {J. Chem. Phys.}\ }\textbf {\bibinfo {volume} {98}},\ \bibinfo
  {pages} {1372} (\bibinfo {year} {1993})}\BibitemShut {NoStop}%
\bibitem [{\citenamefont {Mardirossian}\ and\ \citenamefont
  {Head-Gordon}(2014)}]{Mardirossian2014May}%
  \BibitemOpen
  \bibfield  {author} {\bibinfo {author} {\bibfnamefont {N.}~\bibnamefont
  {Mardirossian}}\ and\ \bibinfo {author} {\bibfnamefont {M.}~\bibnamefont
  {Head-Gordon}},\ }\href {\doibase 10.1039/C3CP54374A} {\bibfield  {journal}
  {\bibinfo  {journal} {Phys. Chem. Chem. Phys.}\ }\textbf {\bibinfo {volume}
  {16}},\ \bibinfo {pages} {9904} (\bibinfo {year} {2014})}\BibitemShut
  {NoStop}%
\bibitem [{\citenamefont {Nusspickel}\ and\ \citenamefont
  {Booth}(2022)}]{nusspickel2022systematic}%
  \BibitemOpen
  \bibfield  {author} {\bibinfo {author} {\bibfnamefont {M.}~\bibnamefont
  {Nusspickel}}\ and\ \bibinfo {author} {\bibfnamefont {G.~H.}\ \bibnamefont
  {Booth}},\ }\href@noop {} {\bibfield  {journal} {\bibinfo  {journal} {Phys.
  Rev. X}\ }\textbf {\bibinfo {volume} {12}},\ \bibinfo {pages} {011046}
  (\bibinfo {year} {2022})}\BibitemShut {NoStop}%
\bibitem [{\citenamefont {Lin}(2016)}]{Lin2016Apr}%
  \BibitemOpen
  \bibfield  {author} {\bibinfo {author} {\bibfnamefont {L.}~\bibnamefont
  {Lin}},\ }\href {\doibase 10.1021/acs.jctc.6b00092} {\bibfield  {journal}
  {\bibinfo  {journal} {J. Chem. Theory Comput.}\ }\textbf {\bibinfo {volume}
  {12}},\ \bibinfo {pages} {2242} (\bibinfo {year} {2016})}\BibitemShut
  {NoStop}%
\bibitem [{\citenamefont {Dreuw}\ and\ \citenamefont
  {Head-Gordon}(2004)}]{AndreasDreuw*2004Mar}%
  \BibitemOpen
  \bibfield  {author} {\bibinfo {author} {\bibfnamefont {A.}~\bibnamefont
  {Dreuw}}\ and\ \bibinfo {author} {\bibfnamefont {M.}~\bibnamefont
  {Head-Gordon}},\ }\href {\doibase 10.1021/ja039556n} {\bibfield  {journal}
  {\bibinfo  {journal} {J. Am. Chem. Soc.}\ }\textbf {\bibinfo {volume}
  {126}},\ \bibinfo {pages} {4007} (\bibinfo {year} {2004})}\BibitemShut
  {NoStop}%
\bibitem [{\citenamefont {Perdew}(1985)}]{Perdew1985Mar}%
  \BibitemOpen
  \bibfield  {author} {\bibinfo {author} {\bibfnamefont {J.~P.}\ \bibnamefont
  {Perdew}},\ }\href {\doibase 10.1002/qua.560280846} {\bibfield  {journal}
  {\bibinfo  {journal} {Int. J. Quantum Chem.}\ }\textbf {\bibinfo {volume}
  {28}},\ \bibinfo {pages} {497} (\bibinfo {year} {1985})}\BibitemShut
  {NoStop}%
\bibitem [{\citenamefont {Heyd}\ and\ \citenamefont
  {Scuseria}(2004)}]{Heyd2004Jul}%
  \BibitemOpen
  \bibfield  {author} {\bibinfo {author} {\bibfnamefont {J.}~\bibnamefont
  {Heyd}}\ and\ \bibinfo {author} {\bibfnamefont {G.~E.}\ \bibnamefont
  {Scuseria}},\ }\href {\doibase 10.1063/1.1760074} {\bibfield  {journal}
  {\bibinfo  {journal} {J. Chem. Phys.}\ }\textbf {\bibinfo {volume} {121}},\
  \bibinfo {pages} {1187} (\bibinfo {year} {2004})}\BibitemShut {NoStop}%
\bibitem [{\citenamefont {Lininger}\ \emph {et~al.}(2021)\citenamefont
  {Lininger}, \citenamefont {Gauthier}, \citenamefont {Li}, \citenamefont
  {Rossomme}, \citenamefont {Welborn}, \citenamefont {Lin}, \citenamefont
  {Head-Gordon}, \citenamefont {Head-Gordon},\ and\ \citenamefont
  {Bell}}]{lininger_challenges_2021}%
  \BibitemOpen
  \bibfield  {author} {\bibinfo {author} {\bibfnamefont {C.~N.}\ \bibnamefont
  {Lininger}}, \bibinfo {author} {\bibfnamefont {J.~A.}\ \bibnamefont
  {Gauthier}}, \bibinfo {author} {\bibfnamefont {W.-L.}\ \bibnamefont {Li}},
  \bibinfo {author} {\bibfnamefont {E.}~\bibnamefont {Rossomme}}, \bibinfo
  {author} {\bibfnamefont {V.~V.}\ \bibnamefont {Welborn}}, \bibinfo {author}
  {\bibfnamefont {Z.}~\bibnamefont {Lin}}, \bibinfo {author} {\bibfnamefont
  {T.}~\bibnamefont {Head-Gordon}}, \bibinfo {author} {\bibfnamefont
  {M.}~\bibnamefont {Head-Gordon}}, \ and\ \bibinfo {author} {\bibfnamefont
  {A.~T.}\ \bibnamefont {Bell}},\ }\href {\doibase 10.1039/D0CP03821K}
  {\bibfield  {journal} {\bibinfo  {journal} {Phys. Chem. Chem. Phys.}\
  }\textbf {\bibinfo {volume} {23}},\ \bibinfo {pages} {9394} (\bibinfo {year}
  {2021})}\BibitemShut {NoStop}%
\bibitem [{\citenamefont {Schmidt}\ and\ \citenamefont
  {Thygesen}(2018)}]{schmidt_benchmark_2018}%
  \BibitemOpen
  \bibfield  {author} {\bibinfo {author} {\bibfnamefont {P.~S.}\ \bibnamefont
  {Schmidt}}\ and\ \bibinfo {author} {\bibfnamefont {K.~S.}\ \bibnamefont
  {Thygesen}},\ }\href {\doibase 10.1021/acs.jpcc.7b12258} {\bibfield
  {journal} {\bibinfo  {journal} {J. Phys. Chem. C}\ }\textbf {\bibinfo
  {volume} {122}},\ \bibinfo {pages} {4381} (\bibinfo {year}
  {2018})}\BibitemShut {NoStop}%
\bibitem [{\citenamefont {Stroppa}\ and\ \citenamefont
  {Kresse}(2008)}]{Stroppa2008Jun}%
  \BibitemOpen
  \bibfield  {author} {\bibinfo {author} {\bibfnamefont {A.}~\bibnamefont
  {Stroppa}}\ and\ \bibinfo {author} {\bibfnamefont {G.}~\bibnamefont
  {Kresse}},\ }\href {\doibase 10.1088/1367-2630/10/6/063020} {\bibfield
  {journal} {\bibinfo  {journal} {New J. Phys.}\ }\textbf {\bibinfo {volume}
  {10}},\ \bibinfo {pages} {063020} (\bibinfo {year} {2008})}\BibitemShut
  {NoStop}%
\bibitem [{\citenamefont {Helgaker}\ \emph {et~al.}(1997)\citenamefont
  {Helgaker}, \citenamefont {Klopper}, \citenamefont {Koch},\ and\
  \citenamefont {Noga}}]{Helgaker1997Jun}%
  \BibitemOpen
  \bibfield  {author} {\bibinfo {author} {\bibfnamefont {T.}~\bibnamefont
  {Helgaker}}, \bibinfo {author} {\bibfnamefont {W.}~\bibnamefont {Klopper}},
  \bibinfo {author} {\bibfnamefont {H.}~\bibnamefont {Koch}}, \ and\ \bibinfo
  {author} {\bibfnamefont {J.}~\bibnamefont {Noga}},\ }\href {\doibase
  10.1063/1.473863} {\bibfield  {journal} {\bibinfo  {journal} {J. Chem.
  Phys.}\ }\textbf {\bibinfo {volume} {106}},\ \bibinfo {pages} {9639}
  (\bibinfo {year} {1997})}\BibitemShut {NoStop}%
\bibitem [{\citenamefont {Bakowies}(2007)}]{Bakowies2007Aug}%
  \BibitemOpen
  \bibfield  {author} {\bibinfo {author} {\bibfnamefont {D.}~\bibnamefont
  {Bakowies}},\ }\href {\doibase 10.1063/1.2749516} {\bibfield  {journal}
  {\bibinfo  {journal} {J. Chem. Phys.}\ }\textbf {\bibinfo {volume} {127}},\
  \bibinfo {pages} {084105} (\bibinfo {year} {2007})}\BibitemShut {NoStop}%
\bibitem [{\citenamefont {Dunning}(1989)}]{Dunning1989Jan}%
  \BibitemOpen
  \bibfield  {author} {\bibinfo {author} {\bibfnamefont {T.~H.}\ \bibnamefont
  {Dunning}},\ }\href {\doibase 10.1063/1.456153} {\bibfield  {journal}
  {\bibinfo  {journal} {J. Chem. Phys.}\ }\textbf {\bibinfo {volume} {90}},\
  \bibinfo {pages} {1007} (\bibinfo {year} {1989})}\BibitemShut {NoStop}%
\bibitem [{\citenamefont {Bertels}\ \emph {et~al.}(2019)\citenamefont
  {Bertels}, \citenamefont {Lee},\ and\ \citenamefont
  {Head-Gordon}}]{Bertels2019Aug}%
  \BibitemOpen
  \bibfield  {author} {\bibinfo {author} {\bibfnamefont {L.~W.}\ \bibnamefont
  {Bertels}}, \bibinfo {author} {\bibfnamefont {J.}~\bibnamefont {Lee}}, \ and\
  \bibinfo {author} {\bibfnamefont {M.}~\bibnamefont {Head-Gordon}},\ }\href
  {\doibase 10.1021/acs.jpclett.9b01641} {\bibfield  {journal} {\bibinfo
  {journal} {J. Phys. Chem. Lett.}\ }\textbf {\bibinfo {volume} {10}},\
  \bibinfo {pages} {4170} (\bibinfo {year} {2019})}\BibitemShut {NoStop}%
\bibitem [{\citenamefont {Loipersberger}\ \emph {et~al.}(2021)\citenamefont
  {Loipersberger}, \citenamefont {Bertels}, \citenamefont {Lee},\ and\
  \citenamefont {Head-Gordon}}]{Loipersberger2021Sep}%
  \BibitemOpen
  \bibfield  {author} {\bibinfo {author} {\bibfnamefont {M.}~\bibnamefont
  {Loipersberger}}, \bibinfo {author} {\bibfnamefont {L.~W.}\ \bibnamefont
  {Bertels}}, \bibinfo {author} {\bibfnamefont {J.}~\bibnamefont {Lee}}, \ and\
  \bibinfo {author} {\bibfnamefont {M.}~\bibnamefont {Head-Gordon}},\ }\href
  {\doibase 10.1021/acs.jctc.1c00469} {\bibfield  {journal} {\bibinfo
  {journal} {J. Chem. Theory Comput.}\ }\textbf {\bibinfo {volume} {17}},\
  \bibinfo {pages} {5582} (\bibinfo {year} {2021})}\BibitemShut {NoStop}%
\bibitem [{\citenamefont {Rettig}\ \emph {et~al.}(2020)\citenamefont {Rettig},
  \citenamefont {Hait}, \citenamefont {Bertels},\ and\ \citenamefont
  {Head-Gordon}}]{Rettig2020Dec}%
  \BibitemOpen
  \bibfield  {author} {\bibinfo {author} {\bibfnamefont {A.}~\bibnamefont
  {Rettig}}, \bibinfo {author} {\bibfnamefont {D.}~\bibnamefont {Hait}},
  \bibinfo {author} {\bibfnamefont {L.~W.}\ \bibnamefont {Bertels}}, \ and\
  \bibinfo {author} {\bibfnamefont {M.}~\bibnamefont {Head-Gordon}},\ }\href
  {\doibase 10.1021/acs.jctc.0c00986} {\bibfield  {journal} {\bibinfo
  {journal} {J. Chem. Theory Comput.}\ }\textbf {\bibinfo {volume} {16}},\
  \bibinfo {pages} {7473} (\bibinfo {year} {2020})}\BibitemShut {NoStop}%
\bibitem [{\citenamefont {Lee}\ and\ \citenamefont
  {Head-Gordon}(2018)}]{Lee2018Oct}%
  \BibitemOpen
  \bibfield  {author} {\bibinfo {author} {\bibfnamefont {J.}~\bibnamefont
  {Lee}}\ and\ \bibinfo {author} {\bibfnamefont {M.}~\bibnamefont
  {Head-Gordon}},\ }\href {\doibase 10.1021/acs.jctc.8b00731} {\bibfield
  {journal} {\bibinfo  {journal} {J. Chem. Theory Comput.}\ }\textbf {\bibinfo
  {volume} {14}},\ \bibinfo {pages} {5203} (\bibinfo {year}
  {2018})}\BibitemShut {NoStop}%
\bibitem [{\citenamefont {Rettig}\ \emph {et~al.}(2022)\citenamefont {Rettig},
  \citenamefont {Shee}, \citenamefont {Lee},\ and\ \citenamefont
  {Head-Gordon}}]{Rettig2022Sep}%
  \BibitemOpen
  \bibfield  {author} {\bibinfo {author} {\bibfnamefont {A.}~\bibnamefont
  {Rettig}}, \bibinfo {author} {\bibfnamefont {J.}~\bibnamefont {Shee}},
  \bibinfo {author} {\bibfnamefont {J.}~\bibnamefont {Lee}}, \ and\ \bibinfo
  {author} {\bibfnamefont {M.}~\bibnamefont {Head-Gordon}},\ }\href {\doibase
  10.1021/acs.jctc.2c00641} {\bibfield  {journal} {\bibinfo  {journal} {J.
  Chem. Theory Comput.}\ }\textbf {\bibinfo {volume} {18}},\ \bibinfo {pages}
  {5382} (\bibinfo {year} {2022})}\BibitemShut {NoStop}%
\bibitem [{\citenamefont {Bartlett}\ and\ \citenamefont
  {Musia{\l}}(2007)}]{Bartlett2007Feb}%
  \BibitemOpen
  \bibfield  {author} {\bibinfo {author} {\bibfnamefont {R.~J.}\ \bibnamefont
  {Bartlett}}\ and\ \bibinfo {author} {\bibfnamefont {M.}~\bibnamefont
  {Musia{\l}}},\ }\href {\doibase 10.1103/RevModPhys.79.291} {\bibfield
  {journal} {\bibinfo  {journal} {Rev. Mod. Phys.}\ }\textbf {\bibinfo {volume}
  {79}},\ \bibinfo {pages} {291} (\bibinfo {year} {2007})}\BibitemShut
  {NoStop}%
\bibitem [{\citenamefont {Thouless}(1960)}]{thouless1960stability}%
  \BibitemOpen
  \bibfield  {author} {\bibinfo {author} {\bibfnamefont {D.~J.}\ \bibnamefont
  {Thouless}},\ }\href@noop {} {\bibfield  {journal} {\bibinfo  {journal}
  {Nucl. Phys.}\ }\textbf {\bibinfo {volume} {21}},\ \bibinfo {pages} {225}
  (\bibinfo {year} {1960})}\BibitemShut {NoStop}%
\bibitem [{\citenamefont {Neufeld}\ and\ \citenamefont
  {Berkelbach}(2023)}]{Neufeld2023Oct}%
  \BibitemOpen
  \bibfield  {author} {\bibinfo {author} {\bibfnamefont {V.~A.}\ \bibnamefont
  {Neufeld}}\ and\ \bibinfo {author} {\bibfnamefont {T.~C.}\ \bibnamefont
  {Berkelbach}},\ }\href {\doibase 10.1103/PhysRevLett.131.186402} {\bibfield
  {journal} {\bibinfo  {journal} {Phys. Rev. Lett.}\ }\textbf {\bibinfo
  {volume} {131}},\ \bibinfo {pages} {186402} (\bibinfo {year}
  {2023})}\BibitemShut {NoStop}%
\bibitem [{\citenamefont {Masios}\ \emph {et~al.}(2023)\citenamefont {Masios},
  \citenamefont {Irmler}, \citenamefont
  {Sch{\ifmmode\ddot{a}\else\"{a}\fi}fer},\ and\ \citenamefont
  {Gr{\ifmmode\ddot{u}\else\"{u}\fi}neis}}]{Masios2023Oct}%
  \BibitemOpen
  \bibfield  {author} {\bibinfo {author} {\bibfnamefont {N.}~\bibnamefont
  {Masios}}, \bibinfo {author} {\bibfnamefont {A.}~\bibnamefont {Irmler}},
  \bibinfo {author} {\bibfnamefont {T.}~\bibnamefont
  {Sch{\ifmmode\ddot{a}\else\"{a}\fi}fer}}, \ and\ \bibinfo {author}
  {\bibfnamefont {A.}~\bibnamefont {Gr{\ifmmode\ddot{u}\else\"{u}\fi}neis}},\
  }\href {\doibase 10.1103/PhysRevLett.131.186401} {\bibfield  {journal}
  {\bibinfo  {journal} {Phys. Rev. Lett.}\ }\textbf {\bibinfo {volume} {131}},\
  \bibinfo {pages} {186401} (\bibinfo {year} {2023})}\BibitemShut {NoStop}%
\bibitem [{\citenamefont {Zhang}\ and\ \citenamefont
  {Grüneis}(2019)}]{zhang_coupled_2019}%
  \BibitemOpen
  \bibfield  {author} {\bibinfo {author} {\bibfnamefont {I.~Y.}\ \bibnamefont
  {Zhang}}\ and\ \bibinfo {author} {\bibfnamefont {A.}~\bibnamefont
  {Grüneis}},\ }\href
  {https://www.frontiersin.org/articles/10.3389/fmats.2019.00123} {\bibfield
  {journal} {\bibinfo  {journal} {Front. Mater.}\ }\textbf {\bibinfo {volume}
  {6}} (\bibinfo {year} {2019})}\BibitemShut {NoStop}%
\bibitem [{\citenamefont {Irmler}\ \emph {et~al.}(2023)\citenamefont {Irmler},
  \citenamefont {Kanakagiri}, \citenamefont {Ohlmann}, \citenamefont
  {Solomonik},\ and\ \citenamefont {Grüneis}}]{Irmler2023Jul}%
  \BibitemOpen
  \bibfield  {author} {\bibinfo {author} {\bibfnamefont {A.}~\bibnamefont
  {Irmler}}, \bibinfo {author} {\bibfnamefont {R.}~\bibnamefont {Kanakagiri}},
  \bibinfo {author} {\bibfnamefont {S.~T.}\ \bibnamefont {Ohlmann}}, \bibinfo
  {author} {\bibfnamefont {E.}~\bibnamefont {Solomonik}}, \ and\ \bibinfo
  {author} {\bibfnamefont {A.}~\bibnamefont {Grüneis}},\ }in\ \href {\doibase
  10.1007/978-3-031-39698-4_48} {\emph {\bibinfo {booktitle} {Euro-Par 2023:
  Parallel Processing}}},\ \bibinfo {series and number} {Lecture Notes in
  Computer Science}\ (\bibinfo  {publisher} {Springer Nature Switzerland},\
  \bibinfo {address} {Cham},\ \bibinfo {year} {2023})\ pp.\ \bibinfo {pages}
  {710--724}\BibitemShut {NoStop}%
\bibitem [{\citenamefont {Solomonik}\ \emph {et~al.}(2013)\citenamefont
  {Solomonik}, \citenamefont {Matthews}, \citenamefont {Hammond},\ and\
  \citenamefont {Demmel}}]{Solomonik}%
  \BibitemOpen
  \bibfield  {author} {\bibinfo {author} {\bibfnamefont {E.}~\bibnamefont
  {Solomonik}}, \bibinfo {author} {\bibfnamefont {D.}~\bibnamefont {Matthews}},
  \bibinfo {author} {\bibfnamefont {J.}~\bibnamefont {Hammond}}, \ and\
  \bibinfo {author} {\bibfnamefont {J.}~\bibnamefont {Demmel}},\ }in\ \href
  {\doibase 10.1109/IPDPS.2013.112} {\emph {\bibinfo {booktitle} {2013 IEEE
  27th International Symposium on Parallel and Distributed Processing}}}\
  (\bibinfo  {publisher} {IEEE},\ \bibinfo {year} {2013})\ pp.\ \bibinfo
  {pages} {20--24}\BibitemShut {NoStop}%
\bibitem [{\citenamefont {Bohm}\ and\ \citenamefont
  {Pines}(1951)}]{Bohm_Pines_1951}%
  \BibitemOpen
  \bibfield  {author} {\bibinfo {author} {\bibfnamefont {D.}~\bibnamefont
  {Bohm}}\ and\ \bibinfo {author} {\bibfnamefont {D.}~\bibnamefont {Pines}},\
  }\href {\doibase 10.1103/PhysRev.82.625} {\bibfield  {journal} {\bibinfo
  {journal} {Phys. Rev.}\ }\textbf {\bibinfo {volume} {82}},\ \bibinfo {pages}
  {625} (\bibinfo {year} {1951})}\BibitemShut {NoStop}%
\bibitem [{\citenamefont {Pines}\ and\ \citenamefont
  {Bohm}(1952)}]{Bohm_Pines_1952}%
  \BibitemOpen
  \bibfield  {author} {\bibinfo {author} {\bibfnamefont {D.}~\bibnamefont
  {Pines}}\ and\ \bibinfo {author} {\bibfnamefont {D.}~\bibnamefont {Bohm}},\
  }\href {\doibase 10.1103/PhysRev.85.338} {\bibfield  {journal} {\bibinfo
  {journal} {Phys. Rev.}\ }\textbf {\bibinfo {volume} {85}},\ \bibinfo {pages}
  {338} (\bibinfo {year} {1952})}\BibitemShut {NoStop}%
\bibitem [{\citenamefont {Bohm}\ and\ \citenamefont
  {Pines}(1953)}]{Bohm_Pines_1953}%
  \BibitemOpen
  \bibfield  {author} {\bibinfo {author} {\bibfnamefont {D.}~\bibnamefont
  {Bohm}}\ and\ \bibinfo {author} {\bibfnamefont {D.}~\bibnamefont {Pines}},\
  }\href {\doibase 10.1103/PhysRev.92.609} {\bibfield  {journal} {\bibinfo
  {journal} {Phys. Rev.}\ }\textbf {\bibinfo {volume} {92}},\ \bibinfo {pages}
  {609} (\bibinfo {year} {1953})}\BibitemShut {NoStop}%
\bibitem [{\citenamefont {Langreth}\ and\ \citenamefont
  {Perdew}(1975)}]{Langreth_1975}%
  \BibitemOpen
  \bibfield  {author} {\bibinfo {author} {\bibfnamefont {D.}~\bibnamefont
  {Langreth}}\ and\ \bibinfo {author} {\bibfnamefont {J.}~\bibnamefont
  {Perdew}},\ }\href {\doibase 10.1016/0038-1098(75)90618-3} {\bibfield
  {journal} {\bibinfo  {journal} {Solid State Commun.}\ }\textbf {\bibinfo
  {volume} {17}},\ \bibinfo {pages} {1425} (\bibinfo {year}
  {1975})}\BibitemShut {NoStop}%
\bibitem [{\citenamefont {Langreth}\ and\ \citenamefont
  {Perdew}(1977)}]{Langreth_1977}%
  \BibitemOpen
  \bibfield  {author} {\bibinfo {author} {\bibfnamefont {D.~C.}\ \bibnamefont
  {Langreth}}\ and\ \bibinfo {author} {\bibfnamefont {J.~P.}\ \bibnamefont
  {Perdew}},\ }\href {\doibase 10.1103/PhysRevB.15.2884} {\bibfield  {journal}
  {\bibinfo  {journal} {Phys. Rev. B}\ }\textbf {\bibinfo {volume} {15}},\
  \bibinfo {pages} {2884} (\bibinfo {year} {1977})}\BibitemShut {NoStop}%
\bibitem [{\citenamefont {Gunnarsson}\ and\ \citenamefont
  {Lundqvist}(1976)}]{Gunnarsson_1976}%
  \BibitemOpen
  \bibfield  {author} {\bibinfo {author} {\bibfnamefont {O.}~\bibnamefont
  {Gunnarsson}}\ and\ \bibinfo {author} {\bibfnamefont {B.~I.}\ \bibnamefont
  {Lundqvist}},\ }\href {\doibase 10.1103/PhysRevB.13.4274} {\bibfield
  {journal} {\bibinfo  {journal} {Phys. Rev. B}\ }\textbf {\bibinfo {volume}
  {13}},\ \bibinfo {pages} {4274} (\bibinfo {year} {1976})}\BibitemShut
  {NoStop}%
\bibitem [{\citenamefont
  {Furche}(2001{\natexlab{a}})}]{Furche_DensityMatrixBased2001}%
  \BibitemOpen
  \bibfield  {author} {\bibinfo {author} {\bibfnamefont {F.}~\bibnamefont
  {Furche}},\ }\href {\doibase 10.1063/1.1353585} {\bibfield  {journal}
  {\bibinfo  {journal} {J. Chem. Phys.}\ }\textbf {\bibinfo {volume} {114}},\
  \bibinfo {pages} {5982} (\bibinfo {year} {2001}{\natexlab{a}})}\BibitemShut
  {NoStop}%
\bibitem [{\citenamefont
  {Furche}(2001{\natexlab{b}})}]{Furche_MolecularTestsRandom2001}%
  \BibitemOpen
  \bibfield  {author} {\bibinfo {author} {\bibfnamefont {F.}~\bibnamefont
  {Furche}},\ }\href {\doibase 10.1103/PhysRevB.64.195120} {\bibfield
  {journal} {\bibinfo  {journal} {Phys. Rev. B}\ }\textbf {\bibinfo {volume}
  {64}},\ \bibinfo {pages} {195120} (\bibinfo {year}
  {2001}{\natexlab{b}})}\BibitemShut {NoStop}%
\bibitem [{\citenamefont {Fuchs}\ and\ \citenamefont
  {Gonze}(2002)}]{Fuchs_AccurateDensityFunctionals2002}%
  \BibitemOpen
  \bibfield  {author} {\bibinfo {author} {\bibfnamefont {M.}~\bibnamefont
  {Fuchs}}\ and\ \bibinfo {author} {\bibfnamefont {X.}~\bibnamefont {Gonze}},\
  }\href {\doibase 10.1103/PhysRevB.65.235109} {\bibfield  {journal} {\bibinfo
  {journal} {Phys. Rev. B}\ }\textbf {\bibinfo {volume} {65}},\ \bibinfo
  {pages} {235109} (\bibinfo {year} {2002})}\BibitemShut {NoStop}%
\bibitem [{\citenamefont {Furche}\ and\ \citenamefont
  {Van~Voorhis}(2005)}]{Furche_FluctuationdissipationTheoremDensityfunctional2005}%
  \BibitemOpen
  \bibfield  {author} {\bibinfo {author} {\bibfnamefont {F.}~\bibnamefont
  {Furche}}\ and\ \bibinfo {author} {\bibfnamefont {T.}~\bibnamefont
  {Van~Voorhis}},\ }\href {\doibase 10.1063/1.1884112} {\bibfield  {journal}
  {\bibinfo  {journal} {J. Chem. Phys.}\ }\textbf {\bibinfo {volume} {122}},\
  \bibinfo {pages} {164106} (\bibinfo {year} {2005})}\BibitemShut {NoStop}%
\bibitem [{\citenamefont {Furche}(2008)}]{Furche_DevelopingRandomPhase2008}%
  \BibitemOpen
  \bibfield  {author} {\bibinfo {author} {\bibfnamefont {F.}~\bibnamefont
  {Furche}},\ }\href {\doibase 10.1063/1.2977789} {\bibfield  {journal}
  {\bibinfo  {journal} {J. Chem. Phys.}\ }\textbf {\bibinfo {volume} {129}},\
  \bibinfo {pages} {114105} (\bibinfo {year} {2008})}\BibitemShut {NoStop}%
\bibitem [{\citenamefont {Eshuis}\ \emph {et~al.}(2010)\citenamefont {Eshuis},
  \citenamefont {Yarkony},\ and\ \citenamefont
  {Furche}}]{Eshuis_FastComputationMolecular2010}%
  \BibitemOpen
  \bibfield  {author} {\bibinfo {author} {\bibfnamefont {H.}~\bibnamefont
  {Eshuis}}, \bibinfo {author} {\bibfnamefont {J.}~\bibnamefont {Yarkony}}, \
  and\ \bibinfo {author} {\bibfnamefont {F.}~\bibnamefont {Furche}},\ }\href
  {\doibase 10.1063/1.3442749} {\bibfield  {journal} {\bibinfo  {journal} {J.
  Chem. Phys.}\ }\textbf {\bibinfo {volume} {132}},\ \bibinfo {pages} {234114}
  (\bibinfo {year} {2010})}\BibitemShut {NoStop}%
\bibitem [{\citenamefont {Scuseria}\ \emph {et~al.}(2008)\citenamefont
  {Scuseria}, \citenamefont {Henderson},\ and\ \citenamefont
  {Sorensen}}]{Scuseria2008Dec}%
  \BibitemOpen
  \bibfield  {author} {\bibinfo {author} {\bibfnamefont {G.~E.}\ \bibnamefont
  {Scuseria}}, \bibinfo {author} {\bibfnamefont {T.~M.}\ \bibnamefont
  {Henderson}}, \ and\ \bibinfo {author} {\bibfnamefont {D.~C.}\ \bibnamefont
  {Sorensen}},\ }\href {\doibase 10.1063/1.3043729} {\bibfield  {journal}
  {\bibinfo  {journal} {J. Chem. Phys.}\ }\textbf {\bibinfo {volume} {129}},\
  \bibinfo {pages} {231101} (\bibinfo {year} {2008})}\BibitemShut {NoStop}%
\bibitem [{\citenamefont {Eshuis}\ \emph {et~al.}(2012)\citenamefont {Eshuis},
  \citenamefont {Bates},\ and\ \citenamefont
  {Furche}}]{Eshuis_ElectronCorrelationMethods2012}%
  \BibitemOpen
  \bibfield  {author} {\bibinfo {author} {\bibfnamefont {H.}~\bibnamefont
  {Eshuis}}, \bibinfo {author} {\bibfnamefont {J.~E.}\ \bibnamefont {Bates}}, \
  and\ \bibinfo {author} {\bibfnamefont {F.}~\bibnamefont {Furche}},\ }\href
  {\doibase 10.1007/s00214-011-1084-8} {\bibfield  {journal} {\bibinfo
  {journal} {Theor. Chem. Acc.}\ }\textbf {\bibinfo {volume} {131}},\ \bibinfo
  {pages} {1084} (\bibinfo {year} {2012})}\BibitemShut {NoStop}%
\bibitem [{\citenamefont {Nguyen}\ \emph {et~al.}(2020)\citenamefont {Nguyen},
  \citenamefont {Chen}, \citenamefont {Agee}, \citenamefont {Burow},
  \citenamefont {Tang},\ and\ \citenamefont
  {Furche}}]{Nguyen_DivergenceManyBodyPerturbation2020}%
  \BibitemOpen
  \bibfield  {author} {\bibinfo {author} {\bibfnamefont {B.~D.}\ \bibnamefont
  {Nguyen}}, \bibinfo {author} {\bibfnamefont {G.~P.}\ \bibnamefont {Chen}},
  \bibinfo {author} {\bibfnamefont {M.~M.}\ \bibnamefont {Agee}}, \bibinfo
  {author} {\bibfnamefont {A.~M.}\ \bibnamefont {Burow}}, \bibinfo {author}
  {\bibfnamefont {M.~P.}\ \bibnamefont {Tang}}, \ and\ \bibinfo {author}
  {\bibfnamefont {F.}~\bibnamefont {Furche}},\ }\href {\doibase
  10.1021/acs.jctc.9b01176} {\bibfield  {journal} {\bibinfo  {journal} {J.
  Chem. Theory Comput.}\ }\textbf {\bibinfo {volume} {16}},\ \bibinfo {pages}
  {2258} (\bibinfo {year} {2020})}\BibitemShut {NoStop}%
\bibitem [{\citenamefont {Chen}\ \emph {et~al.}(2017)\citenamefont {Chen},
  \citenamefont {Voora}, \citenamefont {Agee}, \citenamefont {Balasubramani},\
  and\ \citenamefont {Furche}}]{Chen_RPA2017}%
  \BibitemOpen
  \bibfield  {author} {\bibinfo {author} {\bibfnamefont {G.~P.}\ \bibnamefont
  {Chen}}, \bibinfo {author} {\bibfnamefont {V.~K.}\ \bibnamefont {Voora}},
  \bibinfo {author} {\bibfnamefont {M.~M.}\ \bibnamefont {Agee}}, \bibinfo
  {author} {\bibfnamefont {S.~G.}\ \bibnamefont {Balasubramani}}, \ and\
  \bibinfo {author} {\bibfnamefont {F.}~\bibnamefont {Furche}},\ }\href
  {\doibase 10.1146/annurev-physchem-040215-112308} {\bibfield  {journal}
  {\bibinfo  {journal} {Annu. Rev. Phys. Chem.}\ }\textbf {\bibinfo {volume}
  {68}},\ \bibinfo {pages} {421} (\bibinfo {year} {2017})}\BibitemShut
  {NoStop}%
\bibitem [{\citenamefont {Ren}(2019)}]{Ren_RandomPhaseApproximation2019}%
  \BibitemOpen
  \bibfield  {author} {\bibinfo {author} {\bibfnamefont {X.}~\bibnamefont
  {Ren}},\ }in\ \href@noop {} {\emph {\bibinfo {booktitle} {Many-{{Body
  Methods}} for {{Real Materials Modeling}} and {{Simulation}}}}},\
  Vol.~\bibinfo {volume} {9}\ (\bibinfo  {publisher} {{Forschungszentrum
  Julich}},\ \bibinfo {year} {2019})\BibitemShut {NoStop}%
\bibitem [{\citenamefont {Marini}\ \emph {et~al.}(2006)\citenamefont {Marini},
  \citenamefont {Garc\'{\i}a-Gonz\'alez},\ and\ \citenamefont
  {Rubio}}]{Marini_2006}%
  \BibitemOpen
  \bibfield  {author} {\bibinfo {author} {\bibfnamefont {A.}~\bibnamefont
  {Marini}}, \bibinfo {author} {\bibfnamefont {P.}~\bibnamefont
  {Garc\'{\i}a-Gonz\'alez}}, \ and\ \bibinfo {author} {\bibfnamefont
  {A.}~\bibnamefont {Rubio}},\ }\href {\doibase 10.1103/PhysRevLett.96.136404}
  {\bibfield  {journal} {\bibinfo  {journal} {Phys. Rev. Lett.}\ }\textbf
  {\bibinfo {volume} {96}},\ \bibinfo {pages} {136404} (\bibinfo {year}
  {2006})}\BibitemShut {NoStop}%
\bibitem [{\citenamefont {Lu}\ \emph {et~al.}(2009)\citenamefont {Lu},
  \citenamefont {Li}, \citenamefont {Rocca},\ and\ \citenamefont
  {Galli}}]{Deyu_2009}%
  \BibitemOpen
  \bibfield  {author} {\bibinfo {author} {\bibfnamefont {D.}~\bibnamefont
  {Lu}}, \bibinfo {author} {\bibfnamefont {Y.}~\bibnamefont {Li}}, \bibinfo
  {author} {\bibfnamefont {D.}~\bibnamefont {Rocca}}, \ and\ \bibinfo {author}
  {\bibfnamefont {G.}~\bibnamefont {Galli}},\ }\href {\doibase
  10.1103/PhysRevLett.102.206411} {\bibfield  {journal} {\bibinfo  {journal}
  {Phys. Rev. Lett.}\ }\textbf {\bibinfo {volume} {102}},\ \bibinfo {pages}
  {206411} (\bibinfo {year} {2009})}\BibitemShut {NoStop}%
\bibitem [{\citenamefont {Nguyen}\ and\ \citenamefont
  {de~Gironcoli}(2009)}]{Nguyen_2009}%
  \BibitemOpen
  \bibfield  {author} {\bibinfo {author} {\bibfnamefont {H.-V.}\ \bibnamefont
  {Nguyen}}\ and\ \bibinfo {author} {\bibfnamefont {S.}~\bibnamefont
  {de~Gironcoli}},\ }\href {\doibase 10.1103/PhysRevB.79.205114} {\bibfield
  {journal} {\bibinfo  {journal} {Phys. Rev. B}\ }\textbf {\bibinfo {volume}
  {79}},\ \bibinfo {pages} {205114} (\bibinfo {year} {2009})}\BibitemShut
  {NoStop}%
\bibitem [{\citenamefont {Harl}\ and\ \citenamefont
  {Kresse}(2008)}]{Harl_CohesiveEnergyCurves2008}%
  \BibitemOpen
  \bibfield  {author} {\bibinfo {author} {\bibfnamefont {J.}~\bibnamefont
  {Harl}}\ and\ \bibinfo {author} {\bibfnamefont {G.}~\bibnamefont {Kresse}},\
  }\href {\doibase 10.1103/PhysRevB.77.045136} {\bibfield  {journal} {\bibinfo
  {journal} {Phys. Rev. B}\ }\textbf {\bibinfo {volume} {77}},\ \bibinfo
  {pages} {045136} (\bibinfo {year} {2008})}\BibitemShut {NoStop}%
\bibitem [{\citenamefont {Harl}\ and\ \citenamefont
  {Kresse}(2009)}]{Harl_AccurateBulkProperties2009}%
  \BibitemOpen
  \bibfield  {author} {\bibinfo {author} {\bibfnamefont {J.}~\bibnamefont
  {Harl}}\ and\ \bibinfo {author} {\bibfnamefont {G.}~\bibnamefont {Kresse}},\
  }\href {\doibase 10.1103/PhysRevLett.103.056401} {\bibfield  {journal}
  {\bibinfo  {journal} {Phys. Rev. Lett.}\ }\textbf {\bibinfo {volume} {103}},\
  \bibinfo {pages} {056401} (\bibinfo {year} {2009})}\BibitemShut {NoStop}%
\bibitem [{\citenamefont {Harl}\ \emph {et~al.}(2010)\citenamefont {Harl},
  \citenamefont {Schimka},\ and\ \citenamefont
  {Kresse}}]{Harl_AssessingQualityRandom2010}%
  \BibitemOpen
  \bibfield  {author} {\bibinfo {author} {\bibfnamefont {J.}~\bibnamefont
  {Harl}}, \bibinfo {author} {\bibfnamefont {L.}~\bibnamefont {Schimka}}, \
  and\ \bibinfo {author} {\bibfnamefont {G.}~\bibnamefont {Kresse}},\ }\href
  {\doibase 10.1103/PhysRevB.81.115126} {\bibfield  {journal} {\bibinfo
  {journal} {Phys. Rev. B}\ }\textbf {\bibinfo {volume} {81}},\ \bibinfo
  {pages} {115126} (\bibinfo {year} {2010})}\BibitemShut {NoStop}%
\bibitem [{\citenamefont {Olsen}\ and\ \citenamefont
  {Thygesen}(2013)}]{Olsen_RandomPhaseApproximation2013}%
  \BibitemOpen
  \bibfield  {author} {\bibinfo {author} {\bibfnamefont {T.}~\bibnamefont
  {Olsen}}\ and\ \bibinfo {author} {\bibfnamefont {K.~S.}\ \bibnamefont
  {Thygesen}},\ }\href {\doibase 10.1103/PhysRevB.87.075111} {\bibfield
  {journal} {\bibinfo  {journal} {Phys. Rev. B}\ }\textbf {\bibinfo {volume}
  {87}},\ \bibinfo {pages} {075111} (\bibinfo {year} {2013})}\BibitemShut
  {NoStop}%
\bibitem [{\citenamefont {Del~Ben}\ \emph
  {et~al.}(2013{\natexlab{b}})\citenamefont {Del~Ben}, \citenamefont {Hutter},\
  and\ \citenamefont {VandeVondele}}]{DelBen_ElectronCorrelationCondensed2013}%
  \BibitemOpen
  \bibfield  {author} {\bibinfo {author} {\bibfnamefont {M.}~\bibnamefont
  {Del~Ben}}, \bibinfo {author} {\bibfnamefont {J.}~\bibnamefont {Hutter}}, \
  and\ \bibinfo {author} {\bibfnamefont {J.}~\bibnamefont {VandeVondele}},\
  }\href {\doibase 10.1021/ct4002202} {\bibfield  {journal} {\bibinfo
  {journal} {J. Chem. Theory Comput.}\ }\textbf {\bibinfo {volume} {9}},\
  \bibinfo {pages} {2654} (\bibinfo {year} {2013}{\natexlab{b}})}\BibitemShut
  {NoStop}%
\bibitem [{\citenamefont {Yeh}\ and\ \citenamefont
  {Morales}(2023{\natexlab{b}})}]{Yeh_LowScalingAlgorithmRandom2023}%
  \BibitemOpen
  \bibfield  {author} {\bibinfo {author} {\bibfnamefont {C.-N.}\ \bibnamefont
  {Yeh}}\ and\ \bibinfo {author} {\bibfnamefont {M.~A.}\ \bibnamefont
  {Morales}},\ }\href {\doibase 10.1021/acs.jctc.3c00615} {\bibfield  {journal}
  {\bibinfo  {journal} {J. Chem. Theory Comput.}\ }\textbf {\bibinfo {volume}
  {19}},\ \bibinfo {pages} {6197} (\bibinfo {year}
  {2023}{\natexlab{b}})}\BibitemShut {NoStop}%
\bibitem [{\citenamefont {Kaltak}\ \emph
  {et~al.}(2014{\natexlab{a}})\citenamefont {Kaltak}, \citenamefont {Klime{\v
  s}},\ and\ \citenamefont {Kresse}}]{Kaltak_CubicScalingAlgorithm2014}%
  \BibitemOpen
  \bibfield  {author} {\bibinfo {author} {\bibfnamefont {M.}~\bibnamefont
  {Kaltak}}, \bibinfo {author} {\bibfnamefont {J.}~\bibnamefont {Klime{\v s}}},
  \ and\ \bibinfo {author} {\bibfnamefont {G.}~\bibnamefont {Kresse}},\ }\href
  {\doibase 10.1103/PhysRevB.90.054115} {\bibfield  {journal} {\bibinfo
  {journal} {Phys. Rev. B}\ }\textbf {\bibinfo {volume} {90}},\ \bibinfo
  {pages} {054115} (\bibinfo {year} {2014}{\natexlab{a}})}\BibitemShut
  {NoStop}%
\bibitem [{\citenamefont {Kaltak}\ \emph
  {et~al.}(2014{\natexlab{b}})\citenamefont {Kaltak}, \citenamefont {Klime{\v
  s}},\ and\ \citenamefont {Kresse}}]{Kaltak_LowScalingAlgorithms2014}%
  \BibitemOpen
  \bibfield  {author} {\bibinfo {author} {\bibfnamefont {M.}~\bibnamefont
  {Kaltak}}, \bibinfo {author} {\bibfnamefont {J.}~\bibnamefont {Klime{\v s}}},
  \ and\ \bibinfo {author} {\bibfnamefont {G.}~\bibnamefont {Kresse}},\ }\href
  {\doibase 10.1021/ct5001268} {\bibfield  {journal} {\bibinfo  {journal} {J.
  Chem. Theory Comput.}\ }\textbf {\bibinfo {volume} {10}},\ \bibinfo {pages}
  {2498} (\bibinfo {year} {2014}{\natexlab{b}})}\BibitemShut {NoStop}%
\bibitem [{\citenamefont
  {Casida}(1995)}]{Casida_TimeDependentDensityFunctional1995}%
  \BibitemOpen
  \bibfield  {author} {\bibinfo {author} {\bibfnamefont {M.~E.}\ \bibnamefont
  {Casida}},\ }\enquote {\bibinfo {title} {Time-dependent density functional
  response theory for molecules},}\ in\ \href {\doibase
  10.1142/9789812830586_0005} {\emph {\bibinfo {booktitle} {Recent Advances in
  Density Functional Methods}}}\ (\bibinfo  {publisher} {World Scientific},\
  \bibinfo {year} {1995})\ pp.\ \bibinfo {pages} {155--192}\BibitemShut
  {NoStop}%
\bibitem [{\citenamefont {Hirata}\ \emph {et~al.}(1999)\citenamefont {Hirata},
  \citenamefont {{Head-Gordon}},\ and\ \citenamefont
  {Bartlett}}]{Hirata_ConfigurationInteractionSingles1999}%
  \BibitemOpen
  \bibfield  {author} {\bibinfo {author} {\bibfnamefont {S.}~\bibnamefont
  {Hirata}}, \bibinfo {author} {\bibfnamefont {M.}~\bibnamefont
  {{Head-Gordon}}}, \ and\ \bibinfo {author} {\bibfnamefont {R.~J.}\
  \bibnamefont {Bartlett}},\ }\href {\doibase 10.1063/1.480443} {\bibfield
  {journal} {\bibinfo  {journal} {J. Chem. Phys.}\ }\textbf {\bibinfo {volume}
  {111}},\ \bibinfo {pages} {10774} (\bibinfo {year} {1999})}\BibitemShut
  {NoStop}%
\bibitem [{\citenamefont {Dreuw}\ and\ \citenamefont
  {{Head-Gordon}}(2005)}]{Dreuw_SingleReferenceInitioMethods2005a}%
  \BibitemOpen
  \bibfield  {author} {\bibinfo {author} {\bibfnamefont {A.}~\bibnamefont
  {Dreuw}}\ and\ \bibinfo {author} {\bibfnamefont {M.}~\bibnamefont
  {{Head-Gordon}}},\ }\href {\doibase 10.1021/cr0505627} {\bibfield  {journal}
  {\bibinfo  {journal} {Chem. Rev.}\ }\textbf {\bibinfo {volume} {105}},\
  \bibinfo {pages} {4009} (\bibinfo {year} {2005})}\BibitemShut {NoStop}%
\bibitem [{\citenamefont {Grundei}\ and\ \citenamefont
  {Burow}(2017{\natexlab{b}})}]{Grundei_RandomPhaseApproximation2017}%
  \BibitemOpen
  \bibfield  {author} {\bibinfo {author} {\bibfnamefont {M.~M.~J.}\
  \bibnamefont {Grundei}}\ and\ \bibinfo {author} {\bibfnamefont {A.~M.}\
  \bibnamefont {Burow}},\ }\href {\doibase 10.1021/acs.jctc.6b01146} {\bibfield
   {journal} {\bibinfo  {journal} {J. Chem. Theory Comput.}\ }\textbf {\bibinfo
  {volume} {13}},\ \bibinfo {pages} {1159} (\bibinfo {year}
  {2017}{\natexlab{b}})}\BibitemShut {NoStop}%
\bibitem [{\citenamefont {Wilhelm}\ \emph {et~al.}(2016)\citenamefont
  {Wilhelm}, \citenamefont {Seewald}, \citenamefont {Del~Ben},\ and\
  \citenamefont {Hutter}}]{Wilhelm_LargeScaleCubicScalingRandom2016a}%
  \BibitemOpen
  \bibfield  {author} {\bibinfo {author} {\bibfnamefont {J.}~\bibnamefont
  {Wilhelm}}, \bibinfo {author} {\bibfnamefont {P.}~\bibnamefont {Seewald}},
  \bibinfo {author} {\bibfnamefont {M.}~\bibnamefont {Del~Ben}}, \ and\
  \bibinfo {author} {\bibfnamefont {J.}~\bibnamefont {Hutter}},\ }\href
  {\doibase 10.1021/acs.jctc.6b00840} {\bibfield  {journal} {\bibinfo
  {journal} {J. Chem. Theory Comput.}\ }\textbf {\bibinfo {volume} {12}},\
  \bibinfo {pages} {5851} (\bibinfo {year} {2016})}\BibitemShut {NoStop}%
\bibitem [{\citenamefont {Blankenbecler}\ \emph {et~al.}(1981)\citenamefont
  {Blankenbecler}, \citenamefont {Scalapino},\ and\ \citenamefont
  {Sugar}}]{Blankenbecler1981Oct}%
  \BibitemOpen
  \bibfield  {author} {\bibinfo {author} {\bibfnamefont {R.}~\bibnamefont
  {Blankenbecler}}, \bibinfo {author} {\bibfnamefont {D.~J.}\ \bibnamefont
  {Scalapino}}, \ and\ \bibinfo {author} {\bibfnamefont {R.~L.}\ \bibnamefont
  {Sugar}},\ }\href {\doibase 10.1103/PhysRevD.24.2278} {\bibfield  {journal}
  {\bibinfo  {journal} {Phys. Rev. D}\ }\textbf {\bibinfo {volume} {24}},\
  \bibinfo {pages} {2278} (\bibinfo {year} {1981})}\BibitemShut {NoStop}%
\bibitem [{\citenamefont {Sorella}\ and\ \citenamefont
  {Capriotti}(2000)}]{Sorella2000Jan}%
  \BibitemOpen
  \bibfield  {author} {\bibinfo {author} {\bibfnamefont {S.}~\bibnamefont
  {Sorella}}\ and\ \bibinfo {author} {\bibfnamefont {L.}~\bibnamefont
  {Capriotti}},\ }\href {\doibase 10.1103/PhysRevB.61.2599} {\bibfield
  {journal} {\bibinfo  {journal} {Phys. Rev. B}\ }\textbf {\bibinfo {volume}
  {61}},\ \bibinfo {pages} {2599} (\bibinfo {year} {2000})}\BibitemShut
  {NoStop}%
\bibitem [{\citenamefont {Booth}\ \emph {et~al.}(2009)\citenamefont {Booth},
  \citenamefont {Thom},\ and\ \citenamefont {Alavi}}]{Booth2009Aug}%
  \BibitemOpen
  \bibfield  {author} {\bibinfo {author} {\bibfnamefont {G.~H.}\ \bibnamefont
  {Booth}}, \bibinfo {author} {\bibfnamefont {A.~J.~W.}\ \bibnamefont {Thom}},
  \ and\ \bibinfo {author} {\bibfnamefont {A.}~\bibnamefont {Alavi}},\ }\href
  {\doibase 10.1063/1.3193710} {\bibfield  {journal} {\bibinfo  {journal} {J.
  Chem. Phys.}\ }\textbf {\bibinfo {volume} {131}},\ \bibinfo {pages} {054106}
  (\bibinfo {year} {2009})}\BibitemShut {NoStop}%
\bibitem [{\citenamefont {Motta}\ and\ \citenamefont
  {Zhang}(2018)}]{Motta2018Sep}%
  \BibitemOpen
  \bibfield  {author} {\bibinfo {author} {\bibfnamefont {M.}~\bibnamefont
  {Motta}}\ and\ \bibinfo {author} {\bibfnamefont {S.}~\bibnamefont {Zhang}},\
  }\href {\doibase 10.1002/wcms.1364} {\bibfield  {journal} {\bibinfo
  {journal} {WIREs Comput. Mol. Sci.}\ }\textbf {\bibinfo {volume} {8}},\
  \bibinfo {pages} {e1364} (\bibinfo {year} {2018})}\BibitemShut {NoStop}%
\bibitem [{\citenamefont {Lee}\ \emph {et~al.}(2022{\natexlab{b}})\citenamefont
  {Lee}, \citenamefont {Pham},\ and\ \citenamefont {Reichman}}]{Lee2022AFQMC}%
  \BibitemOpen
  \bibfield  {author} {\bibinfo {author} {\bibfnamefont {J.}~\bibnamefont
  {Lee}}, \bibinfo {author} {\bibfnamefont {H.~Q.}\ \bibnamefont {Pham}}, \
  and\ \bibinfo {author} {\bibfnamefont {D.~R.}\ \bibnamefont {Reichman}},\
  }\href {\doibase 10.1021/acs.jctc.2c00802} {\bibfield  {journal} {\bibinfo
  {journal} {J. Chem. Theory Comput.}\ }\textbf {\bibinfo {volume} {18}},\
  \bibinfo {pages} {7024} (\bibinfo {year} {2022}{\natexlab{b}})}\BibitemShut
  {NoStop}%
\bibitem [{\citenamefont {Zhang}\ and\ \citenamefont
  {Krakauer}(2003)}]{Zhang2003Apr}%
  \BibitemOpen
  \bibfield  {author} {\bibinfo {author} {\bibfnamefont {S.}~\bibnamefont
  {Zhang}}\ and\ \bibinfo {author} {\bibfnamefont {H.}~\bibnamefont
  {Krakauer}},\ }\href {\doibase 10.1103/PhysRevLett.90.136401} {\bibfield
  {journal} {\bibinfo  {journal} {Phys. Rev. Lett.}\ }\textbf {\bibinfo
  {volume} {90}},\ \bibinfo {pages} {136401} (\bibinfo {year}
  {2003})}\BibitemShut {NoStop}%
\bibitem [{\citenamefont {Cleland}\ \emph {et~al.}(2010)\citenamefont
  {Cleland}, \citenamefont {Booth},\ and\ \citenamefont
  {Alavi}}]{Cleland2010Jan}%
  \BibitemOpen
  \bibfield  {author} {\bibinfo {author} {\bibfnamefont {D.}~\bibnamefont
  {Cleland}}, \bibinfo {author} {\bibfnamefont {G.~H.}\ \bibnamefont {Booth}},
  \ and\ \bibinfo {author} {\bibfnamefont {A.}~\bibnamefont {Alavi}},\ }\href
  {\doibase 10.1063/1.3302277} {\bibfield  {journal} {\bibinfo  {journal} {J.
  Chem. Phys.}\ }\textbf {\bibinfo {volume} {132}},\ \bibinfo {pages} {041103}
  (\bibinfo {year} {2010})}\BibitemShut {NoStop}%
\bibitem [{\citenamefont {Sukurma}\ \emph {et~al.}(2023)\citenamefont
  {Sukurma}, \citenamefont {Schlipf}, \citenamefont {Humer}, \citenamefont
  {Taheridehkordi},\ and\ \citenamefont {Kresse}}]{Sukurma2023Aug}%
  \BibitemOpen
  \bibfield  {author} {\bibinfo {author} {\bibfnamefont {Z.}~\bibnamefont
  {Sukurma}}, \bibinfo {author} {\bibfnamefont {M.}~\bibnamefont {Schlipf}},
  \bibinfo {author} {\bibfnamefont {M.}~\bibnamefont {Humer}}, \bibinfo
  {author} {\bibfnamefont {A.}~\bibnamefont {Taheridehkordi}}, \ and\ \bibinfo
  {author} {\bibfnamefont {G.}~\bibnamefont {Kresse}},\ }\href {\doibase
  10.1021/acs.jctc.3c00322} {\bibfield  {journal} {\bibinfo  {journal} {J.
  Chem. Theory Comput.}\ }\textbf {\bibinfo {volume} {19}},\ \bibinfo {pages}
  {4921} (\bibinfo {year} {2023})}\BibitemShut {NoStop}%
\bibitem [{\citenamefont {{Simons Collaboration on the Many-Electron Problem}}\
  \emph {et~al.}(2017)\citenamefont {{Simons Collaboration on the Many-Electron
  Problem}}, \citenamefont {Motta}, \citenamefont {Ceperley}, \citenamefont
  {Chan}, \citenamefont {Gomez}, \citenamefont {Gull}, \citenamefont {Guo},
  \citenamefont {Jim{\ifmmode\acute{e}\else\'{e}\fi}nez-Hoyos}, \citenamefont
  {Lan}, \citenamefont {Li}, \citenamefont {Ma}, \citenamefont {Millis},
  \citenamefont {Prokof{'}ev}, \citenamefont {Ray}, \citenamefont {Scuseria},
  \citenamefont {Sorella}, \citenamefont {Stoudenmire}, \citenamefont {Sun},
  \citenamefont {Tupitsyn}, \citenamefont {White}, \citenamefont {Zgid},\ and\
  \citenamefont {Zhang}}]{SimonsCollaborationontheMany-ElectronProblem2017Sep}%
  \BibitemOpen
  \bibfield  {author} {\bibinfo {author} {\bibnamefont {{Simons Collaboration
  on the Many-Electron Problem}}}, \bibinfo {author} {\bibfnamefont
  {M.}~\bibnamefont {Motta}}, \bibinfo {author} {\bibfnamefont {D.~M.}\
  \bibnamefont {Ceperley}}, \bibinfo {author} {\bibfnamefont {G.~K.-L.}\
  \bibnamefont {Chan}}, \bibinfo {author} {\bibfnamefont {J.~A.}\ \bibnamefont
  {Gomez}}, \bibinfo {author} {\bibfnamefont {E.}~\bibnamefont {Gull}},
  \bibinfo {author} {\bibfnamefont {S.}~\bibnamefont {Guo}}, \bibinfo {author}
  {\bibfnamefont {C.~A.}\ \bibnamefont
  {Jim{\ifmmode\acute{e}\else\'{e}\fi}nez-Hoyos}}, \bibinfo {author}
  {\bibfnamefont {T.~N.}\ \bibnamefont {Lan}}, \bibinfo {author} {\bibfnamefont
  {J.}~\bibnamefont {Li}}, \bibinfo {author} {\bibfnamefont {F.}~\bibnamefont
  {Ma}}, \bibinfo {author} {\bibfnamefont {A.~J.}\ \bibnamefont {Millis}},
  \bibinfo {author} {\bibfnamefont {N.~V.}\ \bibnamefont {Prokof{'}ev}},
  \bibinfo {author} {\bibfnamefont {U.}~\bibnamefont {Ray}}, \bibinfo {author}
  {\bibfnamefont {G.~E.}\ \bibnamefont {Scuseria}}, \bibinfo {author}
  {\bibfnamefont {S.}~\bibnamefont {Sorella}}, \bibinfo {author} {\bibfnamefont
  {E.~M.}\ \bibnamefont {Stoudenmire}}, \bibinfo {author} {\bibfnamefont
  {Q.}~\bibnamefont {Sun}}, \bibinfo {author} {\bibfnamefont {I.~S.}\
  \bibnamefont {Tupitsyn}}, \bibinfo {author} {\bibfnamefont {S.~R.}\
  \bibnamefont {White}}, \bibinfo {author} {\bibfnamefont {D.}~\bibnamefont
  {Zgid}}, \ and\ \bibinfo {author} {\bibfnamefont {S.}~\bibnamefont {Zhang}},\
  }\href {\doibase 10.1103/PhysRevX.7.031059} {\bibfield  {journal} {\bibinfo
  {journal} {Phys. Rev. X}\ }\textbf {\bibinfo {volume} {7}},\ \bibinfo {pages}
  {031059} (\bibinfo {year} {2017})}\BibitemShut {NoStop}%
\bibitem [{\citenamefont {Malone}\ \emph
  {et~al.}(2020{\natexlab{b}})\citenamefont {Malone}, \citenamefont {Benali},
  \citenamefont {Morales}, \citenamefont {Caffarel}, \citenamefont {Kent},\
  and\ \citenamefont {Shulenburger}}]{Malone2020Oct}%
  \BibitemOpen
  \bibfield  {author} {\bibinfo {author} {\bibfnamefont {F.~D.}\ \bibnamefont
  {Malone}}, \bibinfo {author} {\bibfnamefont {A.}~\bibnamefont {Benali}},
  \bibinfo {author} {\bibfnamefont {M.~A.}\ \bibnamefont {Morales}}, \bibinfo
  {author} {\bibfnamefont {M.}~\bibnamefont {Caffarel}}, \bibinfo {author}
  {\bibfnamefont {P.~R.~C.}\ \bibnamefont {Kent}}, \ and\ \bibinfo {author}
  {\bibfnamefont {L.}~\bibnamefont {Shulenburger}},\ }\href {\doibase
  10.1103/PhysRevB.102.161104} {\bibfield  {journal} {\bibinfo  {journal}
  {Phys. Rev. B}\ }\textbf {\bibinfo {volume} {102}},\ \bibinfo {pages}
  {161104} (\bibinfo {year} {2020}{\natexlab{b}})}\BibitemShut {NoStop}%
\bibitem [{\citenamefont {Huggins}\ \emph {et~al.}(2022)\citenamefont
  {Huggins}, \citenamefont {O{'}Gorman}, \citenamefont {Rubin}, \citenamefont
  {Reichman}, \citenamefont {Babbush},\ and\ \citenamefont
  {Lee}}]{Huggins2022Mar}%
  \BibitemOpen
  \bibfield  {author} {\bibinfo {author} {\bibfnamefont {W.~J.}\ \bibnamefont
  {Huggins}}, \bibinfo {author} {\bibfnamefont {B.~A.}\ \bibnamefont
  {O{'}Gorman}}, \bibinfo {author} {\bibfnamefont {N.~C.}\ \bibnamefont
  {Rubin}}, \bibinfo {author} {\bibfnamefont {D.~R.}\ \bibnamefont {Reichman}},
  \bibinfo {author} {\bibfnamefont {R.}~\bibnamefont {Babbush}}, \ and\
  \bibinfo {author} {\bibfnamefont {J.}~\bibnamefont {Lee}},\ }\href {\doibase
  10.1038/s41586-021-04351-z} {\bibfield  {journal} {\bibinfo  {journal}
  {Nature}\ }\textbf {\bibinfo {volume} {603}},\ \bibinfo {pages} {416}
  (\bibinfo {year} {2022})}\BibitemShut {NoStop}%
\bibitem [{\citenamefont {Lee}\ \emph {et~al.}(2021{\natexlab{b}})\citenamefont
  {Lee}, \citenamefont {Malone}, \citenamefont {Morales},\ and\ \citenamefont
  {Reichman}}]{Lee2021Jun}%
  \BibitemOpen
  \bibfield  {author} {\bibinfo {author} {\bibfnamefont {J.}~\bibnamefont
  {Lee}}, \bibinfo {author} {\bibfnamefont {F.~D.}\ \bibnamefont {Malone}},
  \bibinfo {author} {\bibfnamefont {M.~A.}\ \bibnamefont {Morales}}, \ and\
  \bibinfo {author} {\bibfnamefont {D.~R.}\ \bibnamefont {Reichman}},\ }\href
  {\doibase 10.1021/acs.jctc.1c00100} {\bibfield  {journal} {\bibinfo
  {journal} {J. Chem. Theory Comput.}\ }\textbf {\bibinfo {volume} {17}},\
  \bibinfo {pages} {3372} (\bibinfo {year} {2021}{\natexlab{b}})}\BibitemShut
  {NoStop}%
\bibitem [{\citenamefont {Mahan}(2000)}]{mahan_many-particle_2000}%
  \BibitemOpen
  \bibfield  {author} {\bibinfo {author} {\bibfnamefont {G.~D.}\ \bibnamefont
  {Mahan}},\ }\href {\doibase 10.1007/978-1-4757-5714-9} {\emph {\bibinfo
  {title} {Many-{Particle} {Physics}}}},\ \bibinfo {edition} {3rd}\ ed.\
  (\bibinfo  {publisher} {Springer US},\ \bibinfo {address} {Boston, MA},\
  \bibinfo {year} {2000})\BibitemShut {NoStop}%
\bibitem [{\citenamefont {Fetter}\ and\ \citenamefont
  {Walecka}(2003)}]{fetter_quantum_2003}%
  \BibitemOpen
  \bibfield  {author} {\bibinfo {author} {\bibfnamefont {A.~L.}\ \bibnamefont
  {Fetter}}\ and\ \bibinfo {author} {\bibfnamefont {J.~D.}\ \bibnamefont
  {Walecka}},\ }\href@noop {} {\emph {\bibinfo {title} {Quantum {Theory} of
  {Many}-{Particle} {Systems}}}}\ (\bibinfo  {publisher} {Dover Publications,
  Inc.},\ \bibinfo {address} {Mineola, NY},\ \bibinfo {year}
  {2003})\BibitemShut {NoStop}%
\bibitem [{\citenamefont {Landau}(1933)}]{landau_electron_1933}%
  \BibitemOpen
  \bibfield  {author} {\bibinfo {author} {\bibfnamefont {L.~D.}\ \bibnamefont
  {Landau}},\ }\href@noop {} {\bibfield  {journal} {\bibinfo  {journal} {Phys.
  Z. Sowjet.}\ ,\ \bibinfo {pages} {664}} (\bibinfo {year} {1933})}\BibitemShut
  {NoStop}%
\bibitem [{\citenamefont {Giustino}(2017)}]{giustino_electron-phonon_2017}%
  \BibitemOpen
  \bibfield  {author} {\bibinfo {author} {\bibfnamefont {F.}~\bibnamefont
  {Giustino}},\ }\href {\doibase 10.1103/RevModPhys.89.015003} {\bibfield
  {journal} {\bibinfo  {journal} {Rev. Mod. Phys.}\ }\textbf {\bibinfo {volume}
  {89}},\ \bibinfo {pages} {015003} (\bibinfo {year} {2017})}\BibitemShut
  {NoStop}%
\bibitem [{\citenamefont {Baroni}\ \emph
  {et~al.}(2001{\natexlab{a}})\citenamefont {Baroni}, \citenamefont
  {de~Gironcoli}, \citenamefont {Dal~Corso},\ and\ \citenamefont
  {Giannozzi}}]{baroni_phonons_2001}%
  \BibitemOpen
  \bibfield  {author} {\bibinfo {author} {\bibfnamefont {S.}~\bibnamefont
  {Baroni}}, \bibinfo {author} {\bibfnamefont {S.}~\bibnamefont
  {de~Gironcoli}}, \bibinfo {author} {\bibfnamefont {A.}~\bibnamefont
  {Dal~Corso}}, \ and\ \bibinfo {author} {\bibfnamefont {P.}~\bibnamefont
  {Giannozzi}},\ }\href {\doibase 10.1103/RevModPhys.73.515} {\bibfield
  {journal} {\bibinfo  {journal} {Rev. Mod. Phys}\ }\textbf {\bibinfo {volume}
  {73}},\ \bibinfo {pages} {515} (\bibinfo {year}
  {2001}{\natexlab{a}})}\BibitemShut {NoStop}%
\bibitem [{\citenamefont {Stoffel}\ \emph {et~al.}(2010)\citenamefont
  {Stoffel}, \citenamefont {Wessel}, \citenamefont {Lumey},\ and\ \citenamefont
  {Dronskowski}}]{stoffel_ab_2010}%
  \BibitemOpen
  \bibfield  {author} {\bibinfo {author} {\bibfnamefont {R.~P.}\ \bibnamefont
  {Stoffel}}, \bibinfo {author} {\bibfnamefont {C.}~\bibnamefont {Wessel}},
  \bibinfo {author} {\bibfnamefont {M.-W.}\ \bibnamefont {Lumey}}, \ and\
  \bibinfo {author} {\bibfnamefont {R.}~\bibnamefont {Dronskowski}},\
  }\href@noop {} {\bibfield  {journal} {\bibinfo  {journal} {Angew. Chem. Int.
  Ed.}\ }\textbf {\bibinfo {volume} {49}},\ \bibinfo {pages} {5242} (\bibinfo
  {year} {2010})}\BibitemShut {NoStop}%
\bibitem [{\citenamefont {Pokluda}\ \emph {et~al.}(2015)\citenamefont
  {Pokluda}, \citenamefont {Cerny}, \citenamefont {Sob},\ and\ \citenamefont
  {Umeno}}]{pokluda_ab_2015}%
  \BibitemOpen
  \bibfield  {author} {\bibinfo {author} {\bibfnamefont {J.}~\bibnamefont
  {Pokluda}}, \bibinfo {author} {\bibfnamefont {M.}~\bibnamefont {Cerny}},
  \bibinfo {author} {\bibfnamefont {M.}~\bibnamefont {Sob}}, \ and\ \bibinfo
  {author} {\bibfnamefont {Y.}~\bibnamefont {Umeno}},\ }\href@noop {}
  {\bibfield  {journal} {\bibinfo  {journal} {Prog. Mater. Sci.}\ }\textbf
  {\bibinfo {volume} {73}},\ \bibinfo {pages} {127} (\bibinfo {year}
  {2015})}\BibitemShut {NoStop}%
\bibitem [{\citenamefont {Lindsay}\ \emph {et~al.}(2018)\citenamefont
  {Lindsay}, \citenamefont {Hua}, \citenamefont {Ruan},\ and\ \citenamefont
  {Lee}}]{lindsay_survey_2018}%
  \BibitemOpen
  \bibfield  {author} {\bibinfo {author} {\bibfnamefont {L.}~\bibnamefont
  {Lindsay}}, \bibinfo {author} {\bibfnamefont {C.}~\bibnamefont {Hua}},
  \bibinfo {author} {\bibfnamefont {X.}~\bibnamefont {Ruan}}, \ and\ \bibinfo
  {author} {\bibfnamefont {S.}~\bibnamefont {Lee}},\ }\href@noop {} {\bibfield
  {journal} {\bibinfo  {journal} {Mater. Today Phys.}\ }\textbf {\bibinfo
  {volume} {7}},\ \bibinfo {pages} {106} (\bibinfo {year} {2018})}\BibitemShut
  {NoStop}%
\bibitem [{\citenamefont {Giustino}\ \emph
  {et~al.}(2007{\natexlab{a}})\citenamefont {Giustino}, \citenamefont {Cohen},\
  and\ \citenamefont {Louie}}]{giustino_electron-phonon_2007}%
  \BibitemOpen
  \bibfield  {author} {\bibinfo {author} {\bibfnamefont {F.}~\bibnamefont
  {Giustino}}, \bibinfo {author} {\bibfnamefont {M.~L.}\ \bibnamefont {Cohen}},
  \ and\ \bibinfo {author} {\bibfnamefont {S.~G.}\ \bibnamefont {Louie}},\
  }\href {\doibase 10.1103/PhysRevB.76.165108} {\bibfield  {journal} {\bibinfo
  {journal} {Phys. Rev. B}\ }\textbf {\bibinfo {volume} {76}},\ \bibinfo
  {pages} {165108} (\bibinfo {year} {2007}{\natexlab{a}})}\BibitemShut
  {NoStop}%
\bibitem [{\citenamefont {Zhou}\ \emph {et~al.}(2021)\citenamefont {Zhou},
  \citenamefont {Park}, \citenamefont {Lu}, \citenamefont {Maliyov},
  \citenamefont {Tong},\ and\ \citenamefont {Bernardi}}]{zhou_perturbo_2021}%
  \BibitemOpen
  \bibfield  {author} {\bibinfo {author} {\bibfnamefont {J.-J.}\ \bibnamefont
  {Zhou}}, \bibinfo {author} {\bibfnamefont {J.}~\bibnamefont {Park}}, \bibinfo
  {author} {\bibfnamefont {I.-T.}\ \bibnamefont {Lu}}, \bibinfo {author}
  {\bibfnamefont {I.}~\bibnamefont {Maliyov}}, \bibinfo {author} {\bibfnamefont
  {X.}~\bibnamefont {Tong}}, \ and\ \bibinfo {author} {\bibfnamefont
  {M.}~\bibnamefont {Bernardi}},\ }\href {\doibase 10.1016/j.cpc.2021.107970}
  {\bibfield  {journal} {\bibinfo  {journal} {Comput. Phys. Commun.}\ }\textbf
  {\bibinfo {volume} {264}},\ \bibinfo {pages} {107970} (\bibinfo {year}
  {2021})}\BibitemShut {NoStop}%
\bibitem [{\citenamefont {Helgaker}\ and\ \citenamefont
  {Jørgensen}(1988)}]{helgaker1988analytical}%
  \BibitemOpen
  \bibfield  {author} {\bibinfo {author} {\bibfnamefont {T.}~\bibnamefont
  {Helgaker}}\ and\ \bibinfo {author} {\bibfnamefont {P.}~\bibnamefont
  {Jørgensen}}\ }(\bibinfo  {publisher} {Academic Press},\ \bibinfo {year}
  {1988})\ pp.\ \bibinfo {pages} {183--245}\BibitemShut {NoStop}%
\bibitem [{\citenamefont {Lloyd-Williams}\ and\ \citenamefont
  {Monserrat}(2015)}]{Lloyd-Williams2015Nov}%
  \BibitemOpen
  \bibfield  {author} {\bibinfo {author} {\bibfnamefont {J.~H.}\ \bibnamefont
  {Lloyd-Williams}}\ and\ \bibinfo {author} {\bibfnamefont {B.}~\bibnamefont
  {Monserrat}},\ }\href {\doibase 10.1103/PhysRevB.92.184301} {\bibfield
  {journal} {\bibinfo  {journal} {Phys. Rev. B}\ }\textbf {\bibinfo {volume}
  {92}},\ \bibinfo {pages} {184301} (\bibinfo {year} {2015})}\BibitemShut
  {NoStop}%
\bibitem [{\citenamefont {Baroni}\ \emph
  {et~al.}(2001{\natexlab{b}})\citenamefont {Baroni}, \citenamefont
  {De~Gironcoli}, \citenamefont {Dal~Corso},\ and\ \citenamefont
  {Giannozzi}}]{baroni2001phonons}%
  \BibitemOpen
  \bibfield  {author} {\bibinfo {author} {\bibfnamefont {S.}~\bibnamefont
  {Baroni}}, \bibinfo {author} {\bibfnamefont {S.}~\bibnamefont
  {De~Gironcoli}}, \bibinfo {author} {\bibfnamefont {A.}~\bibnamefont
  {Dal~Corso}}, \ and\ \bibinfo {author} {\bibfnamefont {P.}~\bibnamefont
  {Giannozzi}},\ }\href@noop {} {\bibfield  {journal} {\bibinfo  {journal}
  {Rev. Mod. Phys}\ }\textbf {\bibinfo {volume} {73}},\ \bibinfo {pages} {515}
  (\bibinfo {year} {2001}{\natexlab{b}})}\BibitemShut {NoStop}%
\bibitem [{\citenamefont {Giustino}\ \emph
  {et~al.}(2007{\natexlab{b}})\citenamefont {Giustino}, \citenamefont {Cohen},\
  and\ \citenamefont {Louie}}]{Giustino2007Oct}%
  \BibitemOpen
  \bibfield  {author} {\bibinfo {author} {\bibfnamefont {F.}~\bibnamefont
  {Giustino}}, \bibinfo {author} {\bibfnamefont {M.~L.}\ \bibnamefont {Cohen}},
  \ and\ \bibinfo {author} {\bibfnamefont {S.~G.}\ \bibnamefont {Louie}},\
  }\href {\doibase 10.1103/PhysRevB.76.165108} {\bibfield  {journal} {\bibinfo
  {journal} {Phys. Rev. B}\ }\textbf {\bibinfo {volume} {76}},\ \bibinfo
  {pages} {165108} (\bibinfo {year} {2007}{\natexlab{b}})}\BibitemShut
  {NoStop}%
\bibitem [{\citenamefont {Subotnik}\ \emph {et~al.}(2005)\citenamefont
  {Subotnik}, \citenamefont {Dutoi},\ and\ \citenamefont
  {Head-Gordon}}]{Subotnik2005Sep}%
  \BibitemOpen
  \bibfield  {author} {\bibinfo {author} {\bibfnamefont {J.~E.}\ \bibnamefont
  {Subotnik}}, \bibinfo {author} {\bibfnamefont {A.~D.}\ \bibnamefont {Dutoi}},
  \ and\ \bibinfo {author} {\bibfnamefont {M.}~\bibnamefont {Head-Gordon}},\
  }\href {\doibase 10.1063/1.2033687} {\bibfield  {journal} {\bibinfo
  {journal} {J. Chem. Phys.}\ }\textbf {\bibinfo {volume} {123}},\ \bibinfo
  {pages} {114108} (\bibinfo {year} {2005})}\BibitemShut {NoStop}%
\bibitem [{\citenamefont {Agapito}\ and\ \citenamefont
  {Bernardi}(2018)}]{agapito_ab_2018}%
  \BibitemOpen
  \bibfield  {author} {\bibinfo {author} {\bibfnamefont {L.~A.}\ \bibnamefont
  {Agapito}}\ and\ \bibinfo {author} {\bibfnamefont {M.}~\bibnamefont
  {Bernardi}},\ }\href {\doibase 10.1103/PhysRevB.97.235146} {\bibfield
  {journal} {\bibinfo  {journal} {Phys. Rev. B}\ }\textbf {\bibinfo {volume}
  {97}},\ \bibinfo {pages} {235146} (\bibinfo {year} {2018})}\BibitemShut
  {NoStop}%
\bibitem [{\citenamefont {Marini}\ \emph {et~al.}(2015)\citenamefont {Marini},
  \citenamefont {Ponc{\ifmmode\acute{e}\else\'{e}\fi}},\ and\ \citenamefont
  {Gonze}}]{Marini2015Jun}%
  \BibitemOpen
  \bibfield  {author} {\bibinfo {author} {\bibfnamefont {A.}~\bibnamefont
  {Marini}}, \bibinfo {author} {\bibfnamefont {S.}~\bibnamefont
  {Ponc{\ifmmode\acute{e}\else\'{e}\fi}}}, \ and\ \bibinfo {author}
  {\bibfnamefont {X.}~\bibnamefont {Gonze}},\ }\href {\doibase
  10.1103/PhysRevB.91.224310} {\bibfield  {journal} {\bibinfo  {journal} {Phys.
  Rev. B}\ }\textbf {\bibinfo {volume} {91}},\ \bibinfo {pages} {224310}
  (\bibinfo {year} {2015})}\BibitemShut {NoStop}%
\bibitem [{\citenamefont {Li}\ \emph {et~al.}(2019)\citenamefont {Li},
  \citenamefont {Antonius}, \citenamefont {Wu}, \citenamefont {da~Jornada},\
  and\ \citenamefont {Louie}}]{Li2019May}%
  \BibitemOpen
  \bibfield  {author} {\bibinfo {author} {\bibfnamefont {Z.}~\bibnamefont
  {Li}}, \bibinfo {author} {\bibfnamefont {G.}~\bibnamefont {Antonius}},
  \bibinfo {author} {\bibfnamefont {M.}~\bibnamefont {Wu}}, \bibinfo {author}
  {\bibfnamefont {F.~H.}\ \bibnamefont {da~Jornada}}, \ and\ \bibinfo {author}
  {\bibfnamefont {S.~G.}\ \bibnamefont {Louie}},\ }\href {\doibase
  10.1103/PhysRevLett.122.186402} {\bibfield  {journal} {\bibinfo  {journal}
  {Phys. Rev. Lett.}\ }\textbf {\bibinfo {volume} {122}},\ \bibinfo {pages}
  {186402} (\bibinfo {year} {2019})}\BibitemShut {NoStop}%
\bibitem [{\citenamefont {Yang}\ \emph {et~al.}(2022)\citenamefont {Yang},
  \citenamefont {Govoni}, \citenamefont {Kundu},\ and\ \citenamefont
  {Galli}}]{Yang2022Sep}%
  \BibitemOpen
  \bibfield  {author} {\bibinfo {author} {\bibfnamefont {H.}~\bibnamefont
  {Yang}}, \bibinfo {author} {\bibfnamefont {M.}~\bibnamefont {Govoni}},
  \bibinfo {author} {\bibfnamefont {A.}~\bibnamefont {Kundu}}, \ and\ \bibinfo
  {author} {\bibfnamefont {G.}~\bibnamefont {Galli}},\ }\href {\doibase
  10.1021/acs.jctc.2c00579} {\bibfield  {journal} {\bibinfo  {journal} {J.
  Chem. Theory Comput.}\ ,\ \bibinfo {pages} {6031}} (\bibinfo {year}
  {2022})}\BibitemShut {NoStop}%
\bibitem [{\citenamefont {Sio}\ \emph {et~al.}(2019{\natexlab{a}})\citenamefont
  {Sio}, \citenamefont {Verdi}, \citenamefont {Poncé},\ and\ \citenamefont
  {Giustino}}]{sio_ab_2019}%
  \BibitemOpen
  \bibfield  {author} {\bibinfo {author} {\bibfnamefont {W.~H.}\ \bibnamefont
  {Sio}}, \bibinfo {author} {\bibfnamefont {C.}~\bibnamefont {Verdi}}, \bibinfo
  {author} {\bibfnamefont {S.}~\bibnamefont {Poncé}}, \ and\ \bibinfo {author}
  {\bibfnamefont {F.}~\bibnamefont {Giustino}},\ }\href {\doibase
  10.1103/PhysRevB.99.235139} {\bibfield  {journal} {\bibinfo  {journal} {Phys.
  Rev. B}\ }\textbf {\bibinfo {volume} {99}},\ \bibinfo {pages} {235139}
  (\bibinfo {year} {2019}{\natexlab{a}})}\BibitemShut {NoStop}%
\bibitem [{\citenamefont {Sio}\ \emph {et~al.}(2019{\natexlab{b}})\citenamefont
  {Sio}, \citenamefont {Verdi}, \citenamefont {Poncé},\ and\ \citenamefont
  {Giustino}}]{sio_polarons_2019}%
  \BibitemOpen
  \bibfield  {author} {\bibinfo {author} {\bibfnamefont {W.~H.}\ \bibnamefont
  {Sio}}, \bibinfo {author} {\bibfnamefont {C.}~\bibnamefont {Verdi}}, \bibinfo
  {author} {\bibfnamefont {S.}~\bibnamefont {Poncé}}, \ and\ \bibinfo {author}
  {\bibfnamefont {F.}~\bibnamefont {Giustino}},\ }\href {\doibase
  10.1103/PhysRevLett.122.246403} {\bibfield  {journal} {\bibinfo  {journal}
  {Phys. Rev. Lett.}\ }\textbf {\bibinfo {volume} {122}},\ \bibinfo {pages}
  {246403} (\bibinfo {year} {2019}{\natexlab{b}})}\BibitemShut {NoStop}%
\bibitem [{\citenamefont {Lafuente-Bartolome}\ \emph
  {et~al.}(2022{\natexlab{a}})\citenamefont {Lafuente-Bartolome}, \citenamefont
  {Lian}, \citenamefont {Sio}, \citenamefont {Gurtubay}, \citenamefont
  {Eiguren},\ and\ \citenamefont {Giustino}}]{lafuente-bartolome_ab_2022}%
  \BibitemOpen
  \bibfield  {author} {\bibinfo {author} {\bibfnamefont {J.}~\bibnamefont
  {Lafuente-Bartolome}}, \bibinfo {author} {\bibfnamefont {C.}~\bibnamefont
  {Lian}}, \bibinfo {author} {\bibfnamefont {W.~H.}\ \bibnamefont {Sio}},
  \bibinfo {author} {\bibfnamefont {I.~G.}\ \bibnamefont {Gurtubay}}, \bibinfo
  {author} {\bibfnamefont {A.}~\bibnamefont {Eiguren}}, \ and\ \bibinfo
  {author} {\bibfnamefont {F.}~\bibnamefont {Giustino}},\ }\href {\doibase
  10.1103/PhysRevB.106.075119} {\bibfield  {journal} {\bibinfo  {journal}
  {Phys. Rev. B}\ }\textbf {\bibinfo {volume} {106}},\ \bibinfo {pages}
  {075119} (\bibinfo {year} {2022}{\natexlab{a}})}\BibitemShut {NoStop}%
\bibitem [{\citenamefont {Lafuente-Bartolome}\ \emph
  {et~al.}(2022{\natexlab{b}})\citenamefont {Lafuente-Bartolome}, \citenamefont
  {Lian}, \citenamefont {Sio}, \citenamefont {Gurtubay}, \citenamefont
  {Eiguren},\ and\ \citenamefont {Giustino}}]{lafuente-bartolome_unified_2022}%
  \BibitemOpen
  \bibfield  {author} {\bibinfo {author} {\bibfnamefont {J.}~\bibnamefont
  {Lafuente-Bartolome}}, \bibinfo {author} {\bibfnamefont {C.}~\bibnamefont
  {Lian}}, \bibinfo {author} {\bibfnamefont {W.~H.}\ \bibnamefont {Sio}},
  \bibinfo {author} {\bibfnamefont {I.~G.}\ \bibnamefont {Gurtubay}}, \bibinfo
  {author} {\bibfnamefont {A.}~\bibnamefont {Eiguren}}, \ and\ \bibinfo
  {author} {\bibfnamefont {F.}~\bibnamefont {Giustino}},\ }\href {\doibase
  10.1103/PhysRevLett.129.076402} {\bibfield  {journal} {\bibinfo  {journal}
  {Phys. Rev. Lett.}\ }\textbf {\bibinfo {volume} {129}},\ \bibinfo {pages}
  {076402} (\bibinfo {year} {2022}{\natexlab{b}})}\BibitemShut {NoStop}%
\bibitem [{\citenamefont {White}\ \emph {et~al.}(2020)\citenamefont {White},
  \citenamefont {Gao}, \citenamefont {Minnich},\ and\ \citenamefont
  {Chan}}]{White2020Dec}%
  \BibitemOpen
  \bibfield  {author} {\bibinfo {author} {\bibfnamefont {A.~F.}\ \bibnamefont
  {White}}, \bibinfo {author} {\bibfnamefont {Y.}~\bibnamefont {Gao}}, \bibinfo
  {author} {\bibfnamefont {A.~J.}\ \bibnamefont {Minnich}}, \ and\ \bibinfo
  {author} {\bibfnamefont {G.~K.-L.}\ \bibnamefont {Chan}},\ }\href {\doibase
  10.1063/5.0033132} {\bibfield  {journal} {\bibinfo  {journal} {J. Chem.
  Phys.}\ }\textbf {\bibinfo {volume} {153}},\ \bibinfo {pages} {224112}
  (\bibinfo {year} {2020})}\BibitemShut {NoStop}%
\bibitem [{\citenamefont {Lee}\ \emph {et~al.}(2021{\natexlab{c}})\citenamefont
  {Lee}, \citenamefont {Zhang},\ and\ \citenamefont
  {Reichman}}]{lee2021constrained}%
  \BibitemOpen
  \bibfield  {author} {\bibinfo {author} {\bibfnamefont {J.}~\bibnamefont
  {Lee}}, \bibinfo {author} {\bibfnamefont {S.}~\bibnamefont {Zhang}}, \ and\
  \bibinfo {author} {\bibfnamefont {D.~R.}\ \bibnamefont {Reichman}},\
  }\href@noop {} {\bibfield  {journal} {\bibinfo  {journal} {Phys. Rev. B}\
  }\textbf {\bibinfo {volume} {103}},\ \bibinfo {pages} {115123} (\bibinfo
  {year} {2021}{\natexlab{c}})}\BibitemShut {NoStop}%
\bibitem [{\citenamefont {Erba}\ \emph {et~al.}(2023)\citenamefont {Erba},
  \citenamefont {Desmarais}, \citenamefont {Casassa}, \citenamefont
  {Civalleri}, \citenamefont {Donà}, \citenamefont {Bush}, \citenamefont
  {Searle}, \citenamefont {Maschio}, \citenamefont {Edith-Daga}, \citenamefont
  {Cossard}, \citenamefont {Ribaldone}, \citenamefont {Ascrizzi}, \citenamefont
  {Marana}, \citenamefont {Flament},\ and\ \citenamefont
  {Kirtman}}]{Erba2022Dec}%
  \BibitemOpen
  \bibfield  {author} {\bibinfo {author} {\bibfnamefont {A.}~\bibnamefont
  {Erba}}, \bibinfo {author} {\bibfnamefont {J.~K.}\ \bibnamefont {Desmarais}},
  \bibinfo {author} {\bibfnamefont {S.}~\bibnamefont {Casassa}}, \bibinfo
  {author} {\bibfnamefont {B.}~\bibnamefont {Civalleri}}, \bibinfo {author}
  {\bibfnamefont {L.}~\bibnamefont {Donà}}, \bibinfo {author} {\bibfnamefont
  {I.~J.}\ \bibnamefont {Bush}}, \bibinfo {author} {\bibfnamefont
  {B.}~\bibnamefont {Searle}}, \bibinfo {author} {\bibfnamefont
  {L.}~\bibnamefont {Maschio}}, \bibinfo {author} {\bibfnamefont
  {L.}~\bibnamefont {Edith-Daga}}, \bibinfo {author} {\bibfnamefont
  {A.}~\bibnamefont {Cossard}}, \bibinfo {author} {\bibfnamefont
  {C.}~\bibnamefont {Ribaldone}}, \bibinfo {author} {\bibfnamefont
  {E.}~\bibnamefont {Ascrizzi}}, \bibinfo {author} {\bibfnamefont {N.~L.}\
  \bibnamefont {Marana}}, \bibinfo {author} {\bibfnamefont {J.-P.}\
  \bibnamefont {Flament}}, \ and\ \bibinfo {author} {\bibfnamefont
  {B.}~\bibnamefont {Kirtman}},\ }\href {\doibase 10.1021/acs.jctc.2c00958}
  {\bibfield  {journal} {\bibinfo  {journal} {J. Chem. Theory Comput.}\
  }\textbf {\bibinfo {volume} {19}},\ \bibinfo {pages} {6891} (\bibinfo {year}
  {2023})}\BibitemShut {NoStop}%
\bibitem [{\citenamefont {Blum}\ \emph {et~al.}(2009)\citenamefont {Blum},
  \citenamefont {Gehrke}, \citenamefont {Hanke}, \citenamefont {Havu},
  \citenamefont {Havu}, \citenamefont {Ren}, \citenamefont {Reuter},\ and\
  \citenamefont {Scheffler}}]{Blum2009Nov}%
  \BibitemOpen
  \bibfield  {author} {\bibinfo {author} {\bibfnamefont {V.}~\bibnamefont
  {Blum}}, \bibinfo {author} {\bibfnamefont {R.}~\bibnamefont {Gehrke}},
  \bibinfo {author} {\bibfnamefont {F.}~\bibnamefont {Hanke}}, \bibinfo
  {author} {\bibfnamefont {P.}~\bibnamefont {Havu}}, \bibinfo {author}
  {\bibfnamefont {V.}~\bibnamefont {Havu}}, \bibinfo {author} {\bibfnamefont
  {X.}~\bibnamefont {Ren}}, \bibinfo {author} {\bibfnamefont {K.}~\bibnamefont
  {Reuter}}, \ and\ \bibinfo {author} {\bibfnamefont {M.}~\bibnamefont
  {Scheffler}},\ }\href {\doibase 10.1016/j.cpc.2009.06.022} {\bibfield
  {journal} {\bibinfo  {journal} {Comput. Phys. Commun.}\ }\textbf {\bibinfo
  {volume} {180}},\ \bibinfo {pages} {2175} (\bibinfo {year}
  {2009})}\BibitemShut {NoStop}%
\bibitem [{\citenamefont {García}\ \emph {et~al.}(2020)\citenamefont
  {García}, \citenamefont {Papior}, \citenamefont {Akhtar}, \citenamefont
  {Artacho}, \citenamefont {Blum}, \citenamefont {Bosoni}, \citenamefont
  {Brandimarte}, \citenamefont {Brandbyge}, \citenamefont {Cerdá},
  \citenamefont {Corsetti}, \citenamefont {Cuadrado}, \citenamefont {Dikan},
  \citenamefont {Ferrer}, \citenamefont {Gale}, \citenamefont
  {García-Fernández}, \citenamefont {García-Suárez}, \citenamefont
  {García}, \citenamefont {Huhs}, \citenamefont {Illera}, \citenamefont
  {Korytár}, \citenamefont {Koval}, \citenamefont {Lebedeva}, \citenamefont
  {Lin}, \citenamefont {López-Tarifa}, \citenamefont {Mayo}, \citenamefont
  {Mohr}, \citenamefont {Ordejón}, \citenamefont {Postnikov}, \citenamefont
  {Pouillon}, \citenamefont {Pruneda}, \citenamefont {Robles}, \citenamefont
  {Sánchez-Portal}, \citenamefont {Soler}, \citenamefont {Ullah},
  \citenamefont {Yu},\ and\ \citenamefont {Junquera}}]{Garcia2020May}%
  \BibitemOpen
  \bibfield  {author} {\bibinfo {author} {\bibfnamefont {A.}~\bibnamefont
  {García}}, \bibinfo {author} {\bibfnamefont {N.}~\bibnamefont {Papior}},
  \bibinfo {author} {\bibfnamefont {A.}~\bibnamefont {Akhtar}}, \bibinfo
  {author} {\bibfnamefont {E.}~\bibnamefont {Artacho}}, \bibinfo {author}
  {\bibfnamefont {V.}~\bibnamefont {Blum}}, \bibinfo {author} {\bibfnamefont
  {E.}~\bibnamefont {Bosoni}}, \bibinfo {author} {\bibfnamefont
  {P.}~\bibnamefont {Brandimarte}}, \bibinfo {author} {\bibfnamefont
  {M.}~\bibnamefont {Brandbyge}}, \bibinfo {author} {\bibfnamefont {J.~I.}\
  \bibnamefont {Cerdá}}, \bibinfo {author} {\bibfnamefont {F.}~\bibnamefont
  {Corsetti}}, \bibinfo {author} {\bibfnamefont {R.}~\bibnamefont {Cuadrado}},
  \bibinfo {author} {\bibfnamefont {V.}~\bibnamefont {Dikan}}, \bibinfo
  {author} {\bibfnamefont {J.}~\bibnamefont {Ferrer}}, \bibinfo {author}
  {\bibfnamefont {J.}~\bibnamefont {Gale}}, \bibinfo {author} {\bibfnamefont
  {P.}~\bibnamefont {García-Fernández}}, \bibinfo {author} {\bibfnamefont
  {V.~M.}\ \bibnamefont {García-Suárez}}, \bibinfo {author} {\bibfnamefont
  {S.}~\bibnamefont {García}}, \bibinfo {author} {\bibfnamefont
  {G.}~\bibnamefont {Huhs}}, \bibinfo {author} {\bibfnamefont {S.}~\bibnamefont
  {Illera}}, \bibinfo {author} {\bibfnamefont {R.}~\bibnamefont {Korytár}},
  \bibinfo {author} {\bibfnamefont {P.}~\bibnamefont {Koval}}, \bibinfo
  {author} {\bibfnamefont {I.}~\bibnamefont {Lebedeva}}, \bibinfo {author}
  {\bibfnamefont {L.}~\bibnamefont {Lin}}, \bibinfo {author} {\bibfnamefont
  {P.}~\bibnamefont {López-Tarifa}}, \bibinfo {author} {\bibfnamefont {S.~G.}\
  \bibnamefont {Mayo}}, \bibinfo {author} {\bibfnamefont {S.}~\bibnamefont
  {Mohr}}, \bibinfo {author} {\bibfnamefont {P.}~\bibnamefont {Ordejón}},
  \bibinfo {author} {\bibfnamefont {A.}~\bibnamefont {Postnikov}}, \bibinfo
  {author} {\bibfnamefont {Y.}~\bibnamefont {Pouillon}}, \bibinfo {author}
  {\bibfnamefont {M.}~\bibnamefont {Pruneda}}, \bibinfo {author} {\bibfnamefont
  {R.}~\bibnamefont {Robles}}, \bibinfo {author} {\bibfnamefont
  {D.}~\bibnamefont {Sánchez-Portal}}, \bibinfo {author} {\bibfnamefont
  {J.~M.}\ \bibnamefont {Soler}}, \bibinfo {author} {\bibfnamefont
  {R.}~\bibnamefont {Ullah}}, \bibinfo {author} {\bibfnamefont {V.~W.-z.}\
  \bibnamefont {Yu}}, \ and\ \bibinfo {author} {\bibfnamefont {J.}~\bibnamefont
  {Junquera}},\ }\href {\doibase 10.1063/5.0005077} {\bibfield  {journal}
  {\bibinfo  {journal} {J. Chem. Phys.}\ }\textbf {\bibinfo {volume} {152}},\
  \bibinfo {pages} {204108} (\bibinfo {year} {2020})}\BibitemShut {NoStop}%
\bibitem [{\citenamefont {Balasubramani}\ \emph {et~al.}(2020)\citenamefont
  {Balasubramani}, \citenamefont {Chen}, \citenamefont {Coriani}, \citenamefont
  {Diedenhofen}, \citenamefont {Frank}, \citenamefont {Franzke}, \citenamefont
  {Furche}, \citenamefont {Grotjahn}, \citenamefont {Harding}, \citenamefont
  {Hättig}, \citenamefont {Hellweg}, \citenamefont {Helmich-Paris},
  \citenamefont {Holzer}, \citenamefont {Huniar}, \citenamefont {Kaupp},
  \citenamefont {Marefat~Khah}, \citenamefont {Karbalaei~Khani}, \citenamefont
  {Müller}, \citenamefont {Mack}, \citenamefont {Nguyen}, \citenamefont
  {Parker}, \citenamefont {Perlt}, \citenamefont {Rappoport}, \citenamefont
  {Reiter}, \citenamefont {Roy}, \citenamefont {Rückert}, \citenamefont
  {Schmitz}, \citenamefont {Sierka}, \citenamefont {Tapavicza}, \citenamefont
  {Tew}, \citenamefont {van Wüllen}, \citenamefont {Voora}, \citenamefont
  {Weigend}, \citenamefont {Wodyński},\ and\ \citenamefont
  {Yu}}]{Balasubramani2020May}%
  \BibitemOpen
  \bibfield  {author} {\bibinfo {author} {\bibfnamefont {S.~G.}\ \bibnamefont
  {Balasubramani}}, \bibinfo {author} {\bibfnamefont {G.~P.}\ \bibnamefont
  {Chen}}, \bibinfo {author} {\bibfnamefont {S.}~\bibnamefont {Coriani}},
  \bibinfo {author} {\bibfnamefont {M.}~\bibnamefont {Diedenhofen}}, \bibinfo
  {author} {\bibfnamefont {M.~S.}\ \bibnamefont {Frank}}, \bibinfo {author}
  {\bibfnamefont {Y.~J.}\ \bibnamefont {Franzke}}, \bibinfo {author}
  {\bibfnamefont {F.}~\bibnamefont {Furche}}, \bibinfo {author} {\bibfnamefont
  {R.}~\bibnamefont {Grotjahn}}, \bibinfo {author} {\bibfnamefont {M.~E.}\
  \bibnamefont {Harding}}, \bibinfo {author} {\bibfnamefont {C.}~\bibnamefont
  {Hättig}}, \bibinfo {author} {\bibfnamefont {A.}~\bibnamefont {Hellweg}},
  \bibinfo {author} {\bibfnamefont {B.}~\bibnamefont {Helmich-Paris}}, \bibinfo
  {author} {\bibfnamefont {C.}~\bibnamefont {Holzer}}, \bibinfo {author}
  {\bibfnamefont {U.}~\bibnamefont {Huniar}}, \bibinfo {author} {\bibfnamefont
  {M.}~\bibnamefont {Kaupp}}, \bibinfo {author} {\bibfnamefont
  {A.}~\bibnamefont {Marefat~Khah}}, \bibinfo {author} {\bibfnamefont
  {S.}~\bibnamefont {Karbalaei~Khani}}, \bibinfo {author} {\bibfnamefont
  {T.}~\bibnamefont {Müller}}, \bibinfo {author} {\bibfnamefont
  {F.}~\bibnamefont {Mack}}, \bibinfo {author} {\bibfnamefont {B.~D.}\
  \bibnamefont {Nguyen}}, \bibinfo {author} {\bibfnamefont {S.~M.}\
  \bibnamefont {Parker}}, \bibinfo {author} {\bibfnamefont {E.}~\bibnamefont
  {Perlt}}, \bibinfo {author} {\bibfnamefont {D.}~\bibnamefont {Rappoport}},
  \bibinfo {author} {\bibfnamefont {K.}~\bibnamefont {Reiter}}, \bibinfo
  {author} {\bibfnamefont {S.}~\bibnamefont {Roy}}, \bibinfo {author}
  {\bibfnamefont {M.}~\bibnamefont {Rückert}}, \bibinfo {author}
  {\bibfnamefont {G.}~\bibnamefont {Schmitz}}, \bibinfo {author} {\bibfnamefont
  {M.}~\bibnamefont {Sierka}}, \bibinfo {author} {\bibfnamefont
  {E.}~\bibnamefont {Tapavicza}}, \bibinfo {author} {\bibfnamefont {D.~P.}\
  \bibnamefont {Tew}}, \bibinfo {author} {\bibfnamefont {C.}~\bibnamefont {van
  Wüllen}}, \bibinfo {author} {\bibfnamefont {V.~K.}\ \bibnamefont {Voora}},
  \bibinfo {author} {\bibfnamefont {F.}~\bibnamefont {Weigend}}, \bibinfo
  {author} {\bibfnamefont {A.}~\bibnamefont {Wodyński}}, \ and\ \bibinfo
  {author} {\bibfnamefont {J.~M.}\ \bibnamefont {Yu}},\ }\href {\doibase
  10.1063/5.0004635} {\bibfield  {journal} {\bibinfo  {journal} {J. Chem.
  Phys.}\ }\textbf {\bibinfo {volume} {152}},\ \bibinfo {pages} {184107}
  (\bibinfo {year} {2020})}\BibitemShut {NoStop}%
\bibitem [{\citenamefont {Sun}\ \emph {et~al.}(2020{\natexlab{b}})\citenamefont
  {Sun}, \citenamefont {Zhang}, \citenamefont {Banerjee}, \citenamefont {Bao},
  \citenamefont {Barbry}, \citenamefont {Blunt}, \citenamefont {Bogdanov},
  \citenamefont {Booth}, \citenamefont {Chen}, \citenamefont {Cui},
  \citenamefont {Eriksen}, \citenamefont {Gao}, \citenamefont {Guo},
  \citenamefont {Hermann}, \citenamefont {Hermes}, \citenamefont {Koh},
  \citenamefont {Koval}, \citenamefont {Lehtola}, \citenamefont {Li},
  \citenamefont {Liu}, \citenamefont {Mardirossian}, \citenamefont {McClain},
  \citenamefont {Motta}, \citenamefont {Mussard}, \citenamefont {Pham},
  \citenamefont {Pulkin}, \citenamefont {Purwanto}, \citenamefont {Robinson},
  \citenamefont {Ronca}, \citenamefont {Sayfutyarova}, \citenamefont
  {Scheurer}, \citenamefont {Schurkus}, \citenamefont {Smith}, \citenamefont
  {Sun}, \citenamefont {Sun}, \citenamefont {Upadhyay}, \citenamefont {Wagner},
  \citenamefont {Wang}, \citenamefont {White}, \citenamefont {Whitfield},
  \citenamefont {Williamson}, \citenamefont {Wouters}, \citenamefont {Yang},
  \citenamefont {Yu}, \citenamefont {Zhu}, \citenamefont {Berkelbach},
  \citenamefont {Sharma}, \citenamefont {Sokolov},\ and\ \citenamefont
  {Chan}}]{Sun2020Jul}%
  \BibitemOpen
  \bibfield  {author} {\bibinfo {author} {\bibfnamefont {Q.}~\bibnamefont
  {Sun}}, \bibinfo {author} {\bibfnamefont {X.}~\bibnamefont {Zhang}}, \bibinfo
  {author} {\bibfnamefont {S.}~\bibnamefont {Banerjee}}, \bibinfo {author}
  {\bibfnamefont {P.}~\bibnamefont {Bao}}, \bibinfo {author} {\bibfnamefont
  {M.}~\bibnamefont {Barbry}}, \bibinfo {author} {\bibfnamefont {N.~S.}\
  \bibnamefont {Blunt}}, \bibinfo {author} {\bibfnamefont {N.~A.}\ \bibnamefont
  {Bogdanov}}, \bibinfo {author} {\bibfnamefont {G.~H.}\ \bibnamefont {Booth}},
  \bibinfo {author} {\bibfnamefont {J.}~\bibnamefont {Chen}}, \bibinfo {author}
  {\bibfnamefont {Z.-H.}\ \bibnamefont {Cui}}, \bibinfo {author} {\bibfnamefont
  {J.~J.}\ \bibnamefont {Eriksen}}, \bibinfo {author} {\bibfnamefont
  {Y.}~\bibnamefont {Gao}}, \bibinfo {author} {\bibfnamefont {S.}~\bibnamefont
  {Guo}}, \bibinfo {author} {\bibfnamefont {J.}~\bibnamefont {Hermann}},
  \bibinfo {author} {\bibfnamefont {M.~R.}\ \bibnamefont {Hermes}}, \bibinfo
  {author} {\bibfnamefont {K.}~\bibnamefont {Koh}}, \bibinfo {author}
  {\bibfnamefont {P.}~\bibnamefont {Koval}}, \bibinfo {author} {\bibfnamefont
  {S.}~\bibnamefont {Lehtola}}, \bibinfo {author} {\bibfnamefont
  {Z.}~\bibnamefont {Li}}, \bibinfo {author} {\bibfnamefont {J.}~\bibnamefont
  {Liu}}, \bibinfo {author} {\bibfnamefont {N.}~\bibnamefont {Mardirossian}},
  \bibinfo {author} {\bibfnamefont {J.~D.}\ \bibnamefont {McClain}}, \bibinfo
  {author} {\bibfnamefont {M.}~\bibnamefont {Motta}}, \bibinfo {author}
  {\bibfnamefont {B.}~\bibnamefont {Mussard}}, \bibinfo {author} {\bibfnamefont
  {H.~Q.}\ \bibnamefont {Pham}}, \bibinfo {author} {\bibfnamefont
  {A.}~\bibnamefont {Pulkin}}, \bibinfo {author} {\bibfnamefont
  {W.}~\bibnamefont {Purwanto}}, \bibinfo {author} {\bibfnamefont {P.~J.}\
  \bibnamefont {Robinson}}, \bibinfo {author} {\bibfnamefont {E.}~\bibnamefont
  {Ronca}}, \bibinfo {author} {\bibfnamefont {E.~R.}\ \bibnamefont
  {Sayfutyarova}}, \bibinfo {author} {\bibfnamefont {M.}~\bibnamefont
  {Scheurer}}, \bibinfo {author} {\bibfnamefont {H.~F.}\ \bibnamefont
  {Schurkus}}, \bibinfo {author} {\bibfnamefont {J.~E.~T.}\ \bibnamefont
  {Smith}}, \bibinfo {author} {\bibfnamefont {C.}~\bibnamefont {Sun}}, \bibinfo
  {author} {\bibfnamefont {S.-N.}\ \bibnamefont {Sun}}, \bibinfo {author}
  {\bibfnamefont {S.}~\bibnamefont {Upadhyay}}, \bibinfo {author}
  {\bibfnamefont {L.~K.}\ \bibnamefont {Wagner}}, \bibinfo {author}
  {\bibfnamefont {X.}~\bibnamefont {Wang}}, \bibinfo {author} {\bibfnamefont
  {A.}~\bibnamefont {White}}, \bibinfo {author} {\bibfnamefont {J.~D.}\
  \bibnamefont {Whitfield}}, \bibinfo {author} {\bibfnamefont {M.~J.}\
  \bibnamefont {Williamson}}, \bibinfo {author} {\bibfnamefont
  {S.}~\bibnamefont {Wouters}}, \bibinfo {author} {\bibfnamefont
  {J.}~\bibnamefont {Yang}}, \bibinfo {author} {\bibfnamefont {J.~M.}\
  \bibnamefont {Yu}}, \bibinfo {author} {\bibfnamefont {T.}~\bibnamefont
  {Zhu}}, \bibinfo {author} {\bibfnamefont {T.~C.}\ \bibnamefont {Berkelbach}},
  \bibinfo {author} {\bibfnamefont {S.}~\bibnamefont {Sharma}}, \bibinfo
  {author} {\bibfnamefont {A.~Y.}\ \bibnamefont {Sokolov}}, \ and\ \bibinfo
  {author} {\bibfnamefont {G.~K.-L.}\ \bibnamefont {Chan}},\ }\href {\doibase
  10.1063/5.0006074} {\bibfield  {journal} {\bibinfo  {journal} {J. Chem.
  Phys.}\ }\textbf {\bibinfo {volume} {153}},\ \bibinfo {pages} {024109}
  (\bibinfo {year} {2020}{\natexlab{b}})}\BibitemShut {NoStop}%
\bibitem [{\citenamefont {Epifanovsky}\ \emph {et~al.}(2021)\citenamefont
  {Epifanovsky}, \citenamefont {Gilbert}, \citenamefont {Feng}, \citenamefont
  {Lee}, \citenamefont {Mao}, \citenamefont {Mardirossian}, \citenamefont
  {Pokhilko}, \citenamefont {White}, \citenamefont {Coons}, \citenamefont
  {Dempwolff}, \citenamefont {Gan}, \citenamefont {Hait}, \citenamefont {Horn},
  \citenamefont {Jacobson}, \citenamefont {Kaliman}, \citenamefont {Kussmann},
  \citenamefont {Lange}, \citenamefont {Lao}, \citenamefont {Levine},
  \citenamefont {Liu}, \citenamefont {McKenzie}, \citenamefont {Morrison},
  \citenamefont {Nanda}, \citenamefont {Plasser}, \citenamefont {Rehn},
  \citenamefont {Vidal}, \citenamefont {You}, \citenamefont {Zhu},
  \citenamefont {Alam}, \citenamefont {Albrecht}, \citenamefont {Aldossary},
  \citenamefont {Alguire}, \citenamefont {Andersen}, \citenamefont {Athavale},
  \citenamefont {Barton}, \citenamefont {Begam}, \citenamefont {Behn},
  \citenamefont {Bellonzi}, \citenamefont {Bernard}, \citenamefont {Berquist},
  \citenamefont {Burton}, \citenamefont {Carreras}, \citenamefont
  {Carter-Fenk}, \citenamefont {Chakraborty}, \citenamefont {Chien},
  \citenamefont {Closser}, \citenamefont {Cofer-Shabica}, \citenamefont
  {Dasgupta}, \citenamefont {de~Wergifosse}, \citenamefont {Deng},
  \citenamefont {Diedenhofen}, \citenamefont {Do}, \citenamefont {Ehlert},
  \citenamefont {Fang}, \citenamefont {Fatehi}, \citenamefont {Feng},
  \citenamefont {Friedhoff}, \citenamefont {Gayvert}, \citenamefont {Ge},
  \citenamefont {Gidofalvi}, \citenamefont {Goldey}, \citenamefont {Gomes},
  \citenamefont {Gonz{\ifmmode\acute{a}\else\'{a}\fi}lez-Espinoza},
  \citenamefont {Gulania}, \citenamefont {Gunina}, \citenamefont
  {Hanson-Heine}, \citenamefont {Harbach}, \citenamefont {Hauser},
  \citenamefont {Herbst}, \citenamefont
  {Hern{\ifmmode\acute{a}\else\'{a}\fi}ndez~Vera}, \citenamefont {Hodecker},
  \citenamefont {Holden}, \citenamefont {Houck}, \citenamefont {Huang},
  \citenamefont {Hui}, \citenamefont {Huynh}, \citenamefont {Ivanov},
  \citenamefont {J{\ifmmode\acute{a}\else\'{a}\fi}sz}, \citenamefont {Ji},
  \citenamefont {Jiang}, \citenamefont {Kaduk}, \citenamefont
  {K{\ifmmode\ddot{a}\else\"{a}\fi}hler}, \citenamefont {Khistyaev},
  \citenamefont {Kim}, \citenamefont {Kis}, \citenamefont {Klunzinger},
  \citenamefont {Koczor-Benda}, \citenamefont {Koh}, \citenamefont {Kosenkov},
  \citenamefont {Koulias}, \citenamefont {Kowalczyk}, \citenamefont {Krauter},
  \citenamefont {Kue}, \citenamefont {Kunitsa}, \citenamefont {Kus},
  \citenamefont {Ladj{\ifmmode\acute{a}\else\'{a}\fi}nszki}, \citenamefont
  {Landau}, \citenamefont {Lawler}, \citenamefont {Lefrancois}, \citenamefont
  {Lehtola}, \citenamefont {Li}, \citenamefont {Li}, \citenamefont {Liang},
  \citenamefont {Liebenthal}, \citenamefont {Lin}, \citenamefont {Lin},
  \citenamefont {Liu}, \citenamefont {Liu}, \citenamefont {Loipersberger},
  \citenamefont {Luenser}, \citenamefont {Manjanath}, \citenamefont {Manohar},
  \citenamefont {Mansoor}, \citenamefont {Manzer}, \citenamefont {Mao},
  \citenamefont {Marenich}, \citenamefont {Markovich}, \citenamefont {Mason},
  \citenamefont {Maurer}, \citenamefont {McLaughlin}, \citenamefont {Menger},
  \citenamefont {Mewes}, \citenamefont {Mewes}, \citenamefont {Morgante},
  \citenamefont {Mullinax}, \citenamefont {Oosterbaan}, \citenamefont {Paran},
  \citenamefont {Paul}, \citenamefont {Paul}, \citenamefont
  {Pavo{\ifmmode\check{s}\else\v{s}\fi}evi{\ifmmode\acute{c}\else\'{c}\fi}},
  \citenamefont {Pei}, \citenamefont {Prager}, \citenamefont {Proynov},
  \citenamefont {R{\ifmmode\acute{a}\else\'{a}\fi}k}, \citenamefont
  {Ramos-Cordoba}, \citenamefont {Rana}, \citenamefont {Rask}, \citenamefont
  {Rettig}, \citenamefont {Richard}, \citenamefont {Rob}, \citenamefont
  {Rossomme}, \citenamefont {Scheele}, \citenamefont {Scheurer}, \citenamefont
  {Schneider}, \citenamefont {Sergueev}, \citenamefont {Sharada}, \citenamefont
  {Skomorowski}, \citenamefont {Small}, \citenamefont {Stein}, \citenamefont
  {Su}, \citenamefont {Sundstrom}, \citenamefont {Tao}, \citenamefont
  {Thirman}, \citenamefont {Tornai}, \citenamefont {Tsuchimochi}, \citenamefont
  {Tubman}, \citenamefont {Veccham}, \citenamefont {Vydrov}, \citenamefont
  {Wenzel}, \citenamefont {Witte}, \citenamefont {Yamada}, \citenamefont {Yao},
  \citenamefont {Yeganeh}, \citenamefont {Yost}, \citenamefont {Zech},
  \citenamefont {Zhang}, \citenamefont {Zhang}, \citenamefont {Zhang},
  \citenamefont {Zuev}, \citenamefont {Aspuru-Guzik}, \citenamefont {Bell},
  \citenamefont {Besley}, \citenamefont {Bravaya}, \citenamefont {Brooks},
  \citenamefont {Casanova}, \citenamefont {Chai}, \citenamefont {Coriani},
  \citenamefont {Cramer}, \citenamefont {Cserey}, \citenamefont {DePrince},
  \citenamefont {DiStasio}, \citenamefont {Dreuw}, \citenamefont {Dunietz},
  \citenamefont {Furlani}, \citenamefont {Goddard}, \citenamefont
  {Hammes-Schiffer}, \citenamefont {Head-Gordon}, \citenamefont {Hehre},
  \citenamefont {Hsu}, \citenamefont {Jagau}, \citenamefont {Jung},
  \citenamefont {Klamt}, \citenamefont {Kong}, \citenamefont {Lambrecht},
  \citenamefont {Liang}, \citenamefont {Mayhall}, \citenamefont {McCurdy},
  \citenamefont {Neaton}, \citenamefont {Ochsenfeld}, \citenamefont {Parkhill},
  \citenamefont {Peverati}, \citenamefont {Rassolov}, \citenamefont {Shao},
  \citenamefont {Slipchenko}, \citenamefont {Stauch}, \citenamefont {Steele},
  \citenamefont {Subotnik}, \citenamefont {Thom}, \citenamefont {Tkatchenko},
  \citenamefont {Truhlar}, \citenamefont {Van~Voorhis}, \citenamefont
  {Wesolowski}, \citenamefont {Whaley}, \citenamefont {Woodcock}, \citenamefont
  {Zimmerman}, \citenamefont {Faraji}, \citenamefont {Gill}, \citenamefont
  {Head-Gordon}, \citenamefont {Herbert},\ and\ \citenamefont
  {Krylov}}]{Epifanovsky2021Aug}%
  \BibitemOpen
  \bibfield  {author} {\bibinfo {author} {\bibfnamefont {E.}~\bibnamefont
  {Epifanovsky}}, \bibinfo {author} {\bibfnamefont {A.~T.~B.}\ \bibnamefont
  {Gilbert}}, \bibinfo {author} {\bibfnamefont {X.}~\bibnamefont {Feng}},
  \bibinfo {author} {\bibfnamefont {J.}~\bibnamefont {Lee}}, \bibinfo {author}
  {\bibfnamefont {Y.}~\bibnamefont {Mao}}, \bibinfo {author} {\bibfnamefont
  {N.}~\bibnamefont {Mardirossian}}, \bibinfo {author} {\bibfnamefont
  {P.}~\bibnamefont {Pokhilko}}, \bibinfo {author} {\bibfnamefont {A.~F.}\
  \bibnamefont {White}}, \bibinfo {author} {\bibfnamefont {M.~P.}\ \bibnamefont
  {Coons}}, \bibinfo {author} {\bibfnamefont {A.~L.}\ \bibnamefont
  {Dempwolff}}, \bibinfo {author} {\bibfnamefont {Z.}~\bibnamefont {Gan}},
  \bibinfo {author} {\bibfnamefont {D.}~\bibnamefont {Hait}}, \bibinfo {author}
  {\bibfnamefont {P.~R.}\ \bibnamefont {Horn}}, \bibinfo {author}
  {\bibfnamefont {L.~D.}\ \bibnamefont {Jacobson}}, \bibinfo {author}
  {\bibfnamefont {I.}~\bibnamefont {Kaliman}}, \bibinfo {author} {\bibfnamefont
  {J.}~\bibnamefont {Kussmann}}, \bibinfo {author} {\bibfnamefont {A.~W.}\
  \bibnamefont {Lange}}, \bibinfo {author} {\bibfnamefont {K.~U.}\ \bibnamefont
  {Lao}}, \bibinfo {author} {\bibfnamefont {D.~S.}\ \bibnamefont {Levine}},
  \bibinfo {author} {\bibfnamefont {J.}~\bibnamefont {Liu}}, \bibinfo {author}
  {\bibfnamefont {S.~C.}\ \bibnamefont {McKenzie}}, \bibinfo {author}
  {\bibfnamefont {A.~F.}\ \bibnamefont {Morrison}}, \bibinfo {author}
  {\bibfnamefont {K.~D.}\ \bibnamefont {Nanda}}, \bibinfo {author}
  {\bibfnamefont {F.}~\bibnamefont {Plasser}}, \bibinfo {author} {\bibfnamefont
  {D.~R.}\ \bibnamefont {Rehn}}, \bibinfo {author} {\bibfnamefont {M.~L.}\
  \bibnamefont {Vidal}}, \bibinfo {author} {\bibfnamefont {Z.-Q.}\ \bibnamefont
  {You}}, \bibinfo {author} {\bibfnamefont {Y.}~\bibnamefont {Zhu}}, \bibinfo
  {author} {\bibfnamefont {B.}~\bibnamefont {Alam}}, \bibinfo {author}
  {\bibfnamefont {B.~J.}\ \bibnamefont {Albrecht}}, \bibinfo {author}
  {\bibfnamefont {A.}~\bibnamefont {Aldossary}}, \bibinfo {author}
  {\bibfnamefont {E.}~\bibnamefont {Alguire}}, \bibinfo {author} {\bibfnamefont
  {J.~H.}\ \bibnamefont {Andersen}}, \bibinfo {author} {\bibfnamefont
  {V.}~\bibnamefont {Athavale}}, \bibinfo {author} {\bibfnamefont
  {D.}~\bibnamefont {Barton}}, \bibinfo {author} {\bibfnamefont
  {K.}~\bibnamefont {Begam}}, \bibinfo {author} {\bibfnamefont
  {A.}~\bibnamefont {Behn}}, \bibinfo {author} {\bibfnamefont {N.}~\bibnamefont
  {Bellonzi}}, \bibinfo {author} {\bibfnamefont {Y.~A.}\ \bibnamefont
  {Bernard}}, \bibinfo {author} {\bibfnamefont {E.~J.}\ \bibnamefont
  {Berquist}}, \bibinfo {author} {\bibfnamefont {H.~G.~A.}\ \bibnamefont
  {Burton}}, \bibinfo {author} {\bibfnamefont {A.}~\bibnamefont {Carreras}},
  \bibinfo {author} {\bibfnamefont {K.}~\bibnamefont {Carter-Fenk}}, \bibinfo
  {author} {\bibfnamefont {R.}~\bibnamefont {Chakraborty}}, \bibinfo {author}
  {\bibfnamefont {A.~D.}\ \bibnamefont {Chien}}, \bibinfo {author}
  {\bibfnamefont {K.~D.}\ \bibnamefont {Closser}}, \bibinfo {author}
  {\bibfnamefont {V.}~\bibnamefont {Cofer-Shabica}}, \bibinfo {author}
  {\bibfnamefont {S.}~\bibnamefont {Dasgupta}}, \bibinfo {author}
  {\bibfnamefont {M.}~\bibnamefont {de~Wergifosse}}, \bibinfo {author}
  {\bibfnamefont {J.}~\bibnamefont {Deng}}, \bibinfo {author} {\bibfnamefont
  {M.}~\bibnamefont {Diedenhofen}}, \bibinfo {author} {\bibfnamefont
  {H.}~\bibnamefont {Do}}, \bibinfo {author} {\bibfnamefont {S.}~\bibnamefont
  {Ehlert}}, \bibinfo {author} {\bibfnamefont {P.-T.}\ \bibnamefont {Fang}},
  \bibinfo {author} {\bibfnamefont {S.}~\bibnamefont {Fatehi}}, \bibinfo
  {author} {\bibfnamefont {Q.}~\bibnamefont {Feng}}, \bibinfo {author}
  {\bibfnamefont {T.}~\bibnamefont {Friedhoff}}, \bibinfo {author}
  {\bibfnamefont {J.}~\bibnamefont {Gayvert}}, \bibinfo {author} {\bibfnamefont
  {Q.}~\bibnamefont {Ge}}, \bibinfo {author} {\bibfnamefont {G.}~\bibnamefont
  {Gidofalvi}}, \bibinfo {author} {\bibfnamefont {M.}~\bibnamefont {Goldey}},
  \bibinfo {author} {\bibfnamefont {J.}~\bibnamefont {Gomes}}, \bibinfo
  {author} {\bibfnamefont {C.~E.}\ \bibnamefont
  {Gonz{\ifmmode\acute{a}\else\'{a}\fi}lez-Espinoza}}, \bibinfo {author}
  {\bibfnamefont {S.}~\bibnamefont {Gulania}}, \bibinfo {author} {\bibfnamefont
  {A.~O.}\ \bibnamefont {Gunina}}, \bibinfo {author} {\bibfnamefont {M.~W.~D.}\
  \bibnamefont {Hanson-Heine}}, \bibinfo {author} {\bibfnamefont {P.~H.~P.}\
  \bibnamefont {Harbach}}, \bibinfo {author} {\bibfnamefont {A.}~\bibnamefont
  {Hauser}}, \bibinfo {author} {\bibfnamefont {M.~F.}\ \bibnamefont {Herbst}},
  \bibinfo {author} {\bibfnamefont {M.}~\bibnamefont
  {Hern{\ifmmode\acute{a}\else\'{a}\fi}ndez~Vera}}, \bibinfo {author}
  {\bibfnamefont {M.}~\bibnamefont {Hodecker}}, \bibinfo {author}
  {\bibfnamefont {Z.~C.}\ \bibnamefont {Holden}}, \bibinfo {author}
  {\bibfnamefont {S.}~\bibnamefont {Houck}}, \bibinfo {author} {\bibfnamefont
  {X.}~\bibnamefont {Huang}}, \bibinfo {author} {\bibfnamefont
  {K.}~\bibnamefont {Hui}}, \bibinfo {author} {\bibfnamefont {B.~C.}\
  \bibnamefont {Huynh}}, \bibinfo {author} {\bibfnamefont {M.}~\bibnamefont
  {Ivanov}}, \bibinfo {author} {\bibfnamefont
  {{\ifmmode\acute{A}\else\'{A}\fi}.}~\bibnamefont
  {J{\ifmmode\acute{a}\else\'{a}\fi}sz}}, \bibinfo {author} {\bibfnamefont
  {H.}~\bibnamefont {Ji}}, \bibinfo {author} {\bibfnamefont {H.}~\bibnamefont
  {Jiang}}, \bibinfo {author} {\bibfnamefont {B.}~\bibnamefont {Kaduk}},
  \bibinfo {author} {\bibfnamefont {S.}~\bibnamefont
  {K{\ifmmode\ddot{a}\else\"{a}\fi}hler}}, \bibinfo {author} {\bibfnamefont
  {K.}~\bibnamefont {Khistyaev}}, \bibinfo {author} {\bibfnamefont
  {J.}~\bibnamefont {Kim}}, \bibinfo {author} {\bibfnamefont {G.}~\bibnamefont
  {Kis}}, \bibinfo {author} {\bibfnamefont {P.}~\bibnamefont {Klunzinger}},
  \bibinfo {author} {\bibfnamefont {Z.}~\bibnamefont {Koczor-Benda}}, \bibinfo
  {author} {\bibfnamefont {J.~H.}\ \bibnamefont {Koh}}, \bibinfo {author}
  {\bibfnamefont {D.}~\bibnamefont {Kosenkov}}, \bibinfo {author}
  {\bibfnamefont {L.}~\bibnamefont {Koulias}}, \bibinfo {author} {\bibfnamefont
  {T.}~\bibnamefont {Kowalczyk}}, \bibinfo {author} {\bibfnamefont {C.~M.}\
  \bibnamefont {Krauter}}, \bibinfo {author} {\bibfnamefont {K.}~\bibnamefont
  {Kue}}, \bibinfo {author} {\bibfnamefont {A.}~\bibnamefont {Kunitsa}},
  \bibinfo {author} {\bibfnamefont {T.}~\bibnamefont {Kus}}, \bibinfo {author}
  {\bibfnamefont {I.}~\bibnamefont
  {Ladj{\ifmmode\acute{a}\else\'{a}\fi}nszki}}, \bibinfo {author}
  {\bibfnamefont {A.}~\bibnamefont {Landau}}, \bibinfo {author} {\bibfnamefont
  {K.~V.}\ \bibnamefont {Lawler}}, \bibinfo {author} {\bibfnamefont
  {D.}~\bibnamefont {Lefrancois}}, \bibinfo {author} {\bibfnamefont
  {S.}~\bibnamefont {Lehtola}}, \bibinfo {author} {\bibfnamefont {R.~R.}\
  \bibnamefont {Li}}, \bibinfo {author} {\bibfnamefont {Y.-P.}\ \bibnamefont
  {Li}}, \bibinfo {author} {\bibfnamefont {J.}~\bibnamefont {Liang}}, \bibinfo
  {author} {\bibfnamefont {M.}~\bibnamefont {Liebenthal}}, \bibinfo {author}
  {\bibfnamefont {H.-H.}\ \bibnamefont {Lin}}, \bibinfo {author} {\bibfnamefont
  {Y.-S.}\ \bibnamefont {Lin}}, \bibinfo {author} {\bibfnamefont
  {F.}~\bibnamefont {Liu}}, \bibinfo {author} {\bibfnamefont {K.-Y.}\
  \bibnamefont {Liu}}, \bibinfo {author} {\bibfnamefont {M.}~\bibnamefont
  {Loipersberger}}, \bibinfo {author} {\bibfnamefont {A.}~\bibnamefont
  {Luenser}}, \bibinfo {author} {\bibfnamefont {A.}~\bibnamefont {Manjanath}},
  \bibinfo {author} {\bibfnamefont {P.}~\bibnamefont {Manohar}}, \bibinfo
  {author} {\bibfnamefont {E.}~\bibnamefont {Mansoor}}, \bibinfo {author}
  {\bibfnamefont {S.~F.}\ \bibnamefont {Manzer}}, \bibinfo {author}
  {\bibfnamefont {S.-P.}\ \bibnamefont {Mao}}, \bibinfo {author} {\bibfnamefont
  {A.~V.}\ \bibnamefont {Marenich}}, \bibinfo {author} {\bibfnamefont
  {T.}~\bibnamefont {Markovich}}, \bibinfo {author} {\bibfnamefont
  {S.}~\bibnamefont {Mason}}, \bibinfo {author} {\bibfnamefont {S.~A.}\
  \bibnamefont {Maurer}}, \bibinfo {author} {\bibfnamefont {P.~F.}\
  \bibnamefont {McLaughlin}}, \bibinfo {author} {\bibfnamefont {M.~F. S.~J.}\
  \bibnamefont {Menger}}, \bibinfo {author} {\bibfnamefont {J.-M.}\
  \bibnamefont {Mewes}}, \bibinfo {author} {\bibfnamefont {S.~A.}\ \bibnamefont
  {Mewes}}, \bibinfo {author} {\bibfnamefont {P.}~\bibnamefont {Morgante}},
  \bibinfo {author} {\bibfnamefont {J.~W.}\ \bibnamefont {Mullinax}}, \bibinfo
  {author} {\bibfnamefont {K.~J.}\ \bibnamefont {Oosterbaan}}, \bibinfo
  {author} {\bibfnamefont {G.}~\bibnamefont {Paran}}, \bibinfo {author}
  {\bibfnamefont {A.~C.}\ \bibnamefont {Paul}}, \bibinfo {author}
  {\bibfnamefont {S.~K.}\ \bibnamefont {Paul}}, \bibinfo {author}
  {\bibfnamefont {F.}~\bibnamefont
  {Pavo{\ifmmode\check{s}\else\v{s}\fi}evi{\ifmmode\acute{c}\else\'{c}\fi}}},
  \bibinfo {author} {\bibfnamefont {Z.}~\bibnamefont {Pei}}, \bibinfo {author}
  {\bibfnamefont {S.}~\bibnamefont {Prager}}, \bibinfo {author} {\bibfnamefont
  {E.~I.}\ \bibnamefont {Proynov}}, \bibinfo {author} {\bibfnamefont
  {{\ifmmode\acute{A}\else\'{A}\fi}.}~\bibnamefont
  {R{\ifmmode\acute{a}\else\'{a}\fi}k}}, \bibinfo {author} {\bibfnamefont
  {E.}~\bibnamefont {Ramos-Cordoba}}, \bibinfo {author} {\bibfnamefont
  {B.}~\bibnamefont {Rana}}, \bibinfo {author} {\bibfnamefont {A.~E.}\
  \bibnamefont {Rask}}, \bibinfo {author} {\bibfnamefont {A.}~\bibnamefont
  {Rettig}}, \bibinfo {author} {\bibfnamefont {R.~M.}\ \bibnamefont {Richard}},
  \bibinfo {author} {\bibfnamefont {F.}~\bibnamefont {Rob}}, \bibinfo {author}
  {\bibfnamefont {E.}~\bibnamefont {Rossomme}}, \bibinfo {author}
  {\bibfnamefont {T.}~\bibnamefont {Scheele}}, \bibinfo {author} {\bibfnamefont
  {M.}~\bibnamefont {Scheurer}}, \bibinfo {author} {\bibfnamefont
  {M.}~\bibnamefont {Schneider}}, \bibinfo {author} {\bibfnamefont
  {N.}~\bibnamefont {Sergueev}}, \bibinfo {author} {\bibfnamefont {S.~M.}\
  \bibnamefont {Sharada}}, \bibinfo {author} {\bibfnamefont {W.}~\bibnamefont
  {Skomorowski}}, \bibinfo {author} {\bibfnamefont {D.~W.}\ \bibnamefont
  {Small}}, \bibinfo {author} {\bibfnamefont {C.~J.}\ \bibnamefont {Stein}},
  \bibinfo {author} {\bibfnamefont {Y.-C.}\ \bibnamefont {Su}}, \bibinfo
  {author} {\bibfnamefont {E.~J.}\ \bibnamefont {Sundstrom}}, \bibinfo {author}
  {\bibfnamefont {Z.}~\bibnamefont {Tao}}, \bibinfo {author} {\bibfnamefont
  {J.}~\bibnamefont {Thirman}}, \bibinfo {author} {\bibfnamefont {G.~J.}\
  \bibnamefont {Tornai}}, \bibinfo {author} {\bibfnamefont {T.}~\bibnamefont
  {Tsuchimochi}}, \bibinfo {author} {\bibfnamefont {N.~M.}\ \bibnamefont
  {Tubman}}, \bibinfo {author} {\bibfnamefont {S.~P.}\ \bibnamefont {Veccham}},
  \bibinfo {author} {\bibfnamefont {O.}~\bibnamefont {Vydrov}}, \bibinfo
  {author} {\bibfnamefont {J.}~\bibnamefont {Wenzel}}, \bibinfo {author}
  {\bibfnamefont {J.}~\bibnamefont {Witte}}, \bibinfo {author} {\bibfnamefont
  {A.}~\bibnamefont {Yamada}}, \bibinfo {author} {\bibfnamefont
  {K.}~\bibnamefont {Yao}}, \bibinfo {author} {\bibfnamefont {S.}~\bibnamefont
  {Yeganeh}}, \bibinfo {author} {\bibfnamefont {S.~R.}\ \bibnamefont {Yost}},
  \bibinfo {author} {\bibfnamefont {A.}~\bibnamefont {Zech}}, \bibinfo {author}
  {\bibfnamefont {I.~Y.}\ \bibnamefont {Zhang}}, \bibinfo {author}
  {\bibfnamefont {X.}~\bibnamefont {Zhang}}, \bibinfo {author} {\bibfnamefont
  {Y.}~\bibnamefont {Zhang}}, \bibinfo {author} {\bibfnamefont
  {D.}~\bibnamefont {Zuev}}, \bibinfo {author} {\bibfnamefont {A.}~\bibnamefont
  {Aspuru-Guzik}}, \bibinfo {author} {\bibfnamefont {A.~T.}\ \bibnamefont
  {Bell}}, \bibinfo {author} {\bibfnamefont {N.~A.}\ \bibnamefont {Besley}},
  \bibinfo {author} {\bibfnamefont {K.~B.}\ \bibnamefont {Bravaya}}, \bibinfo
  {author} {\bibfnamefont {B.~R.}\ \bibnamefont {Brooks}}, \bibinfo {author}
  {\bibfnamefont {D.}~\bibnamefont {Casanova}}, \bibinfo {author}
  {\bibfnamefont {J.-D.}\ \bibnamefont {Chai}}, \bibinfo {author}
  {\bibfnamefont {S.}~\bibnamefont {Coriani}}, \bibinfo {author} {\bibfnamefont
  {C.~J.}\ \bibnamefont {Cramer}}, \bibinfo {author} {\bibfnamefont
  {G.}~\bibnamefont {Cserey}}, \bibinfo {author} {\bibfnamefont {A.~E.}\
  \bibnamefont {DePrince}}, \bibinfo {author} {\bibfnamefont {R.~A.}\
  \bibnamefont {DiStasio}}, \bibinfo {author} {\bibfnamefont {A.}~\bibnamefont
  {Dreuw}}, \bibinfo {author} {\bibfnamefont {B.~D.}\ \bibnamefont {Dunietz}},
  \bibinfo {author} {\bibfnamefont {T.~R.}\ \bibnamefont {Furlani}}, \bibinfo
  {author} {\bibfnamefont {W.~A.}\ \bibnamefont {Goddard}}, \bibinfo {author}
  {\bibfnamefont {S.}~\bibnamefont {Hammes-Schiffer}}, \bibinfo {author}
  {\bibfnamefont {T.}~\bibnamefont {Head-Gordon}}, \bibinfo {author}
  {\bibfnamefont {W.~J.}\ \bibnamefont {Hehre}}, \bibinfo {author}
  {\bibfnamefont {C.-P.}\ \bibnamefont {Hsu}}, \bibinfo {author} {\bibfnamefont
  {T.-C.}\ \bibnamefont {Jagau}}, \bibinfo {author} {\bibfnamefont
  {Y.}~\bibnamefont {Jung}}, \bibinfo {author} {\bibfnamefont {A.}~\bibnamefont
  {Klamt}}, \bibinfo {author} {\bibfnamefont {J.}~\bibnamefont {Kong}},
  \bibinfo {author} {\bibfnamefont {D.~S.}\ \bibnamefont {Lambrecht}}, \bibinfo
  {author} {\bibfnamefont {W.}~\bibnamefont {Liang}}, \bibinfo {author}
  {\bibfnamefont {N.~J.}\ \bibnamefont {Mayhall}}, \bibinfo {author}
  {\bibfnamefont {C.~W.}\ \bibnamefont {McCurdy}}, \bibinfo {author}
  {\bibfnamefont {J.~B.}\ \bibnamefont {Neaton}}, \bibinfo {author}
  {\bibfnamefont {C.}~\bibnamefont {Ochsenfeld}}, \bibinfo {author}
  {\bibfnamefont {J.~A.}\ \bibnamefont {Parkhill}}, \bibinfo {author}
  {\bibfnamefont {R.}~\bibnamefont {Peverati}}, \bibinfo {author}
  {\bibfnamefont {V.~A.}\ \bibnamefont {Rassolov}}, \bibinfo {author}
  {\bibfnamefont {Y.}~\bibnamefont {Shao}}, \bibinfo {author} {\bibfnamefont
  {L.~V.}\ \bibnamefont {Slipchenko}}, \bibinfo {author} {\bibfnamefont
  {T.}~\bibnamefont {Stauch}}, \bibinfo {author} {\bibfnamefont {R.~P.}\
  \bibnamefont {Steele}}, \bibinfo {author} {\bibfnamefont {J.~E.}\
  \bibnamefont {Subotnik}}, \bibinfo {author} {\bibfnamefont {A.~J.~W.}\
  \bibnamefont {Thom}}, \bibinfo {author} {\bibfnamefont {A.}~\bibnamefont
  {Tkatchenko}}, \bibinfo {author} {\bibfnamefont {D.~G.}\ \bibnamefont
  {Truhlar}}, \bibinfo {author} {\bibfnamefont {T.}~\bibnamefont
  {Van~Voorhis}}, \bibinfo {author} {\bibfnamefont {T.~A.}\ \bibnamefont
  {Wesolowski}}, \bibinfo {author} {\bibfnamefont {K.~B.}\ \bibnamefont
  {Whaley}}, \bibinfo {author} {\bibfnamefont {H.~L.}\ \bibnamefont
  {Woodcock}}, \bibinfo {author} {\bibfnamefont {P.~M.}\ \bibnamefont
  {Zimmerman}}, \bibinfo {author} {\bibfnamefont {S.}~\bibnamefont {Faraji}},
  \bibinfo {author} {\bibfnamefont {P.~M.~W.}\ \bibnamefont {Gill}}, \bibinfo
  {author} {\bibfnamefont {M.}~\bibnamefont {Head-Gordon}}, \bibinfo {author}
  {\bibfnamefont {J.~M.}\ \bibnamefont {Herbert}}, \ and\ \bibinfo {author}
  {\bibfnamefont {A.~I.}\ \bibnamefont {Krylov}},\ }\href {\doibase
  10.1063/5.0055522} {\bibfield  {journal} {\bibinfo  {journal} {J. Chem.
  Phys.}\ }\textbf {\bibinfo {volume} {155}},\ \bibinfo {pages} {084801}
  (\bibinfo {year} {2021})}\BibitemShut {NoStop}%
\end{thebibliography}%
\end{document}